\documentclass[article,twocolumn,onecolappendix]{aastex61}

 \usepackage{amsmath}
  \usepackage{tabularx}
  \usepackage{color}
  \usepackage{graphicx}
  \usepackage{rotating}
  \usepackage{subfigure}

\hypersetup{linkcolor=red,citecolor=blue,filecolor=cyan,urlcolor=magenta}

\received{\today}
\revised{\today}
\accepted{ }

\submitjournal{ApJ}

\begin{document}

\title{Optical spectroscopic survey of a sample of Unidentified \textit{Fermi} objects:~II  \\}

\correspondingauthor{Simona Paiano}
\email{simona.paiano@inaf.it}

\author{Simona Paiano}
\affiliation{INAF, Osservatorio Astronomico di Padova, Vicolo dell'Osservatorio 5 I-35122 Padova - ITALY}
\affiliation{INFN, Sezione di Padova, via Marzolo 8, I-35131 Padova - ITALY}

\author{Renato Falomo}
\affiliation{INAF, Osservatorio Astronomico di Padova, Vicolo dell'Osservatorio 5 I-35122 Padova - ITALY}

\author{Aldo Treves}
\affiliation{Universit\`a degli Studi dell'Insubria, Via Valleggio 11 I-22100 Como - ITALY}
\affiliation{INAF, Osservatorio Astronomico di Brera, Via E. Bianchi 46 I-23807 Merate (Lecco), ITALY}

\author{Alberto Franceschini}
\affiliation{Dipartimento di Fisica e Astronomia, Universit\`a di Padova, Vicolo dell'Osservatorio 3, I-35 Padova - ITALY}

\author{Riccardo Scarpa}
\affiliation{Instituto de Astrofisica de Canarias, C/O Via Lactea, s/n E38205 - La Laguna (Tenerife) - SPAIN}
\affiliation{Universidad de La Laguna, Dpto. Astrofsica, s/n E-38206 La Laguna (Tenerife) - SPAIN}



\begin{abstract}
We report on optical spectroscopy obtained at the 10.4m Gran Telescopio Canarias of 28 \textit{Fermi} $\gamma$-ray sources that completes the study of a sample of 60 targets of unidentified objects for which the detection of an X-ray and/or radio source inside the 3FGL error box is available.
The observations are aimed to characterize the nature and measure the redshift of these sources. 
For all optical counterparts, the observations allow us to establish their AGN nature.
In particular, we found 24 BL Lac objects, one QSO, one NLSy1 and two objects showing spectral features typical of Seyfert~2 galaxies.  
For most of them, we determine a spectroscopic redshift, while for five we can set lower limits based on the lack of stellar features from the host galaxy.
The global properties of the full sample are briefly discussed.

\end{abstract}




\section{Introduction} 
\label{sec:intro} 

The population of jetted Active Galactic Nuclei (AGN) known as blazars are characterized by extreme properties explained by the flux amplification from relativistic bulk motions of the non-thermal emitting plasma inside the jet  oriented at a small angle with respect to the line of sight of the observer. 
A consequence of this is their huge photon emissivity at high, up to very high, photon energies ($>$100~MeV), interpreted as Inverse Compton or even photo-hadronic processes.

From the optical spectroscopic point of view, blazars are divided in two sub-classes, BL Lac Objects (BLLs) and Flat Spectrum Radio Quasars (FSRQs), depending on the strength of the broad emission lines with respect to the continuum: the latter display strong and broad emission lines, while the former are characterized by optical spectra featureless or with very weak emission/absorption lines \citep[see e.g. the review of ][]{falomo2014}.

The \textit{Fermi} observatory, which as produced an all-sky survey at such high-energies \citep{atwood2009}, is then ideally suited to detect blazars in large numbers.
In the release of the \textit{Fermi} survey catalog \citep[3FGL,][]{acero2015}, reporting 3033 detected $\gamma$-ray emitters, and in the the Third \textit{Fermi} catalog of high energy sources \citep[3FHL,][]{3fhl}, more than one thousands of them are already indicated as blazars or blazar candidates resulting as the dominant population of the extragalactic high energy sky.

On the other side, because of the \textit{Fermi} angular resolution (the median error box radius is $\sim$5 arcminutes in the 3FGL catalog), about one third of the 3FGL sources are still unassociated to counterparts in other frequency bands (the unassociated/unidentified gamma-ray sources, henceforth UGS). 
This is due either to the lack of observations at other wavelengths, especially in the X-ray band, or to the presence of many possible counterparts in their $\gamma$-ray error box.
The UGS population then not only represents a fundamental component of the high-energy sky, but is also hiding a substantial fraction of the whole blazar population \citep{mirabal2012x, massaro2012x, dabrusco2013x, acero2013x, doert2014x, landi2015, paiano2017ufo}. 
It is then of utmost interest to perform a full investigation of these sources in order to find their astrophysical counterparts.
In particular, it is of key importance to obtain spectroscopy of the optical counterparts in order to assess their classification and to determine the redshift. Since the population of UGSs is expected to be dominated by objects of BLL-type that are characterized by very weak spectral features, the use of large telescopes is mandatory \citep[see e.g. ][and reference therein]{paiano2017fgl}.

In \citet{paiano2017ufo}, here Paper~I, we defined a sample of 60 UGSs selected from the 2FGL and 3FGL catalogs and with at least one X-ray source detected inside the UGS error box.
In case that more than one X-ray source appears within the \textit{Fermi} error box, we select that coincident with a radio source. 
Once the X-ray counterpart is fixed, the association process then consists to search for counterparts of the X-ray source at radio, infrared, and optical frequencies (see details and example in Fig.~1 of Paper~I).
As variance with the above procedure, in two cases, no X-ray counterpart is available, but one/two radio sources are present. 
For these targets, we consider the likely counterpart of the gamma ray source, only the one that has also a  counterpart both at optical and IR bands. 

In Paper~I, we presented optical spectroscopy of 20 targets that allowed us to provide the source classification and to derive their redshift.
In this work, following the same association strategy of Paper~I, we report on spectra of the optical counterparts for other 28 UGSs (see Tab.~\ref{tab:table1}). The spectroscopic study is thus complete for 80\% of sources of the sample.
With this new effort, we contribute to pose the basis to start identifying interesting trends in the redshift distributions of the different blazar categories, in particular for the BLLs versus FSRQs, that were subject of a long debate \citep[e.g.][]{padovani2017, garofalo2018}. 
In Section \ref{sec:obsdata}, we present the observations and data reduction.
In Section \ref{sec:results}, we summarize our results and we report on the new redshifts. 
Section \ref{sec:notes} is devoted to detailed notes for individual sources and finally in Section \ref{sec:conclusions} we outline our main conclusions.

\section{Observations and data reduction} 
\label{sec:obsdata} 

The observations were carried out in service mode between September 2016 and January 2018 using the spectrograph OSIRIS \citep{cepa2003} at the GTC (see Tab.~\ref{tab:table2}).
The instrument was configured with the grism R1000B, covering the spectral range 4100~-~7500 $\textrm{\AA}$, and a slit width of 1.2".
The strategy of the observations and the data reduction are the same described in the Paper~I and for each source at least three individual exposures were obtained in order to remove cosmic rays and CCD cosmetic defects. All exposures were combined into a single average image. 
Wavelength calibration was performed using the spectra of Hg, Ar, Ne, and Xe lamps. 
Spectra were corrected for atmospheric extinction using the mean La Palma site extinction. 
For each source, during the same observation night, we observed a spectro-photometric standard star in order to perform the relative flux calibration.  
Also an absolute flux calibration was done using the source magnitude (reported in Tab.~\ref{tab:table2} ) estimated by a direct g filter image obtained as part of the target acquisition. 
No significant magnitude variations with respect to the literature values are found (compare Tab.~\ref{tab:table1} and Tab.~\ref{tab:table2}). 
Finally each spectrum has been dereddened for the Galaxy contribution, applying the extinction law by \citet{cardelli1989} and assuming the E(B-V) values taken from the NASA/IPAC Infrared Science Archive 6.

\section{Results} 
\label{sec:results}

Optical images of the UGSs are displayed in Fig.~\ref{fig:fc1} together with the error box of the X-ray and radio counterparts (if available). 
In Fig.~\ref{fig:fig1}, we show the flux-calibrated and de-reddened optical spectra\footnote{All spectra are also available at the online ZBLLAC database: http://www.oapd.inaf.it/zbllac/} of the counterparts of 28 3FGL $\gamma$-ray emitters. 
To emphasize weak emission and/or absorption features, we display also the normalized spectra, obtained dividing the spectrum by its continuum (see details in Paper~I).
From each normalized spectra, we evaluate the signal-to-noise (S/N) value in a number of spectral regions (see Tab.~\ref{tab:table3}). 
All spectra were carefully inspected to find emission and/or absorption lines. 
We check their reliability by comparing the features in individual exposures (see Sec.~\ref{sec:obsdata} for details).

For most of the targets (23 out of 28),  we clearly found spectral lines (see Tab.~\ref{tab:table3} and Tab.~\ref{tab:table4} ). In Fig.~\ref{fig:fig2}, we report the close up of the spectra showing the faintest detected spectral lines.

In the case of 15 objects, we detect absorption lines due to the host galaxies (e.g. Ca~II~3934,3968~$\textrm{\AA}$, G-band~4305~$\textrm{\AA}$, Mg~I~5175~$\textrm{\AA}$ and Na~I~5893~$\textrm{\AA}$), and for 10 sources, we have emission lines, mainly [O~II]~3727~$\textrm{\AA}$ and [O~III]~5007~$\textrm{\AA}$. 
For 7 targets both stellar absorptions and emission lines are detected.
In spectra of 10 objects we have not revealed intrinsic emission/absorption features. 
However for half of them we detected intervening absorption systems of Mg~II 2800~$\textrm{\AA}$ that enable us to set a spectroscopic lower limit of the redshift.
For the remaining  five targets, the optical spectra are featureless. For these sources, assuming that all BLLs are hosted by a massive elliptical galaxy and from the non-detection of the spectral absorption features of the host galaxies \citep{paiano2017tev}, we set a redshift lower limit based on the estimate of the minimum Equivalent Width (EW) (see also Tab.~\ref{tab:table3}).

From these spectroscopic observations, we classify 24 targets as BLL, one object is (3FGL~J2212.5+0703) a QSO, another (3FGL~J0031.6+0938) is Narrow Line Seyfert~1 (NLSy1) (see details in Sec.~\ref{sec:notes}).
Finally two sources (3FGLJ1234.7-0437 and 3FGLJ2358.5+3827) show a spectrum that exhibits prominent and narrow emission lines of [O~II]~3727~$\textrm{\AA}$, H$_{\beta}$, and [O~III]~5007~$\textrm{\AA}$, typical of Seyfert~2 galaxies.

\section{Notes on individual sources} 
\label{sec:notes}

\begin{itemize}
\item[] \textbf{3FGL~J0004.2+0843}: 
In the 3FGL sky region of this source, we found only one object emitting at 1.4~GHz (NVSS~J000359+084137). This coincides with the g~=~20.2 optical counterpart SDSS~J000359+084138. 
The SDSS spectrum appears featureless, although the automatic procedure proposed two tentative values of redshift (z~=~2.057 and z~=~0.187). 
 
In our spectrum, no significant emission or absorption features are found at the redshifts proposed by SDSS. The spectrum is featureless up to 6800~$\textrm{\AA}$. 
An absorption doublet at $\sim$7000~$\textrm{\AA}$ is detected that we identified as an intervening system due to Mg~II~2800~$\textrm{\AA}$ (see Fig.~\ref{fig:fig1} and Fig.~\ref{fig:fig2} ). 
This sets a spectroscopic redshift lower limit of the source at z$~>~$1.5035. 
This is the highest redshift of our UGS sample.

\item[] \textbf{3FGL~J0006.2+0135}: 
For this UGS of the 3FGL catalog, we propose the RASS source 1RXS~J000626.6+01360 as X-ray counterpart (see Fig.~\ref{fig:fc1}), positionally coincident with the radio source NVSS~J000626+013611 and in the optical SDSS~J000626.92+013610.3. 

Our spectrum is dominated by the typical non-thermal power-law continuum, characteristic of BLLs, and we detect an absorption doublet at 7030~$\textrm{\AA}$ and 7092~$\textrm{\AA}$ that we attribute to Ca~II~3934, 3968~$\textrm{\AA}$ absorption of the host galaxy at z~=~0.787. 
This detection is sound since it occurs in the clean spectral region between two atmospheric absorption bands.  
This source was also observed by SDSS and appears in DR14 release where two weak features, consistent with our detection, are present.

\item[] \textbf{3FGL~J0031.6+0938}:
Inside the \textit{Fermi} error box of this UGS, we detect the X-ray source XRT~J003159+093615, coincident with the UV source GALEXASC~J003159.86+093618.7 and the optical object SDSS~J003159.85+093618.4 of magnitude g~=~19.2. No radio counterpart is found and no informations of previous optical spectra are available in literature.

Our spectrum clearly shows prominent emission lines (H$_{\beta}$ and [O~III]~5007~$\textrm{\AA}$) typical of Seyfert QSO , at z~=~0.2207 (see Fig.~\ref{fig:fig1}).
We note that H$_{\beta}$ includes also a broad component and based on the width of this line (FWHM$\sim$1300 km/s) and on the line ratio ([O~III]~5007/~H$_{\beta}$) $\sim$~0.8 \citep[see e.g. ][]{komossa2008}, we propose this source is classified as a NLSy~1.

\item[] \textbf{3FGL~J0158.6+0102}:
The optical source SDSS~J015852.77+010132.8 is coincident with the X-ray source XRTJ015852+010129 and the radio source NVSS~J015852+010133, found inside the positional error box of this unassociated $\gamma$-ray emitter. 
The SDSS survey provides an optical spectrum showing one emission line around 7275~$\textrm{\AA}$ that is attributed to [O~II]~3727~$\textrm{\AA}$ yielding z~=~0.952 (using the automatic procedure). However, a visual inspection of the spectrum does not reveal any other plausible features corresponding to the proposed redshift.
Based on the same SDSS spectrum, \citet{massaro2014} suggested a higher value of the redshift (z~=~1.61), but no line identification was provided.
 
In our spectrum, a faint emission line is clearly detected at 7278~$\textrm{\AA}$  that we attribute to  [O~III]~5007~$\textrm{\AA}$ at z~=~0.4537, since another weaker emission line corresponding to [O~III]~4959~$\textrm{\AA}$ is also present (although it is partially contaminated by telluric absorption).

\item[] \textbf{3FGL~J0234.2-0629}:
Through the \textit{Swift}/XRT data analysis, we found only one X-ray source (XRT~J023410-062829) inside the 3FGL position error box. We propose the spatially coincident object SDSS~J023410.30-062825.7 (g~=~20.5) as the likely optical counterpart. 

Our spectrum exhibits a typical power law of the continuum and a strong (EW~=~6.9~$\textrm{\AA}$) absorption feature is detected at 4564~$\textrm{\AA}$ that we interpret as intervening absorption system due Mg~II 2800~$\textrm{\AA}$ at z~=~0.63. This represents the spectroscopic redshift lower limit of the target. No other emission or absorption lines are detected.
Recently an optical spectrum was obtained by SDSS, but no indication of redshift was given, although the intervening absorption is clearly detected.

\item[] \textbf{3FGLJ0251.1-1829}:
This UGS is also reported in the 3FHL catalog.
By the \textit{Swift}/XRT data analysis, we reveal an X-ray source (XRT~J025111-183141) inside the $\gamma$-ray error box, that is positionally coincident with the radio source NVSS~J025111-183112 and the optical counterpart USNOB1-0714-0029276. 
No previous optical spectra, classification and redshift are in literature.

We observed this source at magnitude g~=~19.1, under modest sky conditions.
Our spectrum is dominated by non thermal power law emission supporting the BLL nature of this target. 
A narrow absorption feature is detected at $\sim$4522~$\textrm{\AA}$ that we attribute as due Mg~II 2800~$\textrm{\AA}$ intervening system and set a spectroscopic redshift lower limit of the target of z~$>$~0.615.

\item[] \textbf{3FGLJ0258.9+0552}:
The possible X-ray counterpart (XRT~J025857+055243) of this 3FHL source is found by the \textit{Swift}/XRT analysis. 
We propose SDSS~J025857.55+055243.9 as optical counterpart, coincident also with the UV and radio source GALEXASC~J025857.49+055244.3 and 1WGA~J0258.9+0552. 
No informations about the nature and the redshift of the source are published in literature.

Our spectrum is featureless and clearly indicate a BLL classification of the target. Several features, visible in the bluer part of the spectrum, are consistent with DIB absorptions.
Under the assumption of an elliptical host galaxy for the BLL, we can set a lower limit of the redshift z~$>$~0.7 based on the minimum equivalent width of absorption features expected from the host galaxy \citep[details in][]{paiano2017tev}.

\item[] \textbf{3FGLJ0414.9-0840}:
The X-ray counterpart is XRT~J041433-084208 and in the optical is coincident with USNOB1-0812-0041101 and the radio source NVSS~J041433-084206.

Our optical spectrum classifies the target as a BLL because the featureless power-law continuum.
Recently another optical spectrum of the target, presented by \citet{pena2017}, confirms our classification. 
As no emission or absorption lines are detected, the redshift of this target remains still unknown.
We can set a lower limit of z~$>$~0.35 based on the non-detection of the absorption features from the starlight of the elliptical host galaxy \citep[details in][]{paiano2017tev}.

\item[] \textbf{3FGLJ0506.9+0321}:
For this 3FHL UGS, we found only the X-ray source XRT~J050650+032358 inside the \textit{Fermi} error-box,  and the optical counterpart is USNOB1-0933-0085213 coincident with the UV object GALEXASC~J050650.15+032358.7 and radio source NVSS~J050650+032401.
No previous optical spectra are available in the literature.

Our spectrum shows a featureless shape, typical of the BLL and the redshift remains unknown. 
We set a lower limit of z~$>$~0.1 according to the method outlined in \citet{paiano2017tev}, based on the lack of stellar features of the host galaxy and on the measure of the minimum EW.

\item[] \textbf{3FGLJ0848.5+7018}: 
Inside the positional $\gamma$-ray error box of this 3FHL object, we found only one radio counterpart NVSS~J084839+701727 and we propose as optical association the object USNOB1-1602-0082223, coincident with the radio and UV objects 87GB~084343.8+702846 and GALEXASC~J084839.45+701726.8.

Our optical spectrum is characterized by a continuum emission that establishes the identification as a BLL.
No significant emission lines are apparent, but we clearly detect three absorption line systems (see also Tab.~\ref{tab:table4}) at $\sim$5344~$\textrm{\AA}$ and  $\sim$5833~$\textrm{\AA}$, that we identified as intervening features due to Fe~II 2382, 2600~$\textrm{\AA}$, and at $\sim$6280~$\textrm{\AA}$  attributed to Mg~II 2800~$\textrm{\AA}$ at z~=~1.2435. 
Therefore this value represents the spectroscopic redshift lower limit of the target.
A similar case was found in our first dataset in Paper~I for the source 3FGL~J1129+3705.

\item[] \textbf{3FGLJ0930.7+5133}:
The X-ray counterpart for this UGS, found by the \textit{Swift}/XRT analysis, is XRT~J093033+513211. 
It coincides with the radio source NVSS~J093033+513216 and in the optical with SDSS~J093033.27+513214.5.
No previous optical spectra are found in literature.

In our optical spectrum, we are able to detect absorption line (see Tab.~\ref{tab:table4}) due to Ca~II~3934, 3968~$\textrm{\AA}$, G-band~3405~$\textrm{\AA}$, Mg~I~5175~$\textrm{\AA}$ and Na~I~5892~$\textrm{\AA}$, typical of the elliptical host galaxy and a weak emission line at $\sim$5955$\textrm{\AA}$ due to [O~III]~5007~$\textrm{\AA}$ that set the source at z~=~0.1893.

\item[] \textbf{3FGLJ1146.1-0640}:
This target is a hard $\gamma$-ray object present also in the 3FHL catalog.
XRT~J114600-063855 is the only X-ray source detected inside the \textit{Fermi} position error box and it is coincident with the UV source GALEXASC~J114600.86-063854.6 and the optical object USNOB1-0833-0250645.

In our spectrum, the continuum is very flat ($\alpha$~=~$-$0.25) and no emission lines are present, but we are able to detect absorption lines identified as the Ca~II~3934, 3968~$\textrm{\AA}$, H$_{\delta}$, G-band 4305~$\textrm{\AA}$ and H$_{\gamma}$, due to the stellar emission of the host galaxy at z~=~0.6407.

\item[] \textbf{3FGLJ1223.3+0818}:
We obtain the optical spectrum of the X-ray counterpart source XRT~J122327+082031, likely associated to the $\gamma$-ray emitter, that establishes the BLL nature for this source. 
The dominant feature is an intervening absorption system at $\sim$4810~$\textrm{\AA}$ (see Tab.~\ref{tab:table4}) due to Mg~II 2800~$\textrm{\AA}$ at z~=~0.7187 which is the spectroscopic redshift lower limit of the target.

\item[] \textbf{3FGLJ1234.7-0437}: 
From the \textit{Swift}/XRT data analysis, for this 3FHL $\gamma$-ray emitter, we found only one bright X-ray source (XRT~J123448-043246) within the $\gamma$-ray error box, coincident with the optical source USNOB1-0854-0232216 (g~=~18.6) and the UV object GALEXASC~J123448.20-043244.6 (F$_{\nu}$$\sim7\times10^{-06}$ Jy).
No radio emission is found at the position of the source.
Based on few lines from an uncalibrated optical spectrum by the 2dF Galaxy Redshift survey, the redshift of 0.277 was proposed. 

Our high quality optical spectrum clearly exhibits emission lines of [O~II]~3727~$\textrm{\AA}$, H$_{\beta}$, and [OIII]~5007~$\textrm{\AA}$ at z~=~0.2765, together with stellar absorption lines of its host galaxy (see also Tab.~\ref{tab:table4}).  
The emission line properties of this object are typical of Seyfert~2 galaxies.

\item[] \textbf{3FGLJ1258.4+2123}:
We found an X-ray source XRT~J125821+212350, within the 3FGL error box that coincides with SDSS~J125821.45+212351.3 (g=20.5).

Our optical spectrum shows the BLL non thermal continuum with superimposed stellar features of its host galaxy: Ca~II~3934, 3968~$\textrm{\AA}$ at 6398~$\textrm{\AA}$ and 6455~$\textrm{\AA}$, and the G-band~4305~$\textrm{\AA}$ at 7001~$\textrm{\AA}$. 
Therefore the redshift of the source is z~=~0.6265.

\item[] \textbf{3FGLJ1525.8-0834}:
The analysis of \textit{Swift}/XRT data reveals one X-ray object (XRT~J152603-083147) in the 3FGL error box, that is spatially coincident with the optical source USNOB1-0814-0287412 (g~=~18.3 ),  the UV object  GALEXASC~J152603.17-083146.0 (F$_{\nu}$$\sim 8\times10^{-06}$ Jy) and the radio source NVSS~J152603-083146 (F$_{\nu}$$\sim 2.5\times10^{-01}$ Jy)). 

Our optical spectrum is featureless and characterized by a power law emission ($\alpha$~=~$-$0.85) and establishes the BLL nature.
A redshift lower limit z~$>$~0.40 can be set on the basis on the minimum equivalent width method \citep{paiano2017tev}.

\item[] \textbf{3FGLJ1541.6+1414}:
Inside the 3FGL sky region of this target, we found an X-ray source XRT~J154150+141442 that is associated to the optical source SDSS~J154150.09+141437.5 (g~=~18.4). 
The SDSS survey provided a spectrum of the optical counterpart characterized by blue continuum emission superposed by a number of relevant absorption features: G-band~(4305), Mg~I~(5175), and the weak doublet of Ca~II~(3934, 3968) at z~=~0.223.

Our spectrum was obtained with a narrower slit of 1.2", compared with that of SDSS, and is more dominated by non thermal emission with relevant blue (at $\lambda<$5000~$\textrm{\AA}$) component. Weak (EW~=~1.1~$\textrm{\AA}$) emission line at 6123$\textrm{\AA}$ attributed to [O~III]~5007~$\textrm{\AA}$ and the absorption doublet of Ca~II~3934, 3968~$\textrm{\AA}$ confirm the redshift z~=~0.223.

\item[] \textbf{3FGLJ2150.5-1754}:
Through the \textit{Swift}/XRT data analysis, we found only one X-ray source (XRT~J215046-174957) within the 3FGL error box. 
We propose the spatially coincident objects USNOB1-0721-1160290 (g~=~17.6) and NVSS~J215046-174954 as optical and radio counterparts. 

Our optical spectrum is dominated by a galactic component with the presence of moderate non thermal emission. Clear absorption lines of the overall stellar population are detected, in particular Ca~II~3934, 3968~$\textrm{\AA}$, G-band~4305~$\textrm{\AA}$, Mg~I~5157~$\textrm{\AA}$ and Na~I~5893~$\textrm{\AA}$ at z~=~0.1855.

\item[] \textbf{3FGLJ2209.8-0450}:
The source XRT~J220941-045108 is proposed as counterpart in the X-ray band of the target and it is associated to the radio source NVSS~J220941-045111 and the optical source SDSS~J220941.7-045110.2 (g~=~18.8).
There are not informations about the classification and redshift in literature.

Our optical spectrum exhibits a typical BLL continuum and is characterized by signature of stellar component from the host galaxy (Ca~II~3934, 3968~$\textrm{\AA}$ doublet and G-band~4305~$\textrm{\AA}$ absorption) at z~=~0.3967. 
In addition a weak possible emission line due to  [O~II]~3727~$\textrm{\AA}$ at 5205~$\textrm{\AA}$ is visible.

\item[] \textbf{3FGLJ2212.5+0703}:
On the basis of the XRT data, we propose that the X-ray and optical counterparts of this UGS are XRT~J221231+070651 and  SDSS~J221230.98+070652.4 (g~=~20.6). 
No radio source is found as counterpart and no previous optical spectra are present in the literature. 

Our optical spectrum clearly shows a prominent and strong (EW~=~116~$\textrm{\AA}$) emission line at 5600~$\textrm{\AA}$. 
If attributed to Mg~II~2800~$\textrm{\AA}$, the redshift is z~=~1.00. 
Based on this spectrum, the target is classified as QSO.

\item[] \textbf{3FGLJ2228.5-1636}:
The X-ray counterpart XRT~J222830-163642 is found by the \textit{Swift}/XRT analysis. 
We propose as optical counterpart USNOB1-0733-0885778, coincident with the UV source GALEXASC~J222830.19-163642.7 and the radio source NVSS~J222830-163643.
No previous spectra or informations are available in literature for this target.

Our optical spectrum shows a power law shape ($\alpha$~=~$-$0.90) confirming the BLL nature of the source. We note a weak signature of Ca~II~3934, 3968~$\textrm{\AA}$ at $\sim$6020~$\textrm{\AA}$ that suggests a tentative redshift of z~=~0.525.

\item[] \textbf{3FGLJ2229.1+2255}:
We found only one X-ray source (XRT~J222911+225458) within the 3FGL error region of the target, that can be associated to the optical source SDSS~J222911.17+225459.7 (g~=~18.9). 
No radio emission is found at the position of the target.

Our optical spectrum shows a characteristic non thermal emission with weak Ca~II~3934, 3968~$\textrm{\AA}$ and G-band~4305~$\textrm{\AA}$ absorption lines  (see Tab.~\ref{tab:table4} and Fig.~\ref{fig:fig2}) at z~=~0.440

\item[] \textbf{3FGLJ2244.6+2503}:
From the XRT analysis, we detect the source XRT~J224436+250343 inside the $\gamma$-ray error box that is coincident with radio and optical objects NVSS~J224436+250345 and SDSS~J224436.70+250342.6.
An optical spectrum was recently provided by SDSS that appears featureless. 

We obtained a good (SNR$\sim$120) optical spectrum that is characterized by a power-law continuum with spectral index ($\alpha$=$-$1.0). 
This supports the BLL classification for the source.  
Two weak absorption features are detected at $\sim$6500$\textrm{\AA}$, partially contaminated by a small telluric absorption. They are consistent with Ca~II~3934, 3968~$\textrm{\AA}$ absorption doublet at z~=~0.650. 
No other absorption or emission lines are found at this redshift. 
We consider this redshift as tentative.

\item[] \textbf{3FGLJ2246.2+1547}:
This source is associated to the radio source NVSS~J224604+154437 and the optical source SDSS~J224604.99+154435.3. 
This source was present in our first dataset Paper~I and was classified as BLL on the basis of the optical spectrum that appeared without evident emission/absorption lines.

The source was re-observed in August 2017 with a better grism resolution and the new spectrum (see Fig.~\ref{fig:fig2}) reveals an emission line (EW~=~1.6~$\textrm{\AA}$) at 5949~$\textrm{\AA}$ identified as due to [O~II]~3727~$\textrm{\AA}$ at z~=~0.5965.  Also a weak absorption doublet at $\sim$6300~$\textrm{\AA}$, that we identified as Ca~II~3934, 3968~$\textrm{\AA}$ at the same redshift, is present although it is contaminated by the telluric absorption.

\item[] \textbf{3FGLJ2250.3+1747}:
We obtain the optical spectrum of the source SDSS~J225032.7+174914.9, coincident with the only X-ray counterpart  XRT~J225031+124916 found inside its  $\gamma$-ray error position. 
This coincides also with the radio source NVSS~J225032+174914.

The spectrum is characterized by several features due to the host galaxy. 
In particular we detect absorption lines due to Ca~II~3934, 3968~$\textrm{\AA}$, H$_{\delta}$, G-band~4305~$\textrm{\AA}$, H$_{\gamma}$ and H$_{\beta}$, together with emission lines attributed to Ne~V~3426~$\textrm{\AA}$, [O~II]~3727~$\textrm{\AA}$ and [O~III]~4959, 5007~$\textrm{\AA}$ (see Tab.~\ref{tab:table4}).
Therefore, the redshift of the source is z~=~0.3437.

\item[] \textbf{3FGLJ2321.6-1619}:
We propose the source XRT~J232136-161926 as X-ray counterpart for this UGS, coincident with the optical object USNOB1-0736-0835813 (g~=~16.9) and the radio NVSS~J232137-161935.
No previous optical spectra are found in the literature and its classification remains uncertain.

Our spectrum establishes the BLL nature of the source and exhibits the characteristic power law continuum ($\alpha$~=~$-$1.15). 
We detect the absorption doublet attributed to Ca~II~3934, 3968~$\textrm{\AA}$ at 6663~$\textrm{\AA}$ and 6722~$\textrm{\AA}$, that locates the source at z~$=$~0.6938.

\item[] \textbf{3FGLJ2358.6-1809}:
XRT~J235836-180718 is the only \textit{Swift} source detected inside the 3FGL position error of the target and is considered the X-ray counterpart. 
It is associated to the optical object USNOB1-0718-1032041 (g~=~17.7) and the radio NVSS~J235836-180718.
On the basis of an optical spectrum secured by 6dF Survey a redshift of z~=~0.0575 was proposed. However, no clear features are visible in that spectrum and the redshift is dubious.

Our optical spectrum (S/N~130) appears featureless, establishing the BLL nature of this source. 
Based on the assumption of the typical BLL host galaxy is a giant elliptical galaxy, by the minimum EW method outlined in \citet{paiano2017tev}, we can set a lower limit of the redshift of $>$~0.25.

\item[] \textbf{3FGLJ2358.5+3827}:
The source is classified as UGS also in the 3FHL catalog. 
Inside the Fermi error box there is only one X-ray source (XRT~J235825+382857)  and the radio source  (NVSS~J235825+382857) that also coincides with the optical object USNOB1-1284-0547072 (g~=~18.4). 
No optical spectra of this object were found available in literature.

Our optical spectrum shows a number of narrow emission lines of [O~II]~3727~$\textrm{\AA}$, [Ne~III]~3869~$\textrm{\AA}$, H$_{\beta}$ and [O~III]~4959, 5007~$\textrm{\AA}$ and the typical absorption features due to the stellar population of its host galaxy (see Tab.~\ref{tab:table4}), yielding the redshift z~=~0.2001.
The emission line properties of this object are typical of Seyfert~2 galaxies.

\end{itemize}

\section{Conclusions} 
\label{sec:conclusions} 

We secured optical spectroscopy of 28 UGSs that, together with other 20 sources previously investigated \citep{paiano2017ufo}, led to almost complete (at 80\% level) the optical investigation of the sample of unidentified $\gamma$-ray sources defined in Paper~I.
This allows us to characterize and to classify all these sources, and to measure the redshift for most of them.

From the analysis of the spectroscopic properties of all observed targets (47 out of 60), we find that:
\begin{itemize}
\item all targets exhibit an AGN optical spectrum: 44 are classified as BLLs, one is QSO, two objects have a Seyfert~2 type spectrum and in one case the source is classified as NLSy1;

\item for 39 objects we measure their spectroscopic redshift: 29 sources exhibit emission and/or absorptions lines, allowing us a firm measurement of the redshift, while for 10 objects we set a spectroscopic lower limit of the redshift on the basis of the detection of absorption lines from intervening systems due to Mg~II~2800~$\textrm{\AA}$ and/or  Fe~II~2382, 2600~$\textrm{\AA}$;

\end{itemize}

The redshift range of our UGS dataset spans from z~$=$~0.171 up to z~$=$~0.787. 
The averaged value is $<$z$>$~$=$~0.45.
For the 10 objects with a spectroscopic lower limit, the range is between 0.38 and 1.5.
We note that the source 3FGL~J0004.2+0843, that is found at z~$>$~1.5035, is among the farthest BLL known detected by the \textit{Fermi} satellite \citep[see also][]{paiano2017fgl} and it is one of the few BLL known at z~$>$~1.

Although the optical spectroscopy confirms that the majority of the sources ($\sim$90\%) are blazars (as expected), in three cases their spectra indicate a different classification. 
In particular, the source 3FGL~J0031.6+0938, showing a broad component of the H$_{\beta}$ emission line, can be classified as a NLSy1 and led to six the total number of known objects that are classified as NLSy1 in the 3FGL catalog.
The other two sources, that are not classified as blazars, show an optical spectrum typical of Seyfert~2 galaxies with only narrow emission lines. 
It is worth to note that in the current \textit{Fermi} catalog, there are only 2 other sources that exhibit a Seyfert~2  optical spectrum \citep{lenain2010, ackermann2012}. The latter (NGC~1068 and NGC~4945) are at much lower redshift (z~=~0.0038 and z~=~0.0019, respectively) with respect to the redshift (z~$\sim$~0.2-0.3) of the Sy2-type sources found in this work.

Only 9 sources of our sample have a featureless optical spectrum and for them we set a redshift lower limits (see Tab.~\ref{tab:table3} of this work and of Paper~I), based on the minimum detectable equivalent width technique and on the lack of detection of host galaxy absorption lines, assuming that a BLL host galaxy with M(R) = -22.9 \citep[see details in ][]{paiano2017tev}.

Finally, we note that 25 targets of our whole sample are also reported in the 3FHL catalog, included the two Seyfert~2-like sources and 16 BLLs with a measurement of the redshift (five with high redshift $>$~0.5), therefore they can be considered as interesting targets for future observations with the next generation TeV Cherenkov telescopes.

\begin{table*}
\caption{THE SAMPLE  }\label{tab:table1}
\begin{tabular}{lllllllcl}
\hline 
3FGL Name  &    Optical Counterpart     &  RA (J2000)      &    DEC (J2000)     &  3FHL &  mag  &  E(B-V)  &  $z$  & Reference \\             
\hline
3FGLJ0004.2+0843 & SDSSJ000359+084138  & 00:03:59.23 & +08:41:38.2  & n   & 20.2 & 0.0614 & 2.06-0.19? &    SDSS \\  
3FGLJ0006.2+0135 & SDSSJ000626+013610  & 00:06:26.92 & +01:36:10.3  & n  & 20.3 & 0.0264 & ?        &  \\
3FGLJ0031.6+0938 & SDSSJ003159+093618  & 00:31:59.86 & +09:36:18.4  & n   & 19.5  & 0.0505  & ?       &  \\
3FGLJ0158.6+0102 & SDSSJ015852+010132  & 01:58:52.77 & +01:01:32.8  & n   & 21.1 & 0.0225  & 0.952? & SDSS  \\ 
3FGLJ0234.2-0629 & SDSSJ023410-062825    & 02:34:10.27 & -06:28:25.6  & n    & 20.5 & 0.0225 &  ? & SDSS \\
3FGLJ0251.1-1829 & USNOB1-0714-0029276  & 02:51:11.55 & -18:31:14.3  & y  & 19.5 & 0.0237 &  ? &        \\
3FGLJ0258.9+0552 & SDSSJ025857+055243 & 02:58:57.55 & +05:52:43.9 & y  & 18.9 & 0.1196  & ? &       \\              
3FGLJ0414.9-0840 & USNOB1-0812-0041101 & 04:14:33.08 & -08:42:06.7 & n  &  19.3 & 0.0405 & ? &       \\
3FGLJ0506.9+0321 & USNOB1-0933-0085213 & 05:06:50.14 & +03:23:58.6 &  y  &   18.4 & 0.0593 & ? &        \\
3FGLJ0848.5+7018 & USNOB1-1602-0082223 & 08:48:39.52 & +70:17:28.0 & y  &   19.7 & 0.0274 & ? &       \\
3FGLJ0930.7+5133 & SDSSJ093033+513214 &  09:30:33.36 & +51:32:14.6 &  n  &   20.0 & 0.0149 &  ? &    \\
3FGLJ1146.1-0640 & USNOB1-0833-0250645 & 11:46:00.96 & -06:38:54.8 & y   & 19.7 & 0.0285 & ? &       \\
3FGLJ1223.3+0818 & SDSSJ122327+082030 &  12:23:27.49 & +08:20:30.4 & n  &   19.8 & 0.0199 &  ? &     \\ 
3FGLJ1234.7-0437 & USNOB1-0854-0232216 & 12:34:48.00 & -04:32:46.2 & y &  19.8 & 0.0322 & 0.277 &   NED \\
3FGLJ1258.4+2123 & SDSSJ125821+212351 & 12:58:21.45 & +21:23:51.0 &  n  &     20.6 & 0.0342 &  ? &       \\ 
3FGLJ1525.8-0834 & USNOB1-0814-0287412 & 15:26:03.17 & -08:31:46.4 &  n  &  18.8 & 0.0796 & ? &        \\
3FGLJ1541.6+1414 & SDSSJ154150+141437 &  15:41:50.16 & +14:14:37.6  & y  & 17.2 & 0.0407 & 0.2230 &   SDSS  \\
3FGLJ2150.5-1754 & USNOB1-0721-1160290 & 21:50:46.43 & -17:49:54.5 & n &   18.2 & 0.0435 & ? &       \\
3FGLJ2209.8-0450 & SDSSJ220941-045110  & 22:09:41.70 & -04:51:10.3 & y &    18.8 & 0.0613 & ? &       \\
3FGLJ2212.5+0703 & SDSSJ221230+070652 & 22:12:30.98 & +07:06:52.5 &  n  &  19.9 & 0.0642 & ?        \\
3FGLJ2228.5-1636 & USNOB1-0733-0885778 & 22:28:30.18 & -16:36:43.0 &  n  &  18.7 & 0.0295  & ?        \\
3FGLJ2229.1+2255 & SDSSJ222911+225459 & 22:29:11.18 & +22:54:59.9 &  n  &   18.9 & 0.0455 & ?        \\
3FGLJ2244.6+2503 & SDSSJ224436+250342 & 22:44:36.70 & +25:03:42.6 & n  &  19.3 & 0.0573  & ?        \\
3FGLJ2246.2+1547* & SDSSJ224604+154435 &  22:46:04.90 &+15:44:35.5  & y &  19.3 & 0.0673 &  ?    & \citet{paiano2017ufo} \\   
3FGLJ2250.3+1747 & SDSSJ225032+174914 & 22:50:32.88 & +17:49:14.8 &  n &  19.9 & 0.0675 &  ? &        \\
3FGLJ2321.6-1619 & USNOB1-0736-0835813 & 23:21:37.01 & -16:19:28.5 & n  &   17.9 & 0.0239 &  ? &       \\
3FGLJ2358.6-1809 & USNOB1-0718-1032041 & 23:58:36.72 & -18:07:17.4 & y   & 17.45 & 0.0214 & 0.0575? & NED \\
3FGLJ2358.5+3827 & USNOB1-1284-0547072 &  23:58:25.17 & +38:28:56.4 & y  &  19.0 & 0.1051 &  ? &        \\
\hline
\end{tabular}
\tablenotetext{}{
\raggedright
\footnotesize \texttt{Col.1}: Name of the target; \texttt{Col.2}: Optical counterpart of the target; \texttt{Col.3 - 4 }: Right ascension and declination of the optical counterpart; \texttt{Col.5}: Source reported in the 3FHL catalog; \texttt{Col.6}: g magnitude from SDSS (for the SDSS sources) or B-band magnitudes taken from USNOB1.0 catalog; \texttt{Col.7}: $E(B-V)$ taken from the NASA/IPAC Infrared Science Archive (https://irsa.ipac.caltech.edu/applications/DUST/); \texttt{Col.8}: Redshift reported in literature; \texttt{Col.9}: Reference for the redshift. \\
(*) This source was also discussed in \citet{paiano2017ufo} } 
\tablenotetext{}{
\raggedright
 } 
\end{table*}

\newpage
\begin{table*}
\caption{LOG OF THE OBSERVATIONS }\label{tab:table2}
\centering
\begin{tabular}{lllll}
\hline
\hline
Obejct          &  Obs. date  & t$_{Exp}$ (s)  &   Seeing ('') & g  \\
\hline
3FGLJ0004.2+0843    &  05-10-2016     &  3600     &    1.3  & 20.2   \\  
3FGLJ0006.2+0135    &  06-10-2016     &  7200     &    2.5    & 20.3  \\  
3FGLJ0031.6+0938    &  03-12-2017     &  3600     &   1.6 & 19.2      \\  
3FGLJ0158.6+0102    &  25-09-2016     &  8750     &   1.5   &  21.4   \\  
3FGLJ0234.2-0629     &  05-10-2016     & 10200    &    2.5 &  19.7    \\  
3FGLJ0251.1-1829     &  26-09-2017     & 2000      &    1.3 &   19.1   \\  
3FGLJ0258.9+0552    &  15-09-2016     & 3600      &    1.5 &  19.2    \\  
3FGLJ0414.9-0840     &  16-09-2016     & 9000      &    2.0  &  19.2   \\  
3FGLJ0506.9+0321    &  15-09-2016     & 1500      &    1.3  &   18.8  \\  
3FGLJ0848.5+7018    &  10-12-2016     & 5100      &   1.9 &   19.9    \\  
3FGLJ0930.7+5133    &   27-01-2018    & 6000      &   1.3  &  20.1    \\  
3FGLJ1146.1-0640     &   29-12-2016    & 15000    &  1.8  &  19.3     \\  
3FGLJ1223.3+0818    &   24-03-2017    & 7500     &  1.5  &  18.9     \\  
3FGLJ1234.7-0437     &   12-04-2017    & 7500     &  3.0 &  18.6      \\  
3FGLJ1258.4+2123    &   03-04-2017    & 7500     &  1.6  & 20.5      \\  
3FGLJ1525.8-0834     &   02-04-2017    & 3000     &  1.8  &  18.3     \\  
3FGLJ1541.6+1414    &   13-04-2017    & 1500     &   0.8 &   18.4   \\  
3FGLJ2150.5-1754     &   08-12-2017    & 1800     &   2.5 &  17.6    \\  
3FGLJ2209.8-0450     &   22-09-2017    & 3000     &   1.6   &  18.8  \\           
3FGLJ2212.5+0703    &   26-09-2017    & 2400     &   0.9   & 20.6   \\  
3FGLJ2228.5-1636     &   22-06-2017    & 2500     &  1.2  &  18.8   \\  
3FGLJ2229.1+2255    &   29-09-2017    & 3000     &  0.9  &   18.9  \\  
3FGLJ2244.6+2503    &   22-06-2017    & 8500     &  1.3  & 18.9   \\  
3FGLJ2246.2+1547    &   11-08-2017    & 7200     &  0.9  &  19.1 \\  
3FGLJ2250.3+1747    &   20-09-2017    & 3600     &  1.8  &   20.2 \\ 
3FGLJ2321.6-1619     &   26-09-2017    & 7200     &  1.1   &  16.9 \\  
3FGLJ2358.6-1809     &   29-09-2017    & 9000     &  2.0  &  17.7 \\  
3FGLJ2358.5+3827    &   24-09-2017    & 4500     &   2.0 &   18.4 \\  
\hline
\end{tabular}
\tablenotetext{}{
\raggedright
\footnotesize \texttt{Col.1}: Name of the target; \texttt{Col.2}: Date of observation;  \texttt{Col.3}: Total integration time; \texttt{Col.4}: Seeing during the observation; \texttt{Col.5}: g magnitude measured from the acquisition image.}
\end{table*}

\newpage
\begin{table*}
\caption{PROPERTIES OF THE OPTICAL SPECTRA }\label{tab:table3}
\centering
\begin{tabular}{lclcll} 
\hline
OBJECT           & $\alpha$  &   SNR       &   EW$_{min}$            &  Class & z           \\
\hline                                                                                                                                        
3FGLJ0004.2+0843    &  -0.95    & 50     &  0.40 - 0.80   &  BLL &    $>$1.5035$^{a}$  \\ 
3FGLJ0006.2+0135    &  -0.90    & 35     & 0.75 - 1.05    &  BLL  & 0.787$^{g}$\\  
3FGLJ0031.6+0938    &   -1.90   & 30    & 0.55 - 1.85   &  NLSY1 & 0.2207$^{e}$ \\  
3FGLJ0158.6+0102    &  -1.00    & 15      &   2.2 - 3.5     &  BLL  & 0.4537$^{e}$ \\  
3FGLJ0234.2-0629     &  -1.25    &  90  &  0.35 - 0.50  &  BLL & $>$0.63$^{a}$\\  
3FGLJ0251.1-1829     & -1.05     &  30    &  0.85 - 1.70  &  BLL &  $>$0.615$^{a}$\\  
3FGLJ0258.9+0552    &   -1.45   & 150   &  0.20 - 0.40  & BLL & $>$0.7$^{h}$ \\  
3FGLJ0414.9-0840     &  -0.70    & 40      &  0.70 - 1.70  &  BLL & $>$0.35$^{h}$\\  
3FGLJ0506.9+0321    &   -1.15   &   30    &   0.80 - 2.80  & BLL  & $>$0.1$^{h}$ \\  
3FGLJ0848.5+7018    &   -1.10   & 45     &  0.50 - 0.75  &  BLL  & $>$1.2435$^{a*}$\\  
3FGLJ0930.7+5133    &   *          &  30     &  0.80 - 3.00  &   BLL & 0.1893$^{eg}$\\  
3FGLJ1146.1-0640     &    -0.25   &  40   &  0.75 - 1.05  &  BLL  & 0.6407$^{g}$\\  
3FGLJ1223.3+0818    &   -1.15   & 135  &  0.20 - 0.30  &   BLL & $>$0.7187$^{a}$\\  
3FGLJ1234.7-0437     &   -0.40   &   30   & 1.25 - 3.75   &  Sy2 & 0.2765$^{eg}$\\  
3FGLJ1258.4+2123    &  -0.40    & 40      &  0.65 - 0.80  &  BLL & 0.6265$^{g}$\\  
3FGLJ1525.8-0834     &   -0.85   &  105   & 0.25 - 0.30   & BLL  & $>$0.40$^{h}$ \\  
3FGLJ1541.6+1414    &    -0.60  &  60     & 0.45 - 0.70   & BLL & 0.223$^{eg}$\\  
3FGLJ2150.5-1754     & *            & 50      &  0.55 - 1.25  &  BLL & 0.1855$^{g}$\\  
3FGLJ2209.8-0450     &  -0.80    &  75   & 0.40 - 0.65   & BLL & 0.3967$^{eg}$ \\    
3FGLJ2212.5+0703    &    -1.15   &  15    &  1.35 - 1.90  & QSO & 1.00$^{e}:$ \\  
3FGLJ2228.5-1636     &    -0.90  &  55     &  0.50 - 1.10  & BLL & $\sim$0.525$^{g}$ :\\  
3FGLJ2229.1+2255    &  -1.30    &  80     &  0.35 - 0.40  &  BLL & 0.440$^{g}$\\  
3FGLJ2244.6+2503    &   -1.05   & 160  &  0.20 - 0.25  &  BLL & 0.650$^{g}$:\\  
3FGLJ2246.2+1547    &   -0.80   &  60     &  0.45 - 1.50  & BLL & 0.5965$^{eg}$ \\  
3FGLJ2250.3+1747    &    *         &  25     &  1.30 - 2.25  &  BLL & 0.3437$^{eg}$\\  
3FGLJ2321.6-1619     &   -1.15   &   135 &  0.20 - 0.35  &  BLL & 0.6938$^{g}$\\  
3FGLJ2358.6-1809     &   -1.40   &   130  &   0.20 - 0.35 &  BLL &  $>$0.25$^{h}$\\  
3FGLJ2358.5+3827    &   *          &   80  &  0.40 - 0.75  &  Sy2 & 0.2001$^{eg}$ \\  
\hline
\end{tabular}
\tablenotetext{}{
\raggedright
\footnotesize \texttt{Col.1}: Name of the target; \texttt{Col.2}: Optical spectral index derived from a power law fit of the continuum (the spectral index is estimated only for the sources that exhibit an optical spectrum dominated by the non-thermal emission); \texttt{Col.3}: Averaged S/N of the spectrum; \texttt{Col.4}: Range of the minimum equivalent width (EW$_{min}$) derived from different regions of the spectrum; \texttt{Col.5}: Classification derived by our optical spectrum; \texttt{Col.6}: Spectroscopic redshift. The superscript letters are: \textit{e} = emission line, \textit{g} = galaxy absorption line, \textit{a}= intervening absorption assuming Mg~II 2800$\textrm{\AA}$ identification, \textit{h}= lower limit derived on the lack of detection of host galaxy absorption lines assuming a BLL elliptical host galaxy with M(R) = -22.9 \citep[see details in][]{paiano2017tev}.\\
(:) This marker indicates that the redshift is tentative. \\(*) For this source we found other two absorption line systems due to Fe~II~(2382, 2600) (See details text). }
\end{table*}


\setcounter{table}{3}                                   
\begin{table*}
\caption{MEASUREMENTS OF THE SPECTRAL LINES } \label{tab:table4}
\centering
\begin{tabular}{lllll}
\hline
Object name          &  $\lambda$    &    EW   &     Line ID    &   z   \\
\hline                                
3FGLJ0004.2+0843    &  7000   &  0.9   & Mg~II 2796~$\textrm{\AA}$   &   $>$1.5035  \\  
                                    &  7016   &  0.5 &  Mg~II 2803~$\textrm{\AA}$   &   $>$1.5035  \\  
\hline
3FGLJ0006.2+0135    &  7030   &  3.1  & Ca~II 3934~$\textrm{\AA}$    & 0.787\\   
                                    &  7092   &  3.3  & Ca~II 3968~$\textrm{\AA}$    & 0.787\\  
\hline
3FGLJ0031.6+0938    &  4549   &  1.7     &  [O~II] 3727~$\textrm{\AA}$  & 0.2207 \\  
                                    &  4723   &  3.9     &  [Ne~III] 3869~$\textrm{\AA}$   & 0.2207 \\  
                                    &  4846   &  3.1     &  H$_{\epsilon}$ 3970~$\textrm{\AA}$ & 0.2207 \\  
                                    &  5007   &  6.7     &  H$_{\delta}$  4102~$\textrm{\AA}$ & 0.2207 \\  
                                    &  5298   &  19.6   &  H$_{\gamma}$  4340~$\textrm{\AA}$ & 0.2207 \\  
                                    &  5934  & 64.8     &  H$_{\beta}$  4861~$\textrm{\AA}$ & 0.2207 \\  
                                    &  6053  & 13.5     &  [O~III] 4959~$\textrm{\AA}$   & 0.2207 \\  
                                    &  6112   & 36.8    &  [O~III] 5007~$\textrm{\AA}$  & 0.2207 \\  
                                    &  7172   &  5.0   &  He~I   5876~$\textrm{\AA}$ & 0.2207 \\  
\hline
3FGLJ0158.6+0102    &  7213*  & 3.5    &  [O~III] 4959~$\textrm{\AA}$  & 0.454\\ 
                                    &  7278   & 7.9    &  [O~III] 5007~$\textrm{\AA}$  & 0.4537\\
\hline
3FGLJ0234.2-0629     &  4564    & 6.9  &  Mg~II 2800~$\textrm{\AA}$   &  $>$0.63\\  
\hline 
3FGLJ0251.1-1829     &  4522   & 3.5  &  Mg~II 2800~$\textrm{\AA}$  &   $>$0.615 \\  
\hline
3FGLJ0848.5+7018    &  5344    &  2.2 &  Fe~II 2382~$\textrm{\AA}$  & $>$1.2435  \\  
                                    &  5833    &  2.1  & Fe~II 2600~$\textrm{\AA}$    &  $>$1.2435 \\ 
                                    &  6273    &  6.0  &  Mg~II 2796~$\textrm{\AA}$  & $>$1.2435 \\ 
                                    &  6289    &  5.3  &  Mg~II 2803~$\textrm{\AA}$  & $>$1.2435 \\ 
\hline
3FGLJ0930.7+5133    &  4678    &  8.9  & Ca~II 3934~$\textrm{\AA}$   &  0.1893\\
                                    &  4720    &  6.9  & Ca~II 3968~$\textrm{\AA}$   &  0.1893\\
                                    &  5119    &  4.1 & G-band 4305~$\textrm{\AA}$  &  0.1893\\
                                    &  5955   &  1.4 & [O~III] 5007~$\textrm{\AA}$  &  0.1893\\
                                    &  6155    & 2.1   & Mg~I 5175~$\textrm{\AA}$   &  0.1893\\
                                    &   7008   &  1.2  & Na~I 5892~$\textrm{\AA}$   &  0.1893\\
\hline  
3FGLJ1146.1-0640     & 6454     &   2.2   &  Ca~II 3934~$\textrm{\AA}$  & 0.6407\\  
                                    & 6511     &   3.2  &  Ca~II 3968~$\textrm{\AA}$  & 0.6407\\  
                                    & 6730     &   0.8  &  H$_{\delta}$ 4102~$\textrm{\AA}$   & 0.6407\\  
                                    & 7062     &   1.5  &    G-band 4305~$\textrm{\AA}$ & 0.6407\\  
\hline
3FGLJ1223.3+0818    & 4805    & 0.5    & Mg~II 2796~$\textrm{\AA}$    &  $>$0.7187\\  
                                    & 4818    & 0.4  &  Mg~II 2803~$\textrm{\AA}$  & $>$0.7187 \\ 
\hline
\end{tabular}
\tablenotetext{}{
\raggedright
\footnotesize \texttt{Col.1}: Name of the target; \texttt{Col.2}: Barycenter of the detected line; \texttt{Col.3}: Measured equivalent width; \texttt{Col.4}: Line identification; \texttt{Col.5}: Spectroscopic redshift.\\
(*) The marker indicates that the line is partially contaminated by telluric band.}
\end{table*}

\newpage
\setcounter{table}{3}                                   
\begin{table*}
\caption{MEASUREMENTS OF THE SPECTRAL LINES \textit{(continued)}} 
\centering
\begin{tabular}{lllll}
\hline
Object name          &  $\lambda$    &    EW   &     Line ID    &   z   \\
\hline
3FGLJ1234.7-0437      & 4373    & 7.7     &   [Ne~V]  3426~$\textrm{\AA}$   & 0.2765\\  
                                     & 4756    & 18.1    &   [O~II] 3727~$\textrm{\AA}$  & 0.277\\  
                                     & 4938    &  7.4    &   [Ne~III] 3869~$\textrm{\AA}$ & 0.2765\\  
                                     & 5023    &  3.1    &   Ca~II 3934~$\textrm{\AA}$ & 0.277\\  
                                     & 5068    &   5.6   &   Ca~II 3968~$\textrm{\AA}$ & 0.277\\  
                                     & 5497    &  1.2    &   G-band 4305~$\textrm{\AA}$  & 0.2775\\  
                                     & 6205   &   7.0   &   H$_{\beta}$ 4861~$\textrm{\AA}$ & 0.2765\\  
                                     & 6328   & 11.4     &   [O~III] 4959~$\textrm{\AA}$ & 0.2765\\  
                                     & 6391    &  28.7    &   [O~III] 5007~$\textrm{\AA}$ & 0.2765\\  
                                     & 6605    & 2.5     &   Mg~I 5175~$\textrm{\AA}$ & 0.2765\\
                                     & 6727    & 1.4     &   Ca+Fe 5269~$\textrm{\AA}$ & 0.2765\\
                                     & 7522   &  3.0    &   Na~I 5892~$\textrm{\AA}$ & 0.2765\\
\hline  
3FGLJ1258.4+2123    &  6398   &  3.2   &   Ca~II 3934~$\textrm{\AA}$ & 0.6265\\  
                                    &  6455   &  4.8  &   Ca~II 3968~$\textrm{\AA}$ & 0.6265\\   
                                    &  7001   &  2.3    &   G-band 4305~$\textrm{\AA}$  & 0.6265\\
\hline                           
3FGLJ1541.6+1414    &  4811   & 1.2   &   Ca~II 3934~$\textrm{\AA}$ & 0.223\\  
                                    &  4853   & 1.2  &   Ca~II 3968~$\textrm{\AA}$ & 0.223\\   
                                    &  6123    &  1.1    &   [O~III] 5007~$\textrm{\AA}$ & 0.223\\ 
\hline
3FGLJ2150.5-1754     &  4663   & 3.8   &   Ca~II 3934~$\textrm{\AA}$ & 0.1855\\  
                                    &  4704   & 3.6   &   Ca~II 3968~$\textrm{\AA}$ & 0.1855\\  
                                    &  5103  &  1.4   &   G-band 4305~$\textrm{\AA}$  & 0.1855\\
                                    &  5765*  &  0.9  &  H$_{\beta}$  4861~$\textrm{\AA}$ & 0.1855 \\
                                    &  6135  &  1.3   &   Mg~I 5175~$\textrm{\AA}$ & 0.1855\\
                                    &  6986  &   1.4  &   Na~I 5892~$\textrm{\AA}$ & 0.1855\\
 \hline                                   
3FGLJ2209.8-0450     &  5204    &  0.9  & [O~II] 3727~$\textrm{\AA}$ &   0.3967 \\    
                                    &  5495    &  1.3   & Ca~II 3934~$\textrm{\AA}$ &   0.3967 \\ 
                                    &  5543    &  1.7 & Ca~II 3968~$\textrm{\AA}$ &   0.3967 \\ 
                                    &  6012    &  1.0  & G-band 4305~$\textrm{\AA}$ &   0.3967 \\ 
\hline       
3FGLJ2212.5+0703    & 5600    & 116  &  Mg~II 2800~$\textrm{\AA}$  & 1.00 \\ 
\hline
3FGLJ2228.5-1636     & $\sim$6020     &  -  & Ca~II break   &  $\sim$0.525\\  
\hline
3FGLJ2229.1+2255    &  5664   & 0.8   &   Ca~II 3934~$\textrm{\AA}$ & 0.440\\  
                                    &  5715   & 1.0  &   Ca~II 3968~$\textrm{\AA}$ & 0.440\\  
                                    &  6198  &  1.1   &   G-band 4305~$\textrm{\AA}$  & 0.440\\
\hline
3FGLJ2244.6+2503    &  6491*   & 0.5  & Ca~II 3934~$\textrm{\AA}$   &   0.650\\  
                                    &  6548   & 0.4  & Ca~II 3968~$\textrm{\AA}$   &   0.650\\  
\hline
3FGLJ2246.2+1547    &  5949    &   1.6  &  [O~II] 3727~$\textrm{\AA}$ & 0.5965 \\  
                                    &  6280*    &   1.5  &  Ca~II 3934~$\textrm{\AA}$  & 0.5965 \\ 
                                    &  6336*    &   0.8  &  Ca~II 3968~$\textrm{\AA}$   & 0.5965 \\ 
\hline
\end{tabular}
\tablenotetext{}{
\footnotesize \textit{}}
\end{table*}

\newpage
\setcounter{table}{3}                                   
\begin{table*}
\caption{MEASUREMENTS OF THE SPECTRAL LINES \textit{(continued)}} 
\centering
\begin{tabular}{lllll}
\hline
Object name          &  $\lambda$    &    EW   &     Line ID    &   z   \\
\hline
3FGLJ2250.3+1747     &   4604  & 2.5   &   [Ne~V]  3426~$\textrm{\AA}$   & 0.3437\\  
                                     &   5008  & 3.5  &   [O~II] 3727~$\textrm{\AA}$  & 0.344\\  
                                     &   5286  &  7.6    &   Ca~II 3934~$\textrm{\AA}$ & 0.3437\\  
                                     &   5332  &  9.0    &   Ca~II 3968~$\textrm{\AA}$ & 0.3437\\  
                                     &   5512  &   1.3   &   H$_{\delta}$  4102~$\textrm{\AA}$ & 0.3437\\  
                                     &   5784   &  1.7   &   G-band 4305~$\textrm{\AA}$  & 0.3437\\  
                                     & 5833 &  0.8   &   H$_{\gamma}$ 4340~$\textrm{\AA}$ & 0.3437\\  
                                     & 6663  &  2.1   &   [O~III] 4959~$\textrm{\AA}$ & 0.3437\\  
                                     & 6728   &  6.5  &   [O~III] 5007~$\textrm{\AA}$ & 0.3437\\  
\hline
3FGLJ2321.6-1619     &  6663   &  0.5 & Ca~II 3934~$\textrm{\AA}$   &   0.6938\\  
                                    &  6722   &  0.3  & Ca~II 3968~$\textrm{\AA}$   &   0.6938\\  
\hline  
3FGLJ2358.5+3827    &   4472   &  11.4   &  [O~II] 3727~$\textrm{\AA}$  & 0.2001 \\   
                                    &   4642  &   1.3  & [Ne~III] 3869~$\textrm{\AA}$   & 0.2000 \\   
                                    &  4721    &  0.8   &  Ca~II 3934~$\textrm{\AA}$  & 0.2001 \\  
                                    &  5166    &  1.1   &  G-band 4305~$\textrm{\AA}$  & 0.2001 \\  
                                    &  5209    &  1.0   &  H$_{\gamma}$  4340~$\textrm{\AA}$  & 0.2001 \\  
                                    &  5834    &   2.7  &  H$_{\beta}$  4861~$\textrm{\AA}$  & 0.2001 \\   
                                    &  5951    &   4.9  &  [O~III] 4959~$\textrm{\AA}$ & 0.2001 \\   
                                    &  6009   &   16.2  &  [O~III] 5007~$\textrm{\AA}$  & 0.2001 \\   
                                    &  6211    &  1.1   &  Mg~I 5175~$\textrm{\AA}$  & 0.2001 \\   
                                    &  7072    &  1.7   &  Na~I 5892~$\textrm{\AA}$  & 0.2001 \\   
\hline
\end{tabular}
\tablenotetext{}{
\footnotesize \textit{}}
\end{table*}

\newpage
\setcounter{figure}{0}
\begin{figure*}
\centering%
\subfigure[\protect\url{3FGLJ0004.2+0843 }\label{fig:0004opt}]%
{\includegraphics[width=0.3\textwidth]{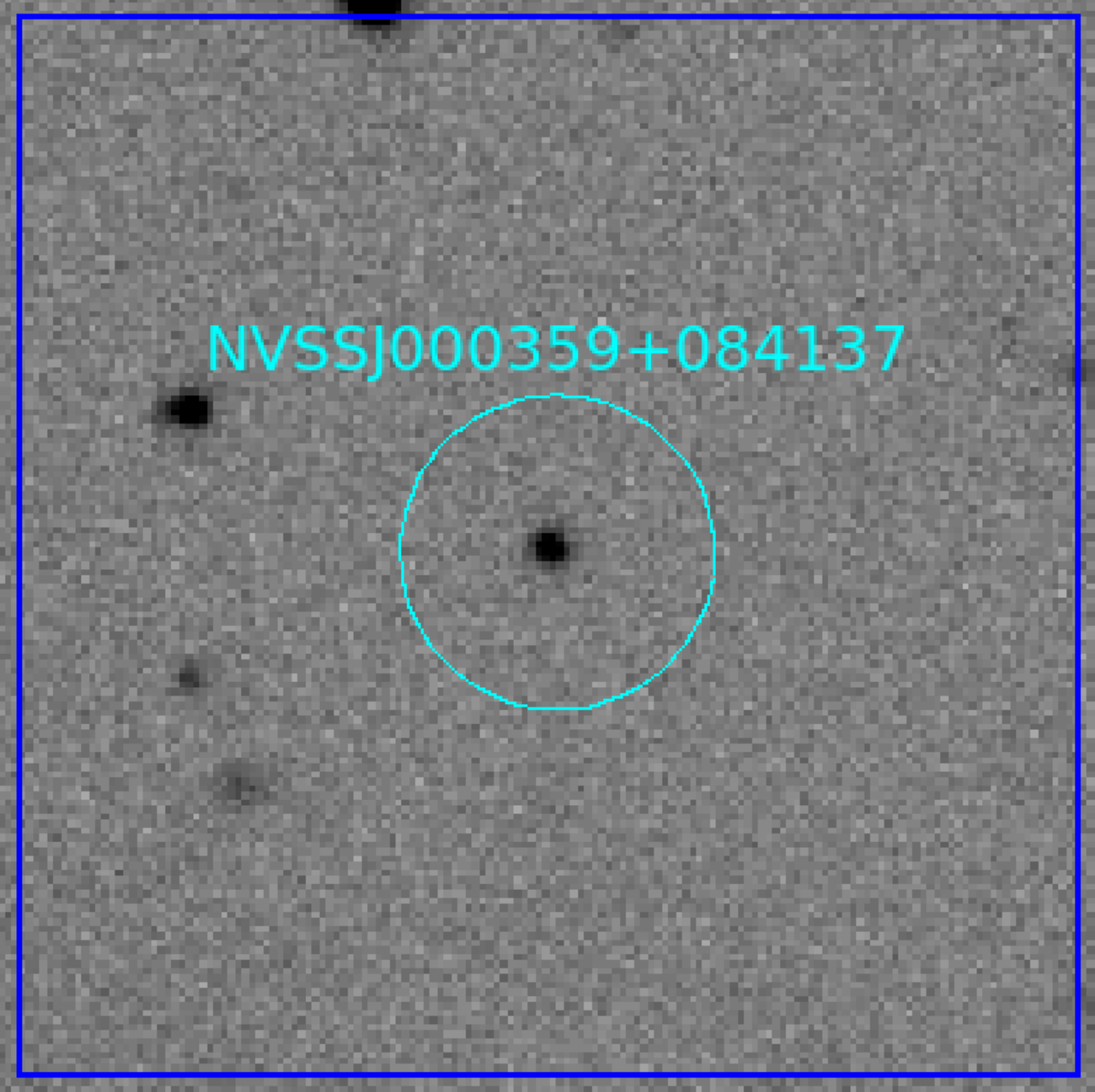}}
\subfigure[\protect\url{3FGLJ0006.2+0135}\label{fig:0006opt}]%
{\includegraphics[width=0.3\textwidth]{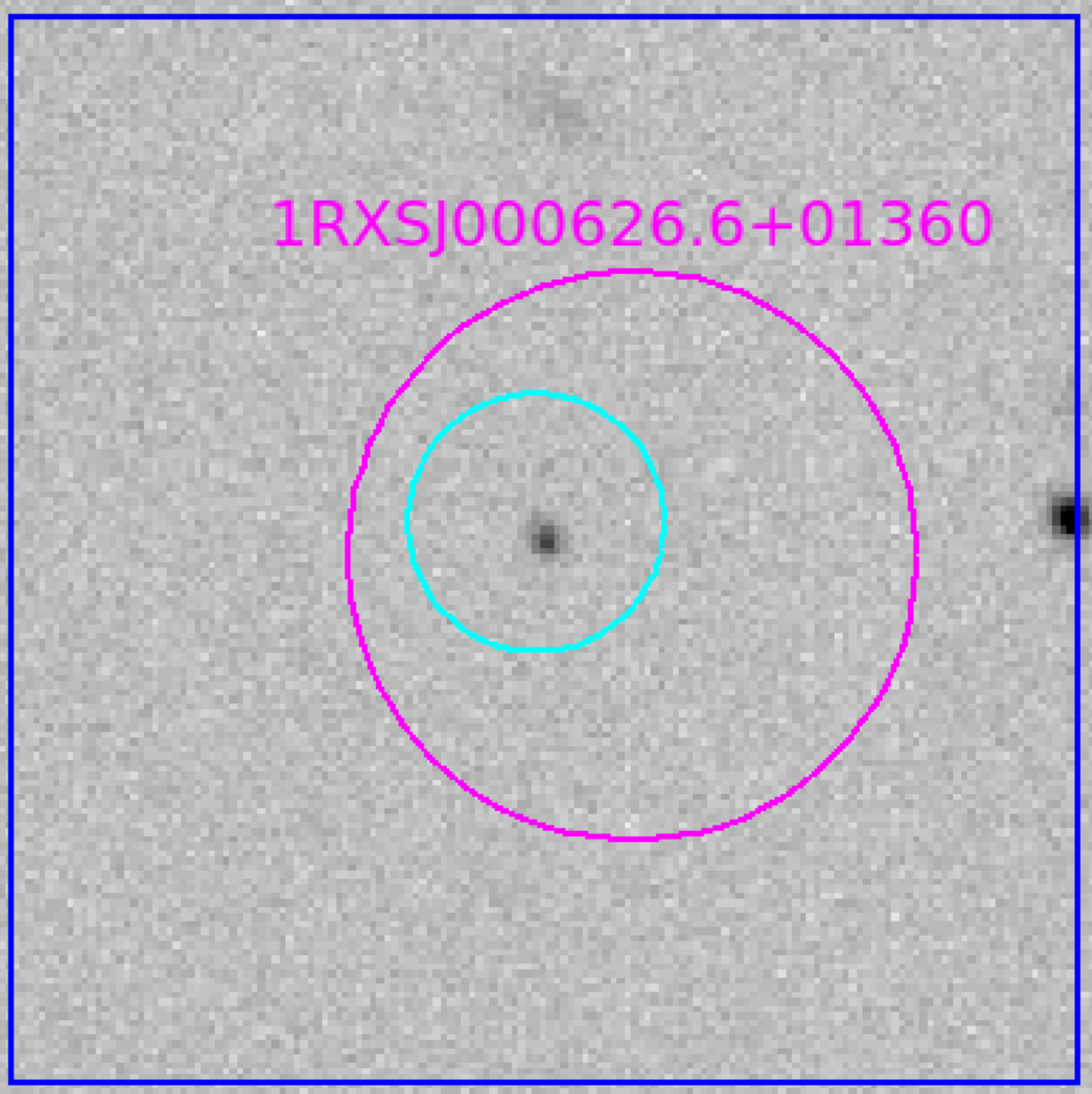}}
\subfigure[\protect\url{3FGLJ0031.6+0938}\label{fig:0031opt}]%
{\includegraphics[width=0.3\textwidth]{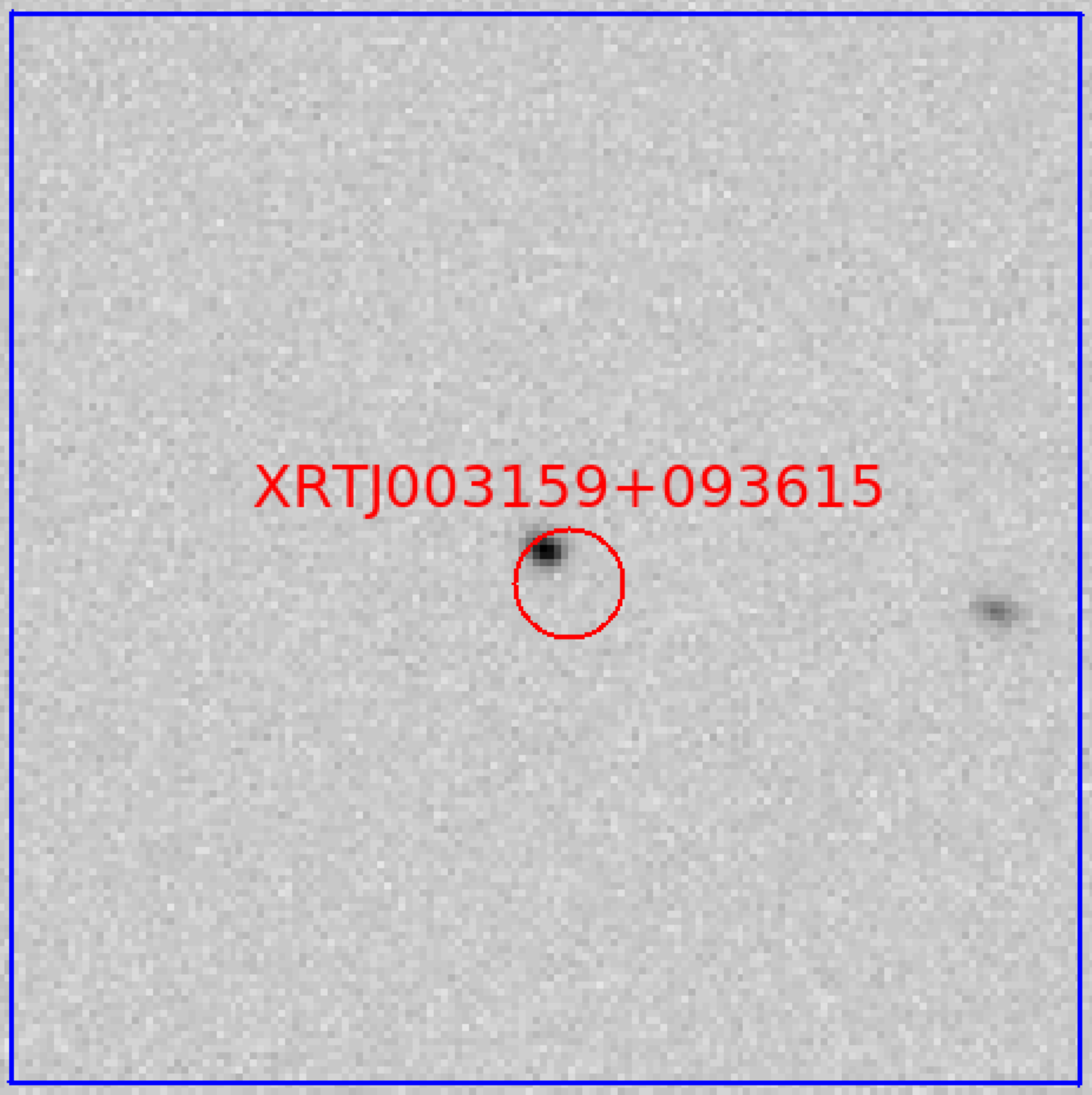}}
\subfigure[\protect\url{3FGLJ0158.6+0102 }\label{fig:0158opt }]%
{\includegraphics[width=0.3\textwidth]{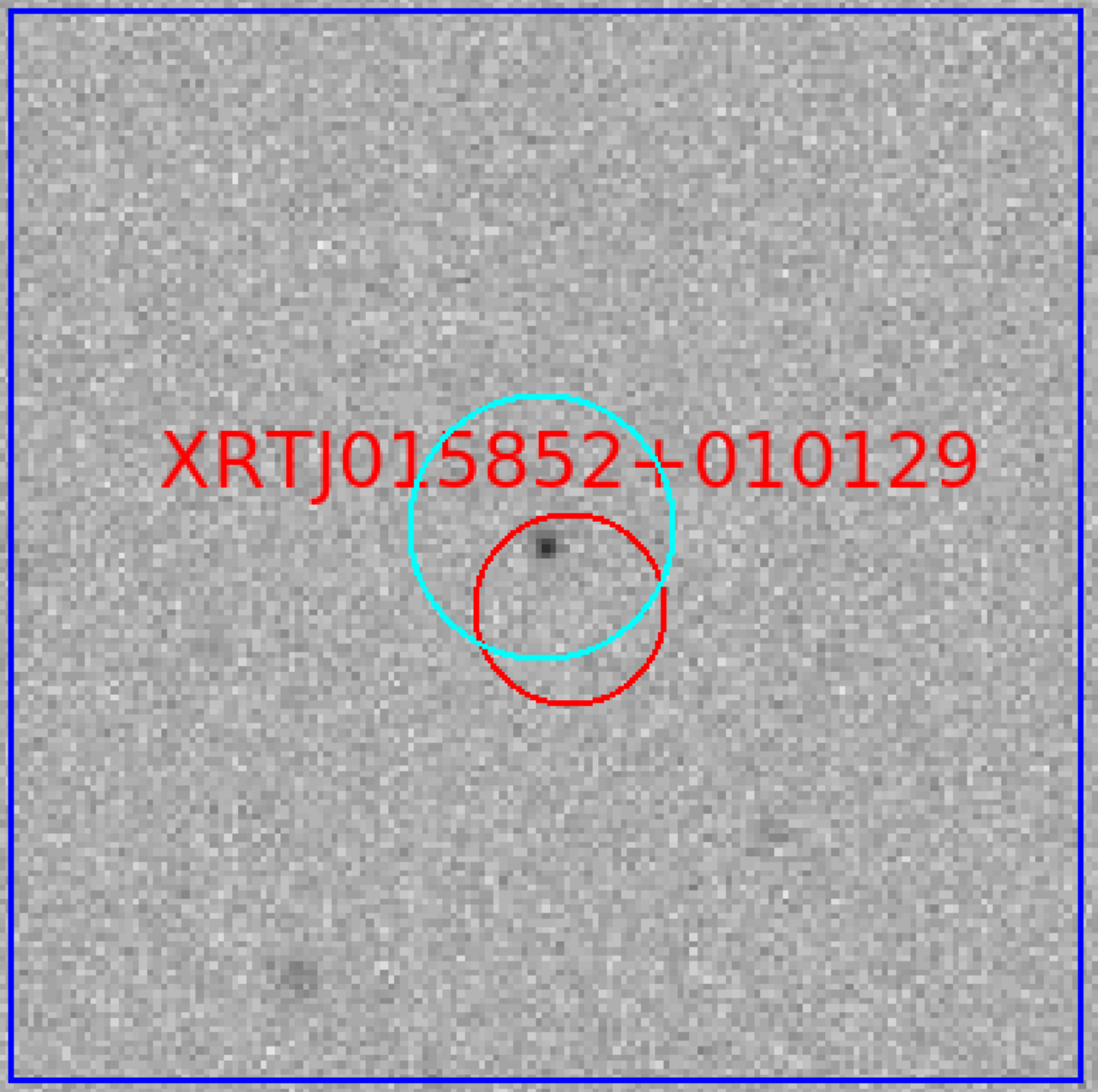}}
\subfigure[\protect\url{ 3FGLJ0234.2-0629}\label{fig:0234opt }]%
{\includegraphics[width=0.3\textwidth]{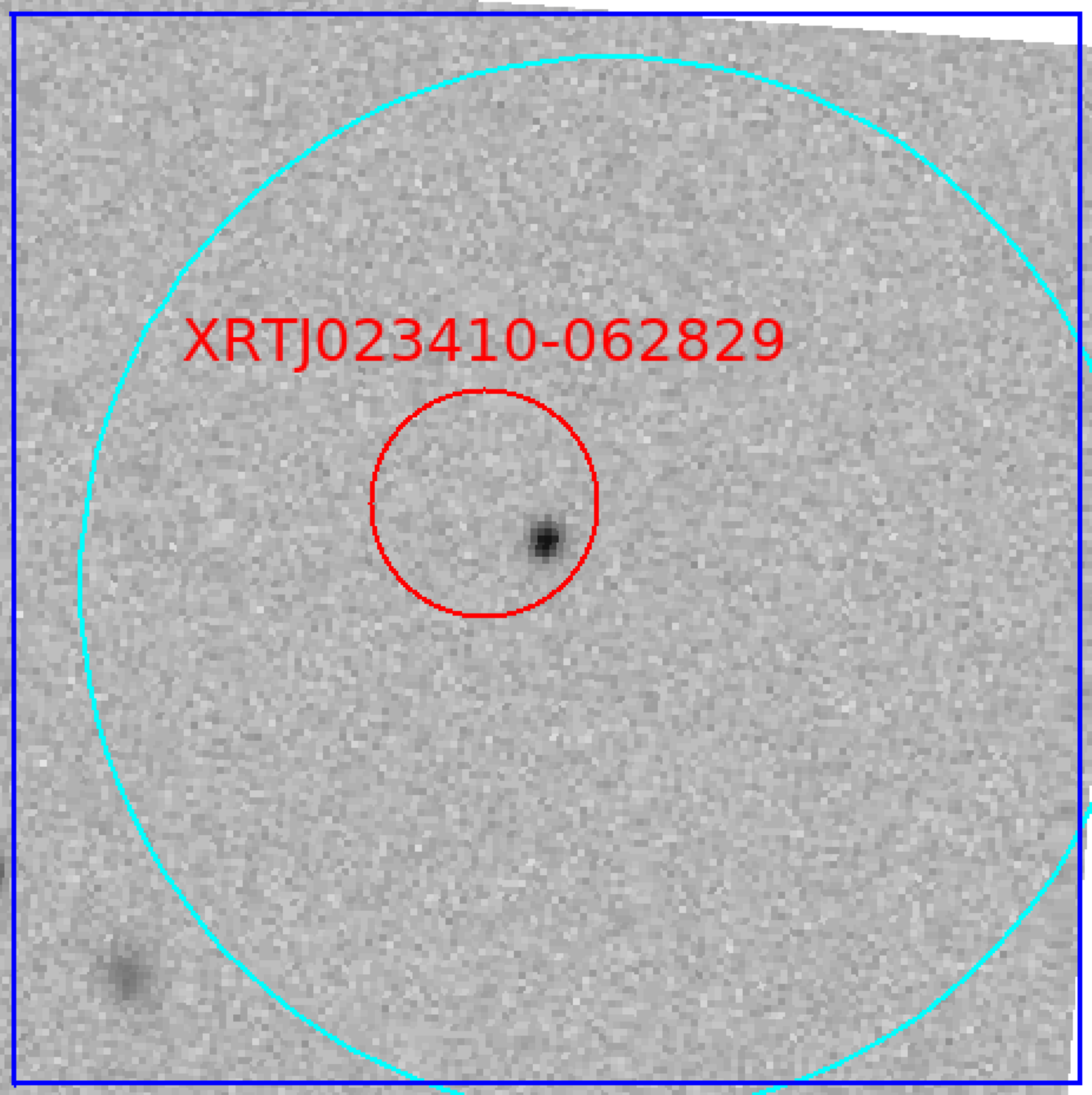}}
\subfigure[\protect\url{3FGLJ0251.1-1829 }\label{fig:0251opt }]%
{\includegraphics[width=0.3\textwidth]{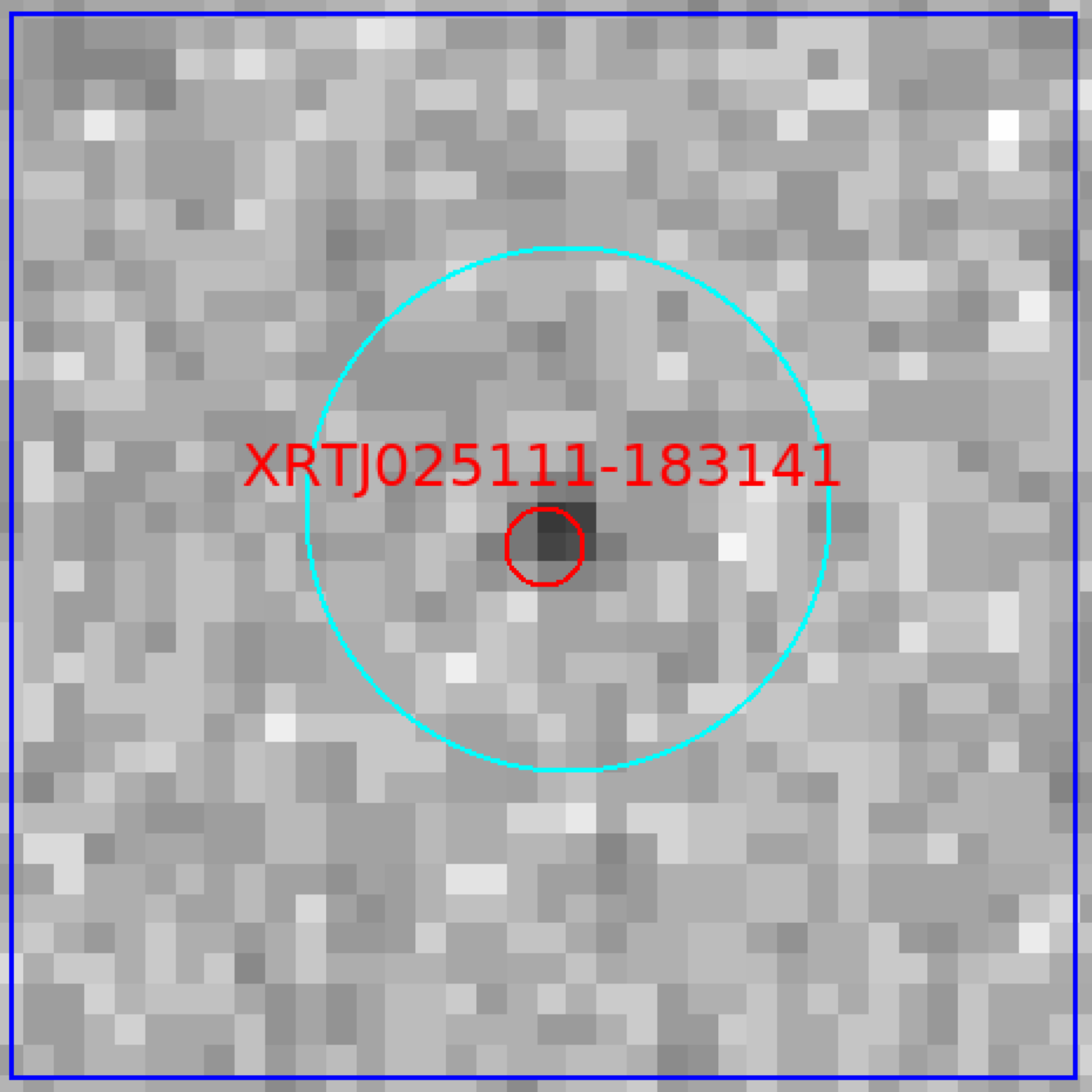}}
\subfigure[\protect\url{3FGLJ0258.9+0552 }\label{fig:0258opt }]%
{\includegraphics[width=0.3\textwidth]{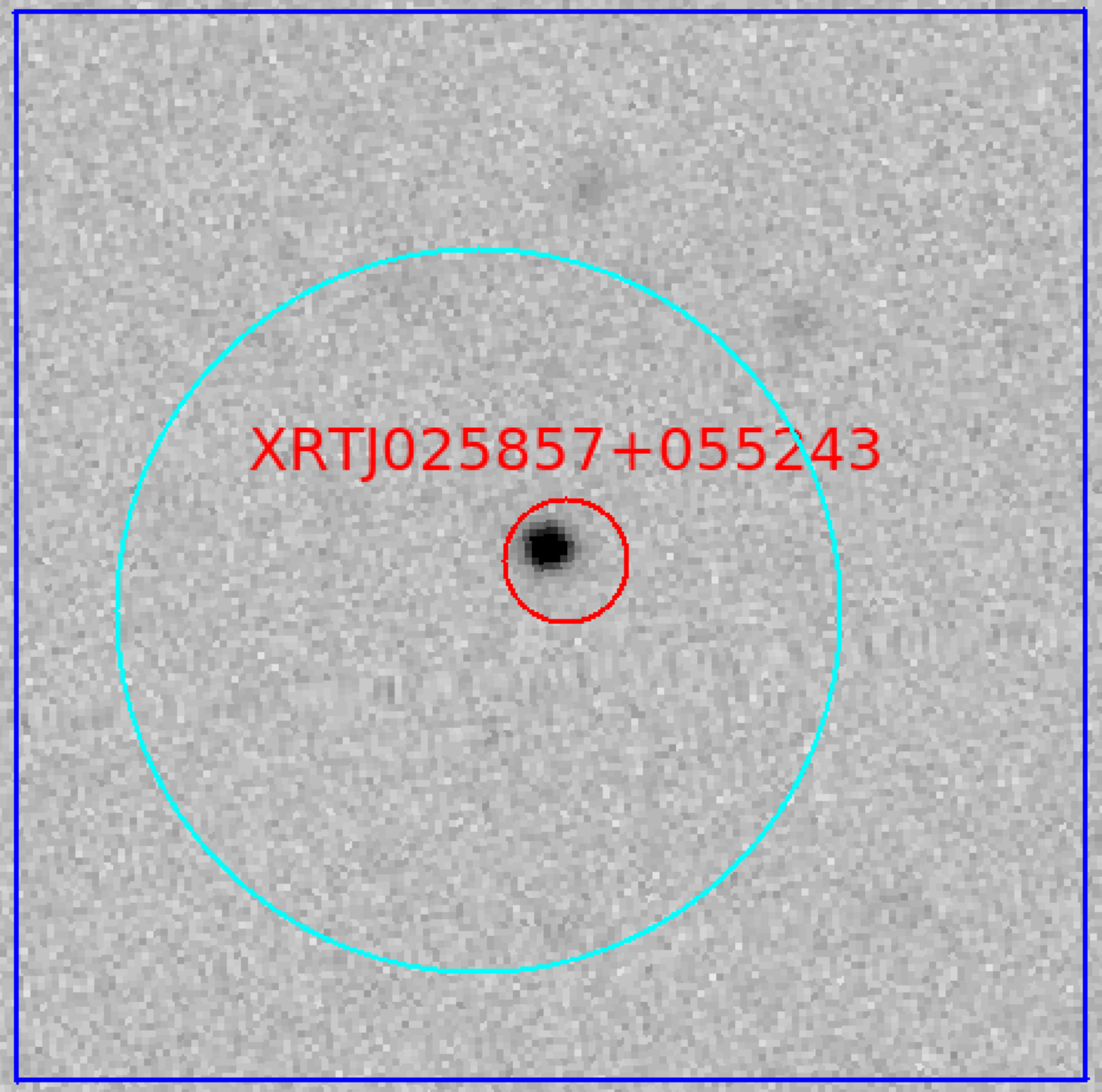}}
\subfigure[\protect\url{3FGLJ0414.9-0840}\label{fig:0414opt}]%
{\includegraphics[width=0.3\textwidth]{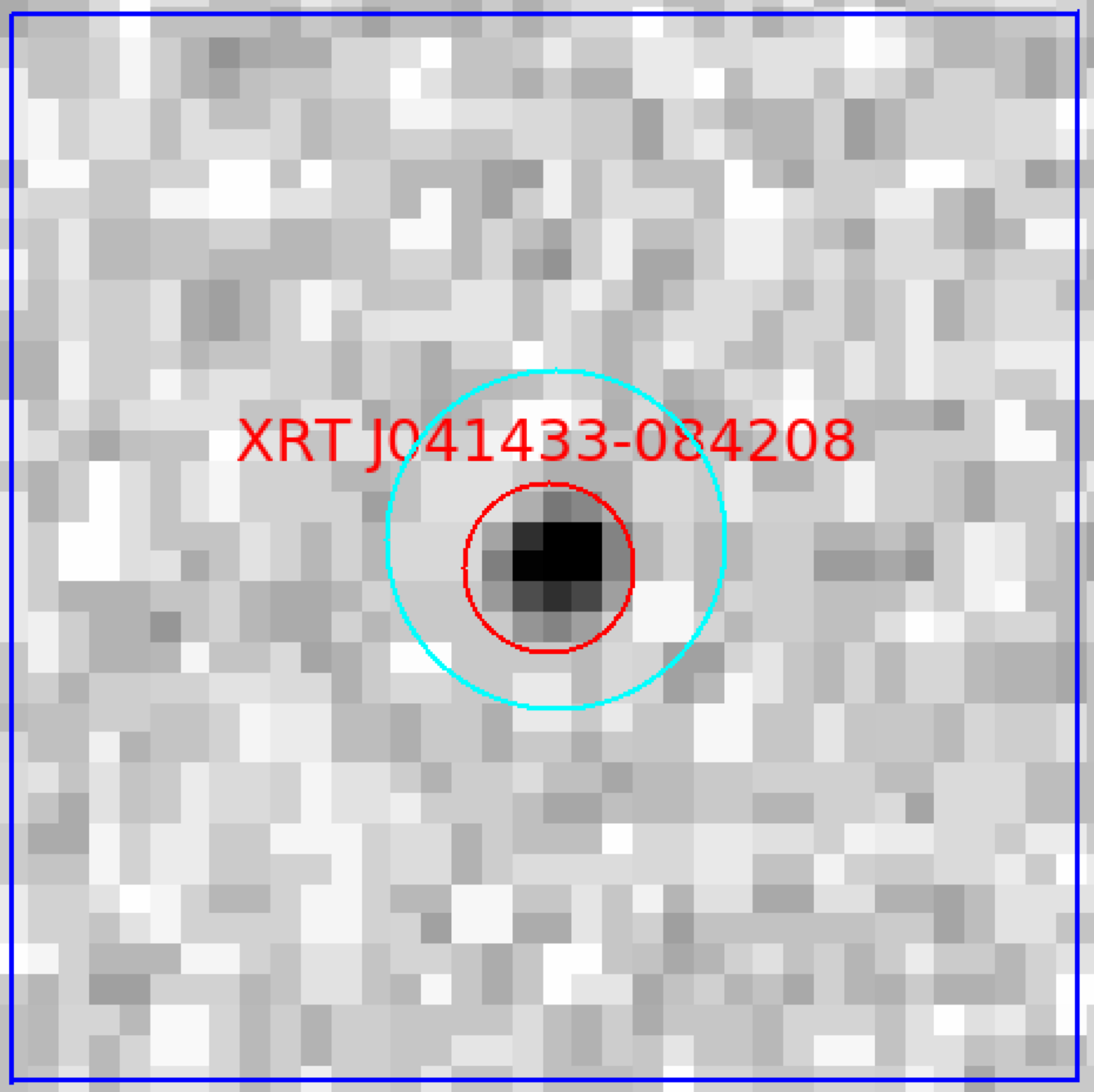}}
\subfigure[\protect\url{3FGLJ0506.9+0321}\label{fig:0506opt }]%
{\includegraphics[width=0.3\textwidth]{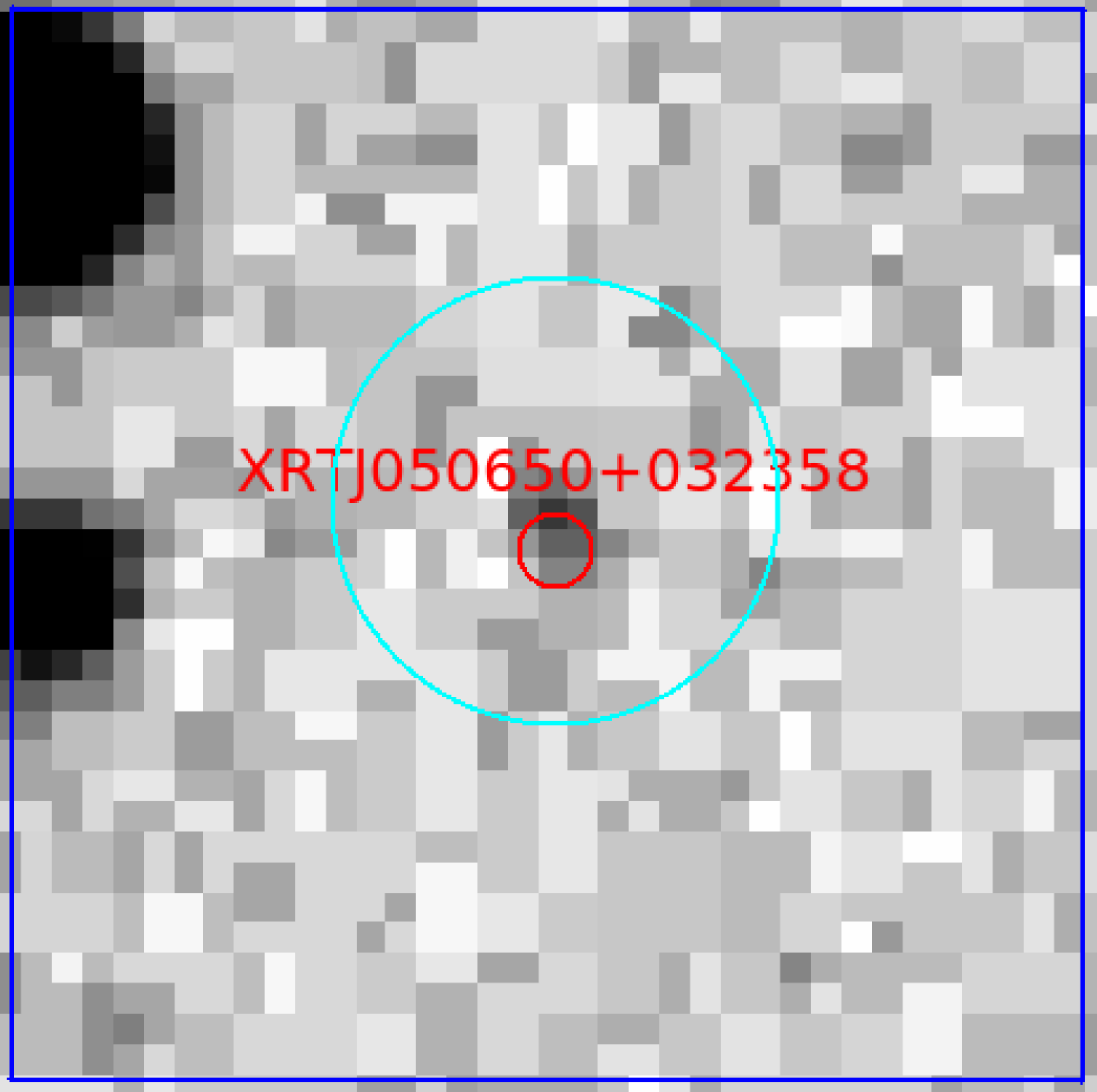}}
\caption{Optical images of the proposed counterparts of the $\gamma$-ray sources. The images are taken from SDSS (g filter) or DSS survey (B filter), field= 60 arcsec x 60 arcsec, North up, East left. Details of the optical counterparts are reported in Tab. \ref{tab:table1}.  The red circles represent the error box of the proposed X-ray counterparts of the $\gamma$-ray sources found with \textit{Swift}/XRT data (magenta circle indicate a X-ray source of the ROSAT catalog),  the cyan circles the radio counterparts (see details in Sec.~1) and the green crosses the WISE IR counterparts.\\
The optical image of 3FGLJ2246.2+1547 is displayed in Paper~I.} 
\label{fig:fc1}
\end{figure*}

\setcounter{figure}{0}
\begin{figure*}
\centering%
\subfigure[\protect\url{3FGLJ0848.5+7018}\label{fig:0848opt}]%
{\includegraphics[width=0.3\textwidth]{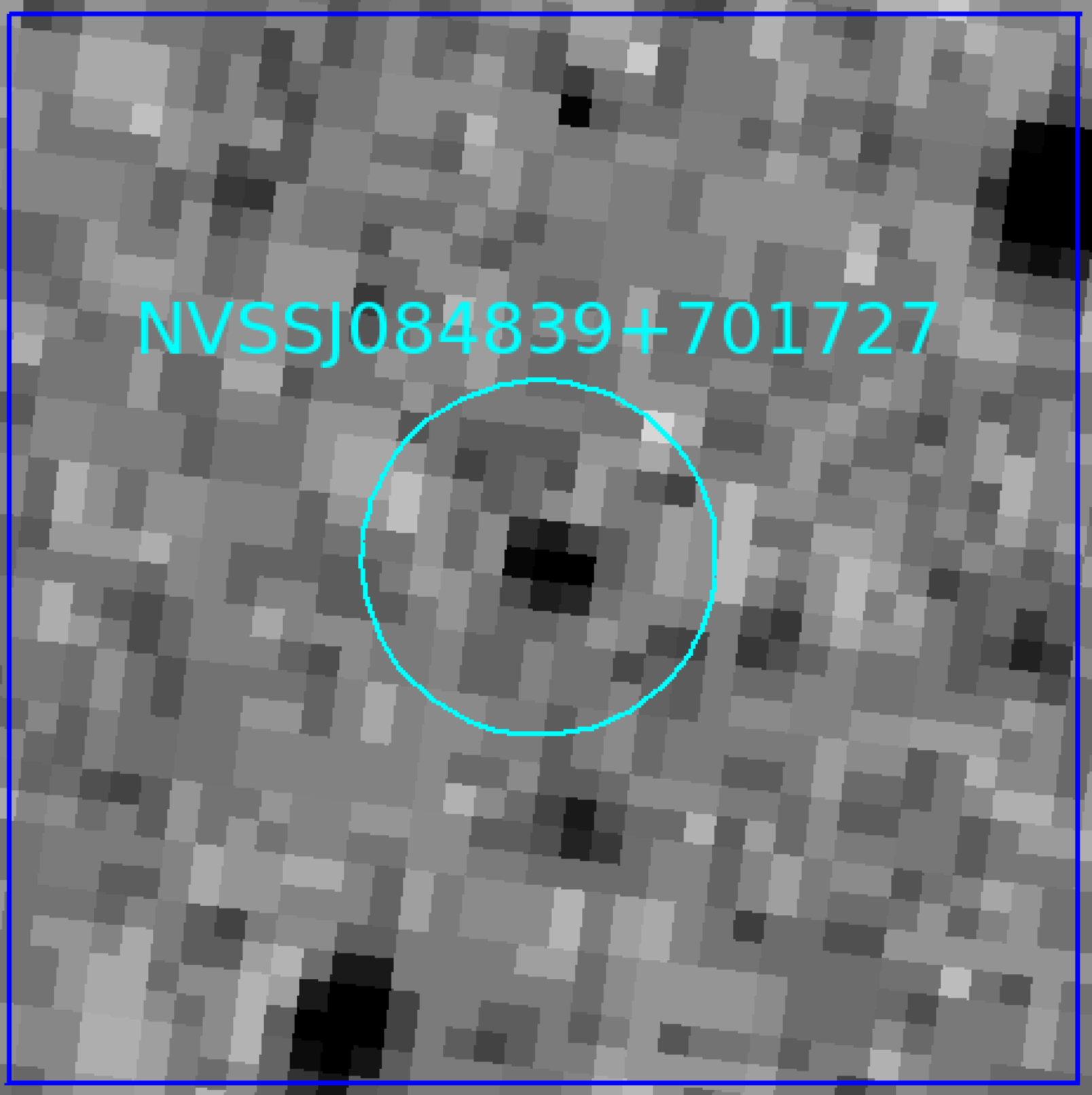}}
\subfigure[\protect\url{3FGLJ0930.7+5133}\label{fig:0930opt}]%
{\includegraphics[width=0.3\textwidth]{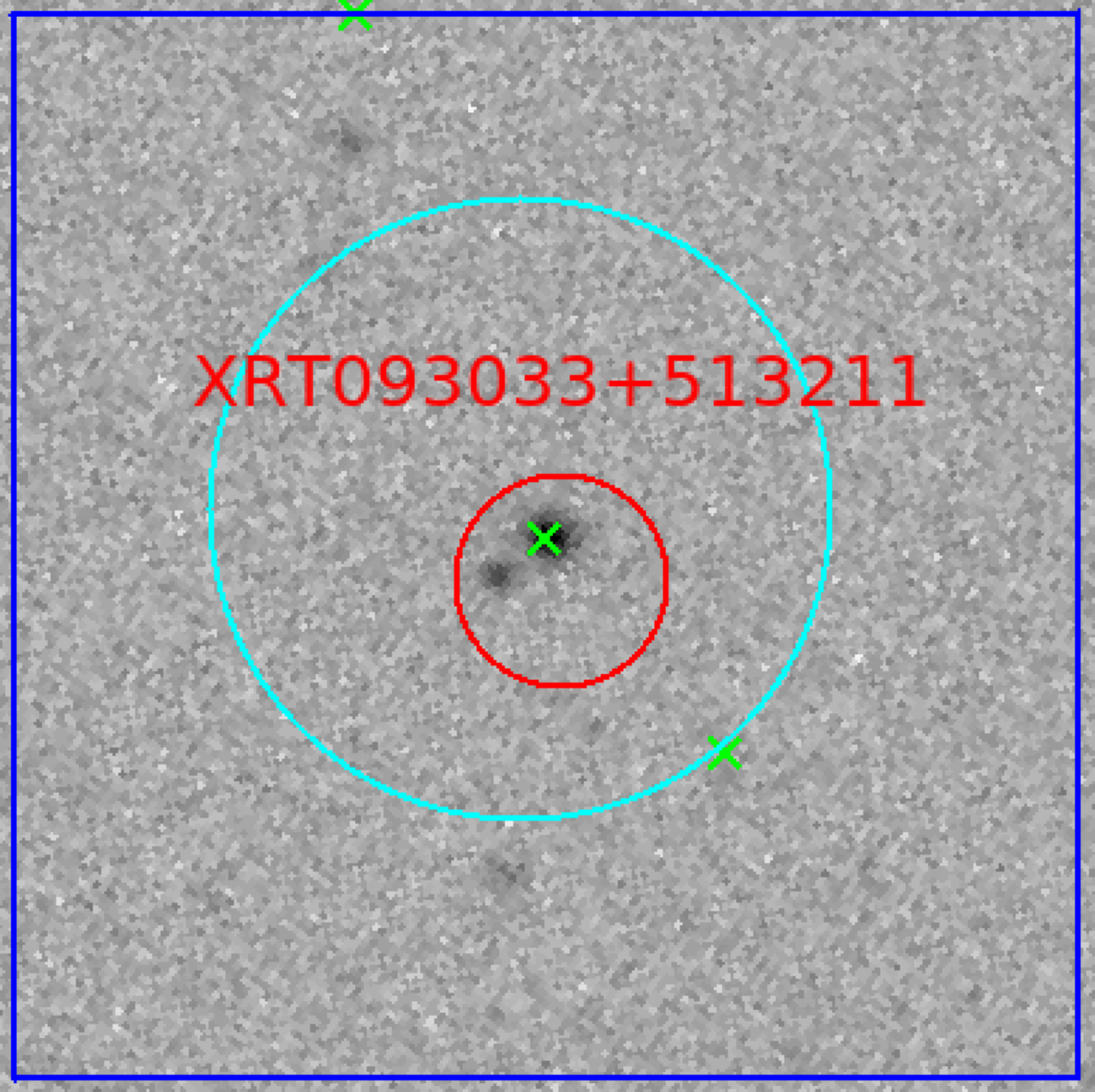}}
\subfigure[\protect\url{3FGLJ1146.1-0640}\label{fig:1146opt}]%
{\includegraphics[width=0.3\textwidth]{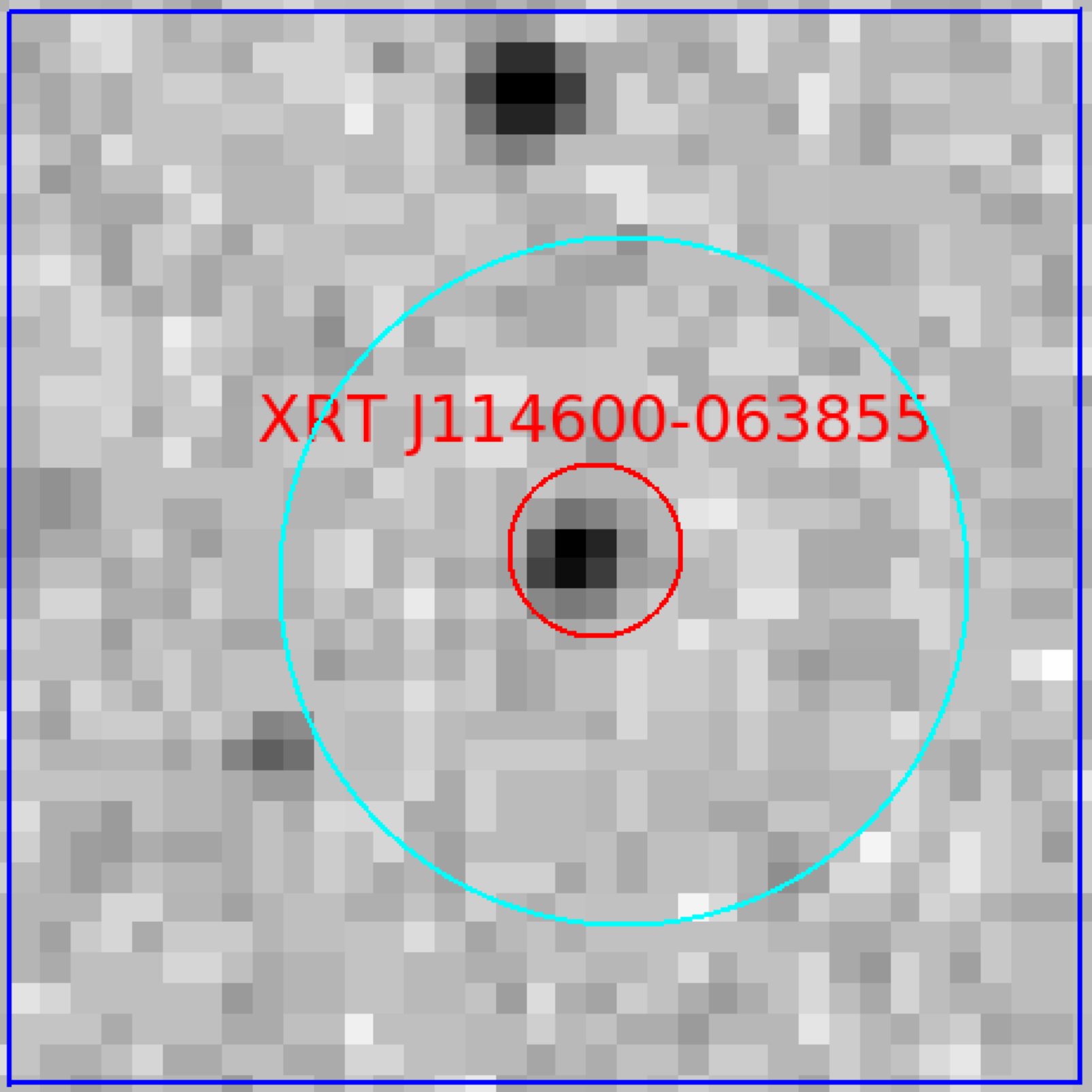}}
\subfigure[\protect\url{3FGLJ1223.3+0818}\label{fig:1223opt}]%
{\includegraphics[width=0.3\textwidth]{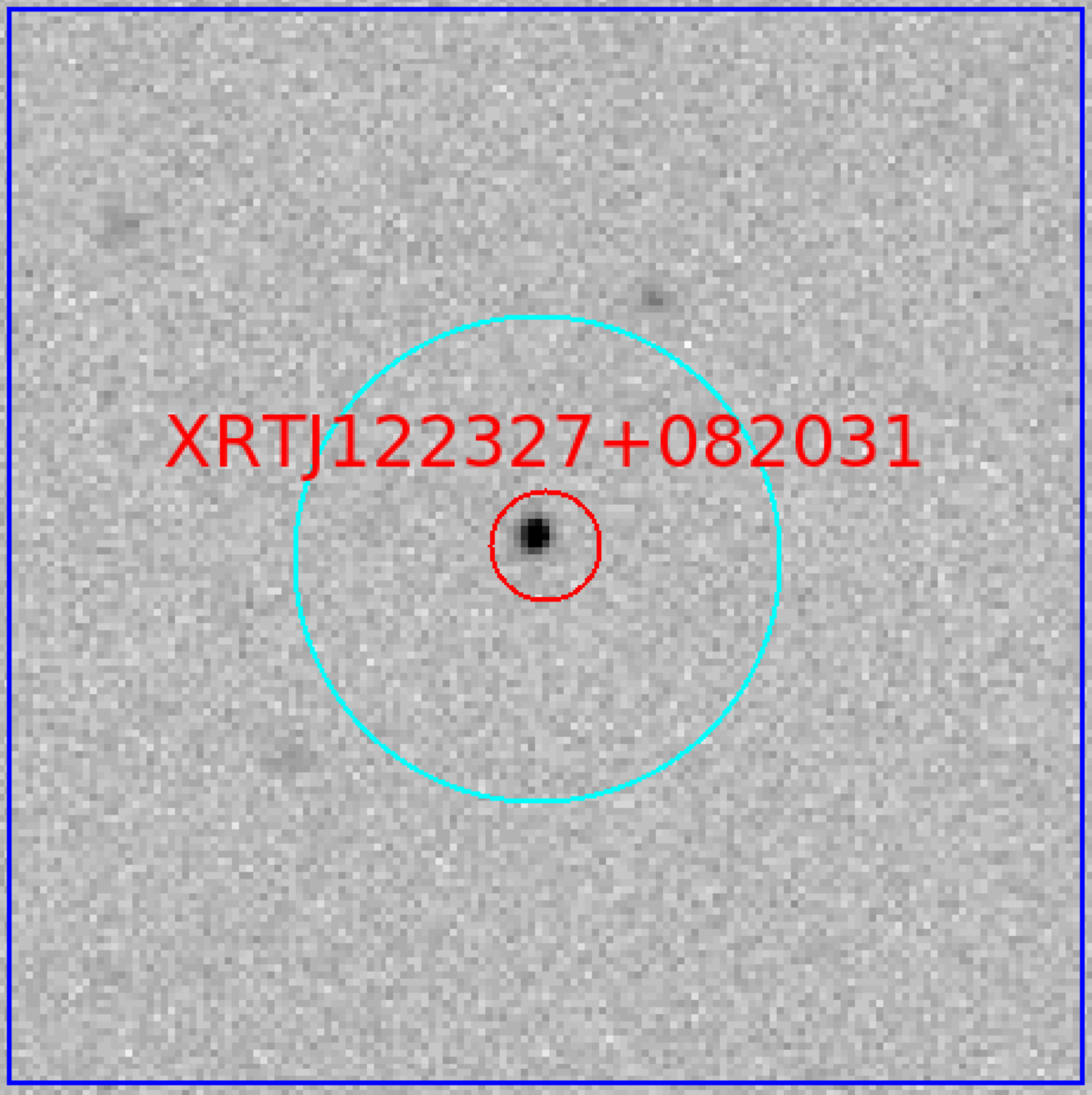}}
\subfigure[\protect\url{3FGLJ1234.7-0437}\label{fig:1234opt}]%
{\includegraphics[width=0.3\textwidth]{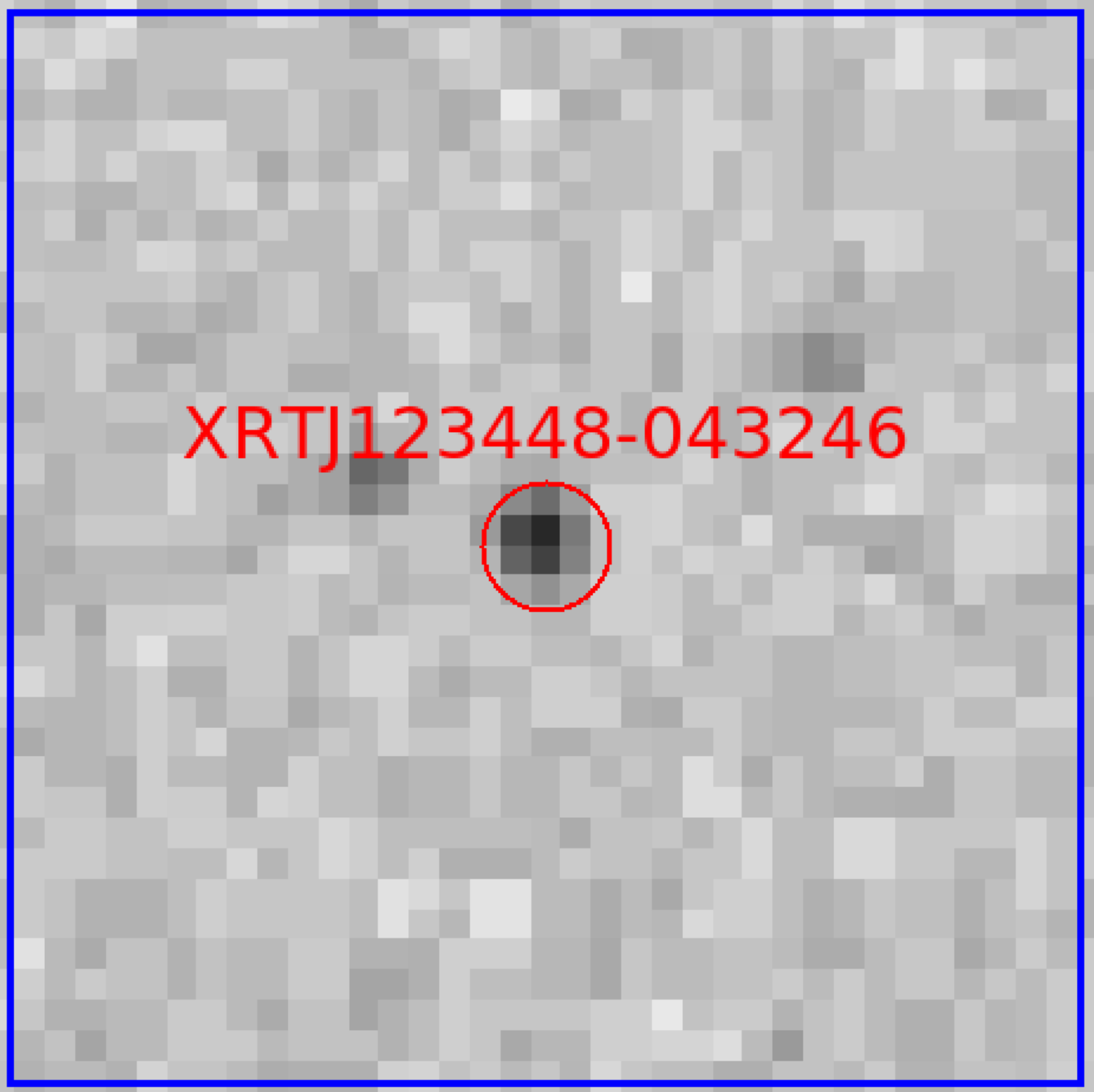}}
\subfigure[\protect\url{3FGLJ1258.4+2123}\label{fig:1258opt}]%
{\includegraphics[width=0.3\textwidth]{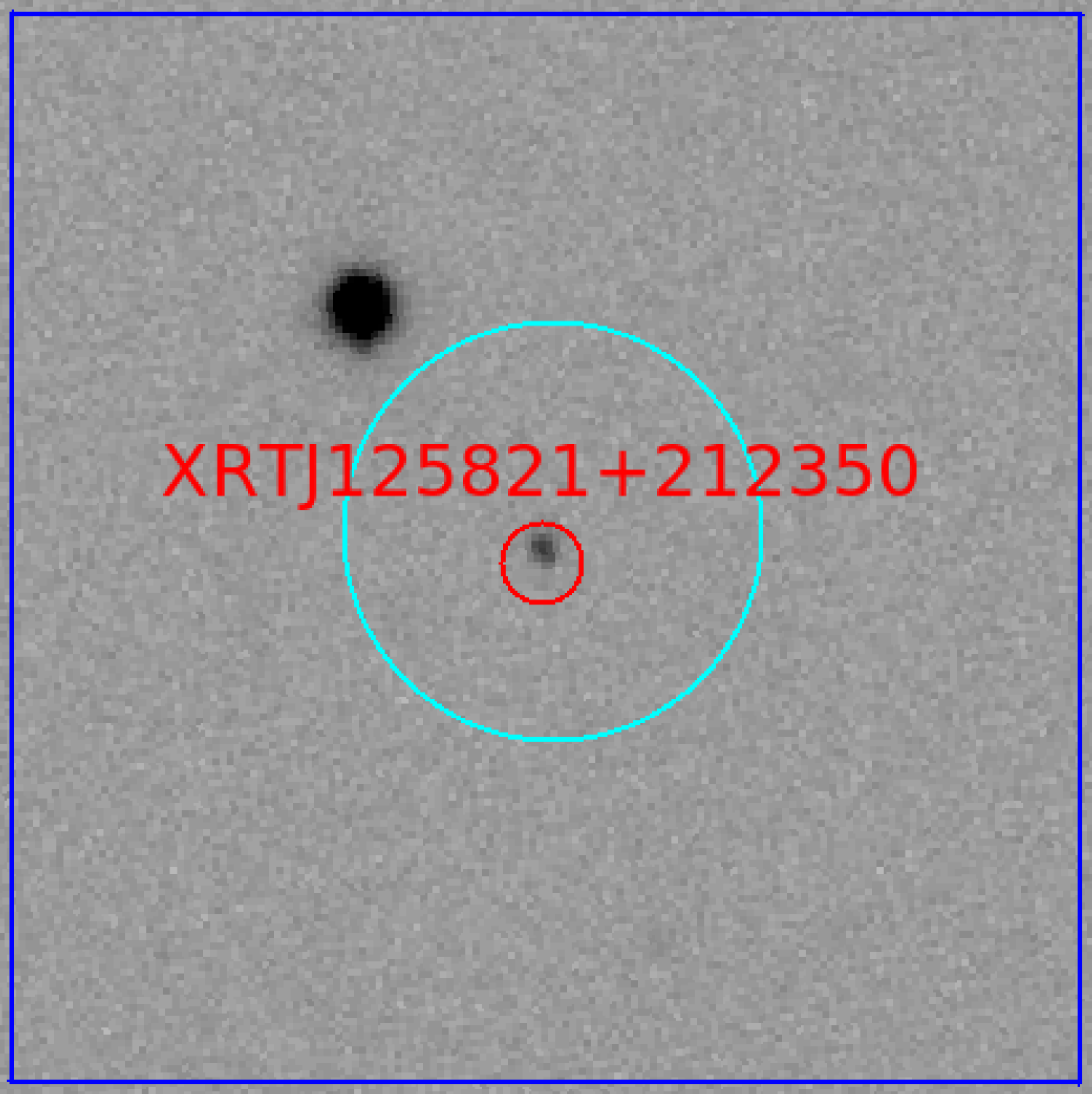}}
\subfigure[\protect\url{3FGLJ1525.8-0834}\label{fig:1525opt}]%
{\includegraphics[width=0.3\textwidth]{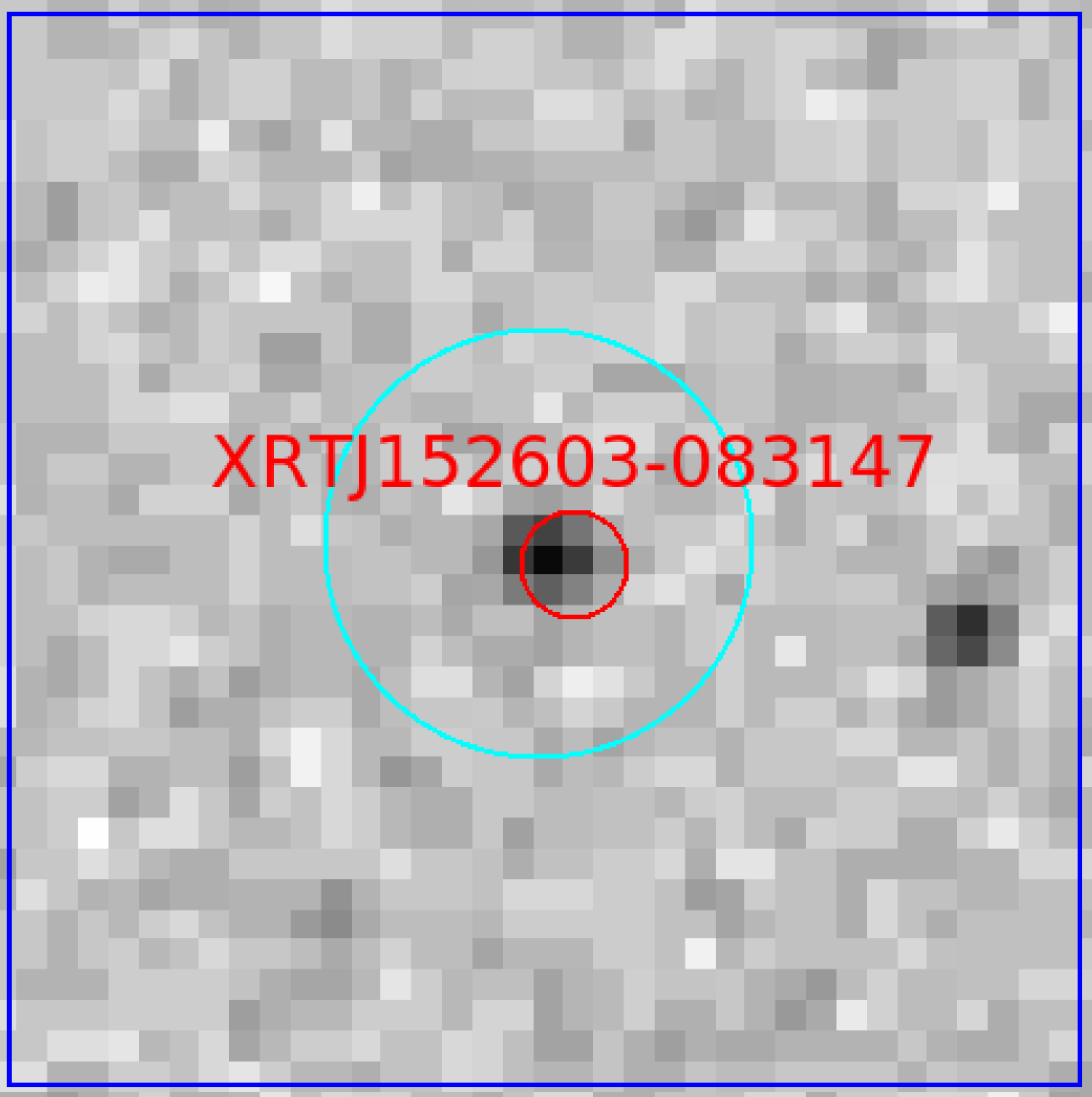}}
\subfigure[\protect\url{3FGLJ1541.6+1414}\label{fig:1541opt }]%
{\includegraphics[width=0.3\textwidth]{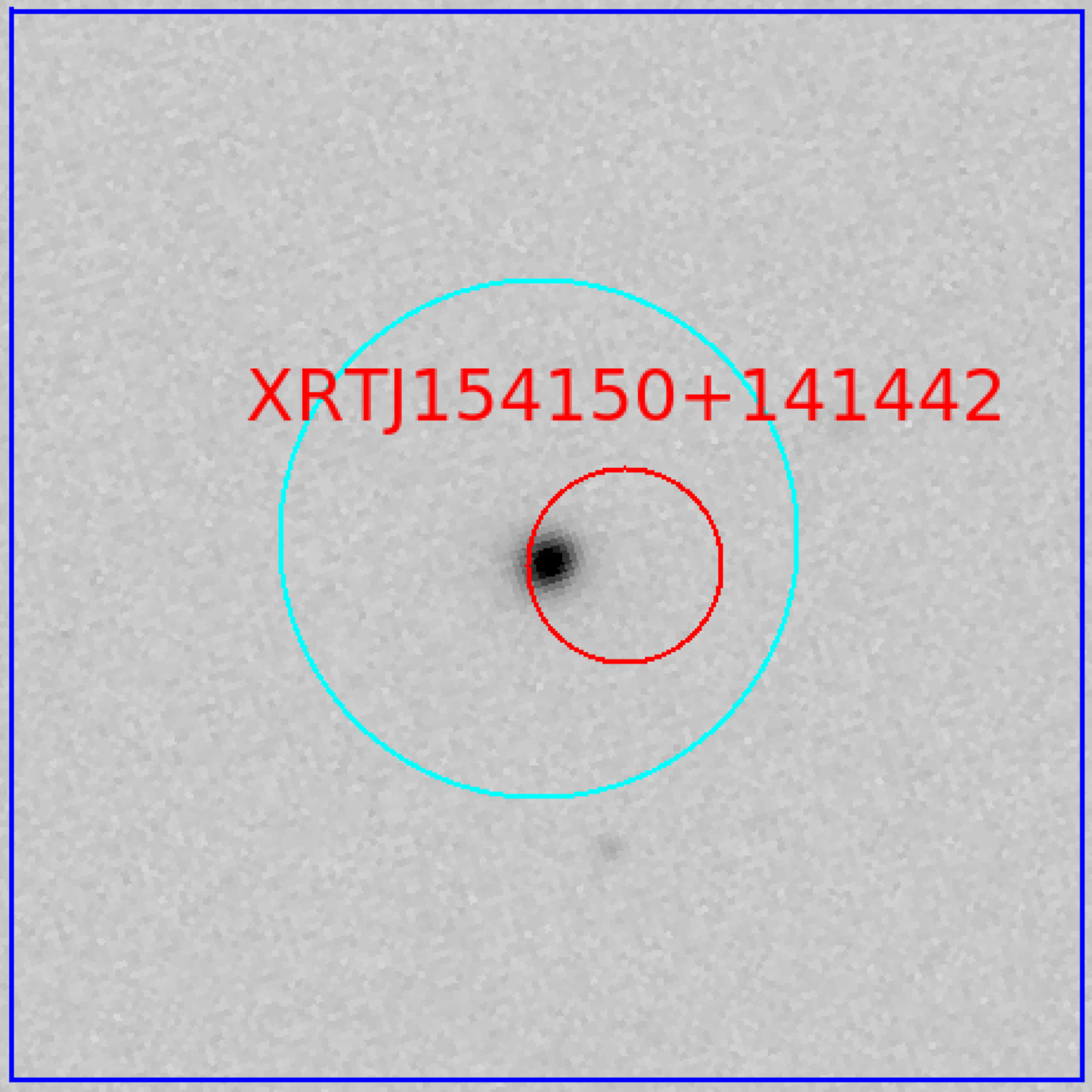}}
\subfigure[\protect\url{3FGLJ2150.5-1754}\label{fig:2150opt}]%
{\includegraphics[width=0.3\textwidth]{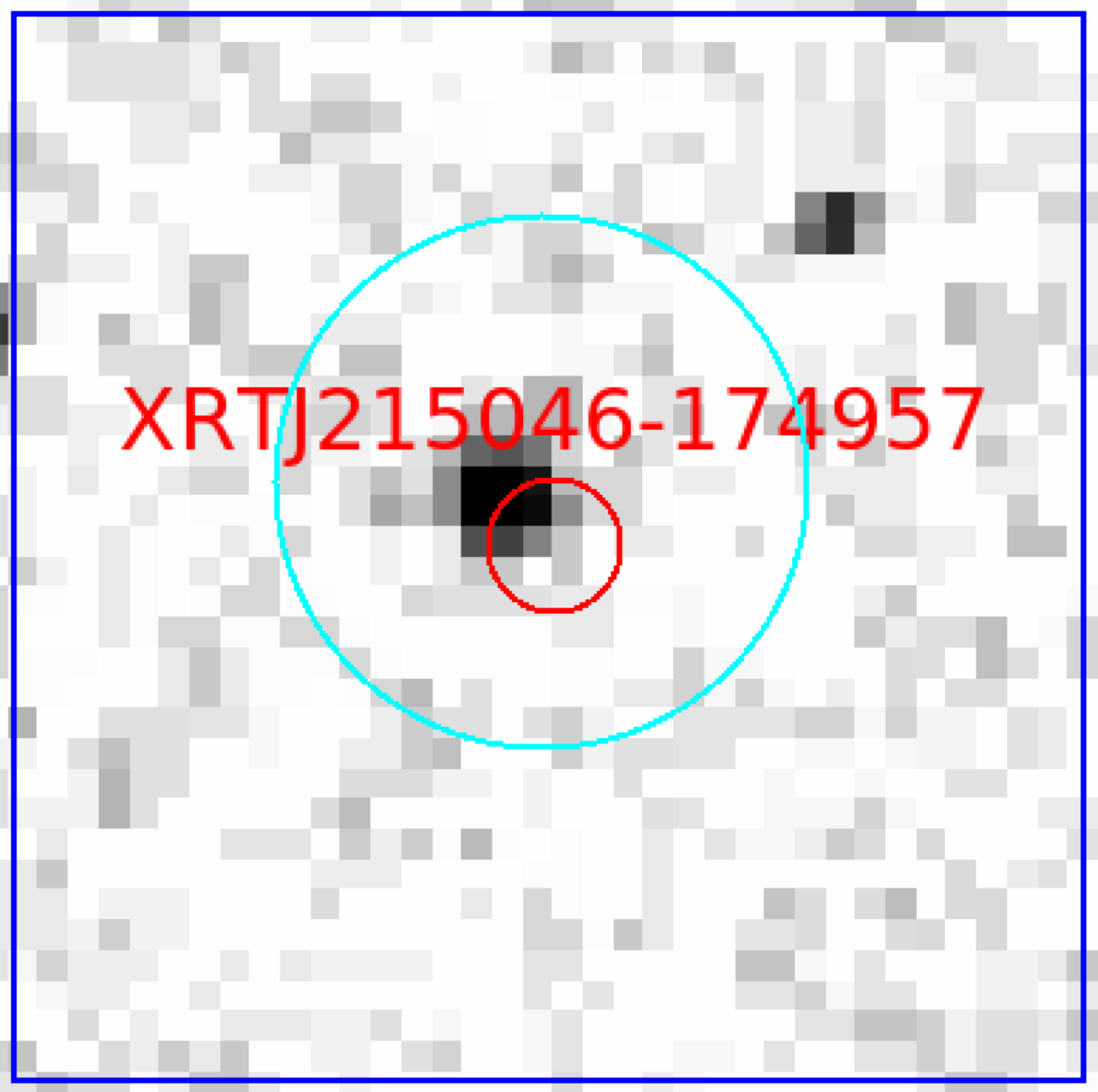}}
\caption{Continued from Fig. \ref{fig:fc1}.}
\end{figure*}

\setcounter{figure}{0}
\begin{figure*}
\centering%
\subfigure[\protect\url{3FGLJ2209.8-0450}\label{fig:2209opt}]%
{\includegraphics[width=0.3\textwidth]{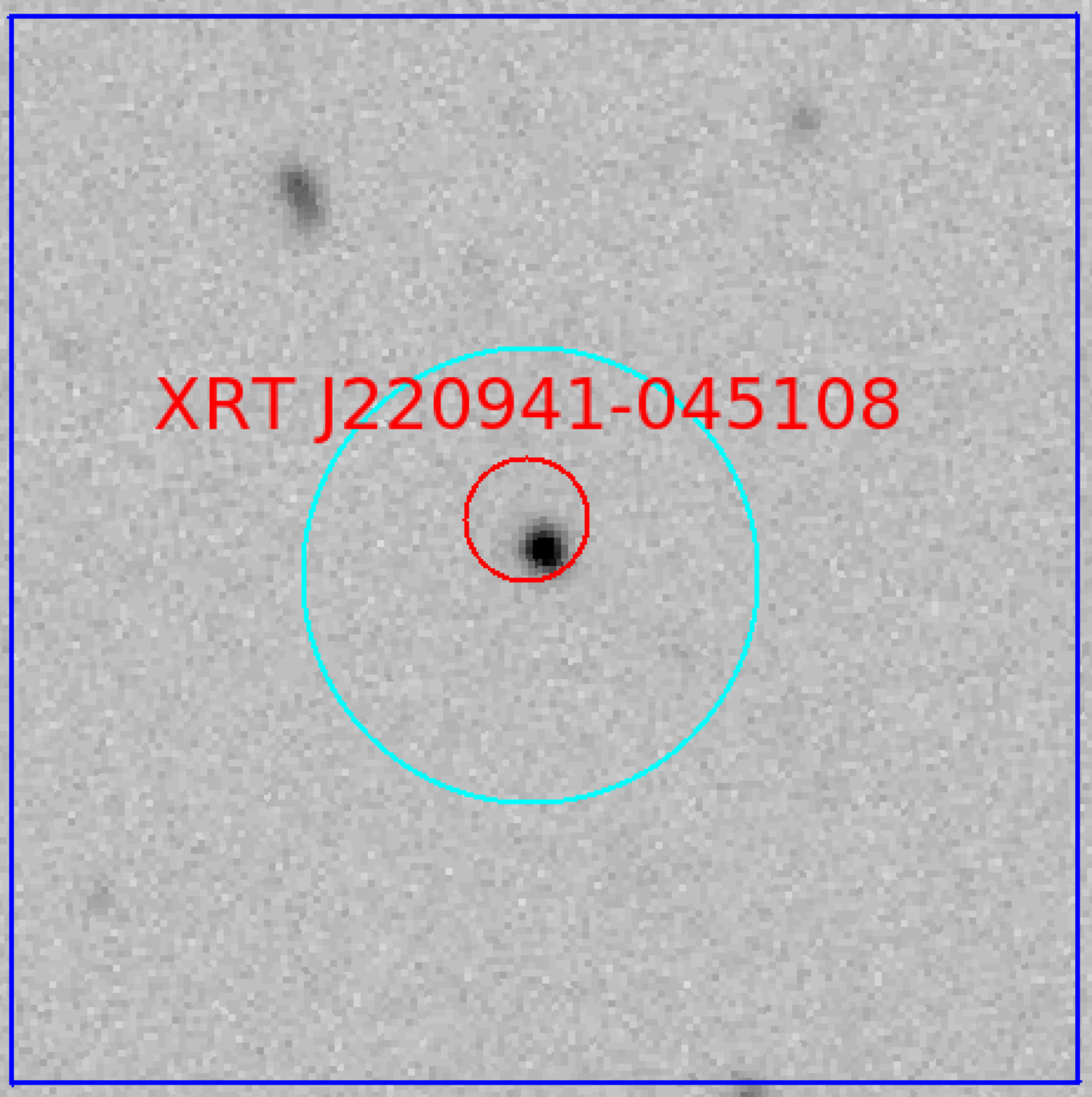}}
\subfigure[\protect\url{3FGLJ2212.5+0703}\label{fig:2212}]%
{\includegraphics[width=0.3\textwidth]{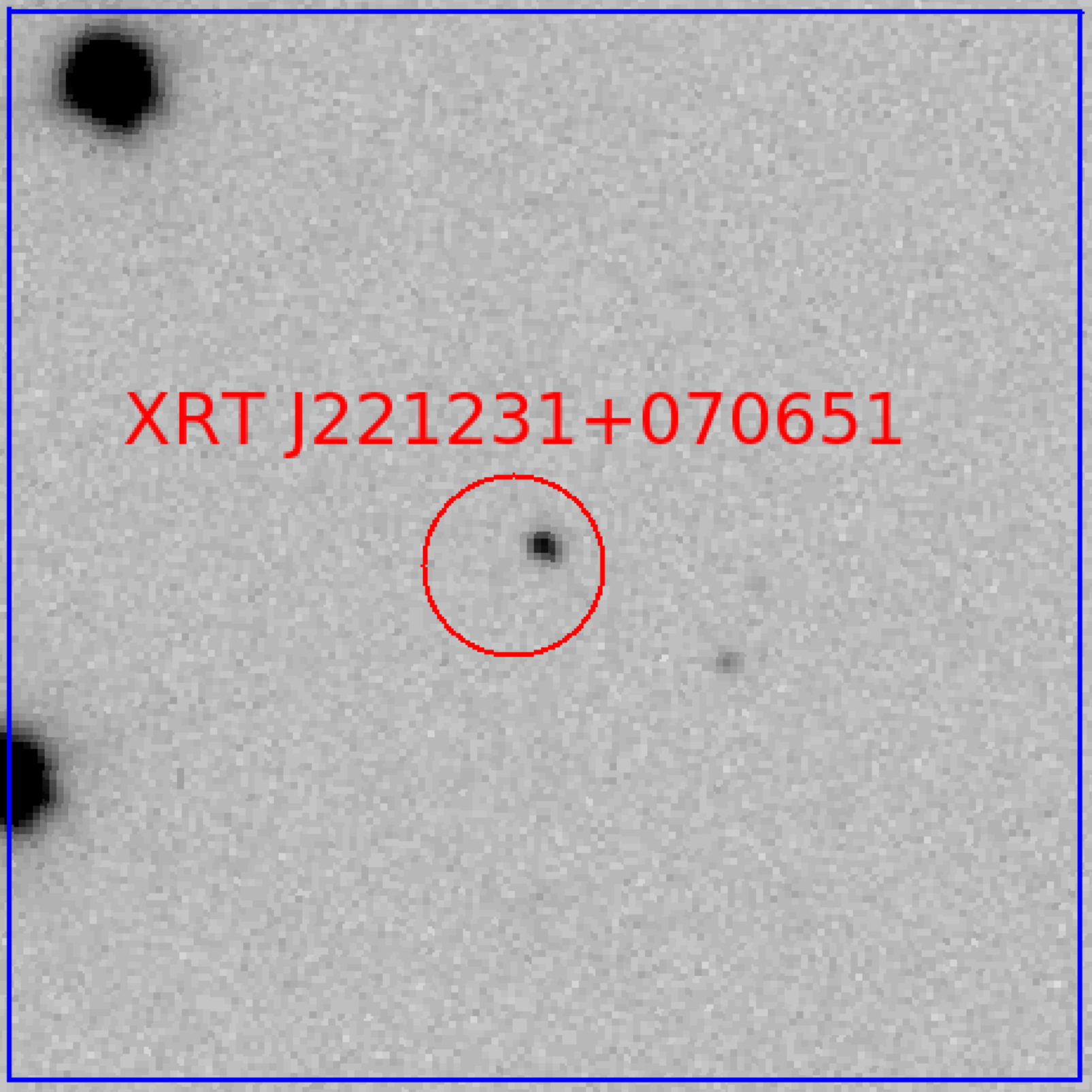}}
\subfigure[\protect\url{3FGLJ2228.5-1636}\label{fig:2228opt}]%
{\includegraphics[width=0.3\textwidth]{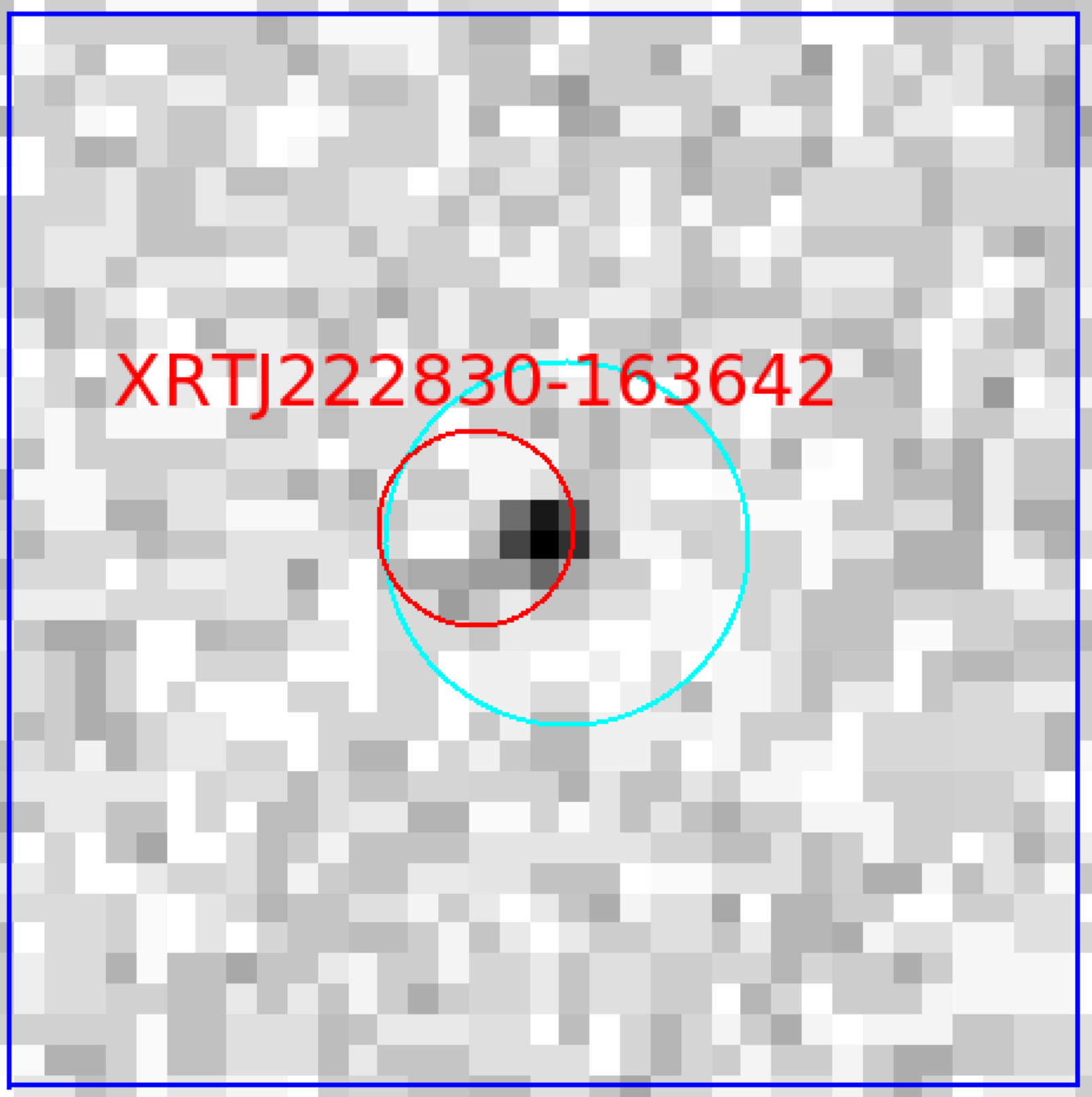}}
\subfigure[\protect\url{3FGLJ2229.1+2255}\label{fig:2229opt }]%
{\includegraphics[width=0.3\textwidth]{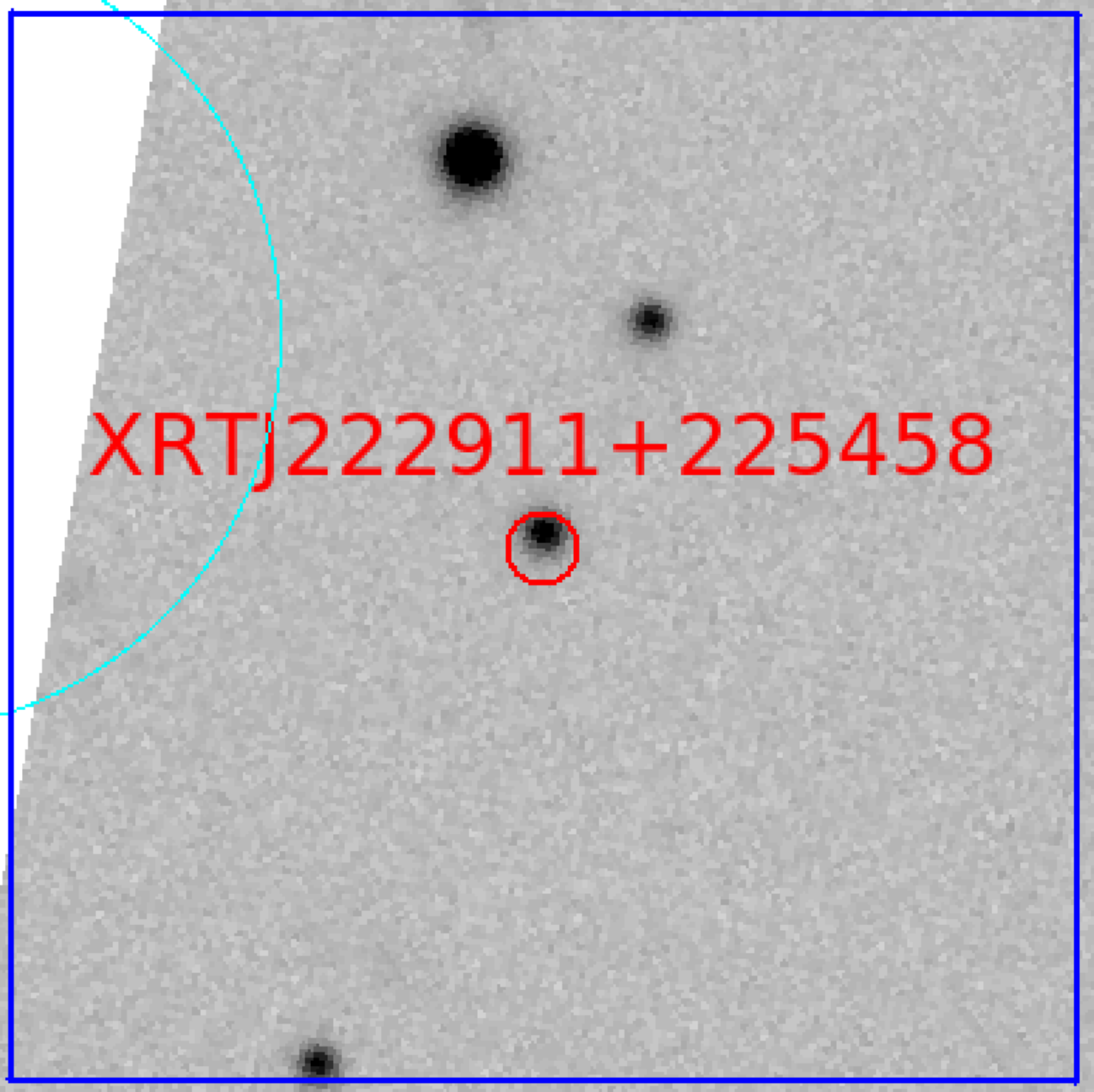}}
\subfigure[\protect\url{3FGLJ2244.6+2503}\label{fig:2244opt}]%
{\includegraphics[width=0.3\textwidth]{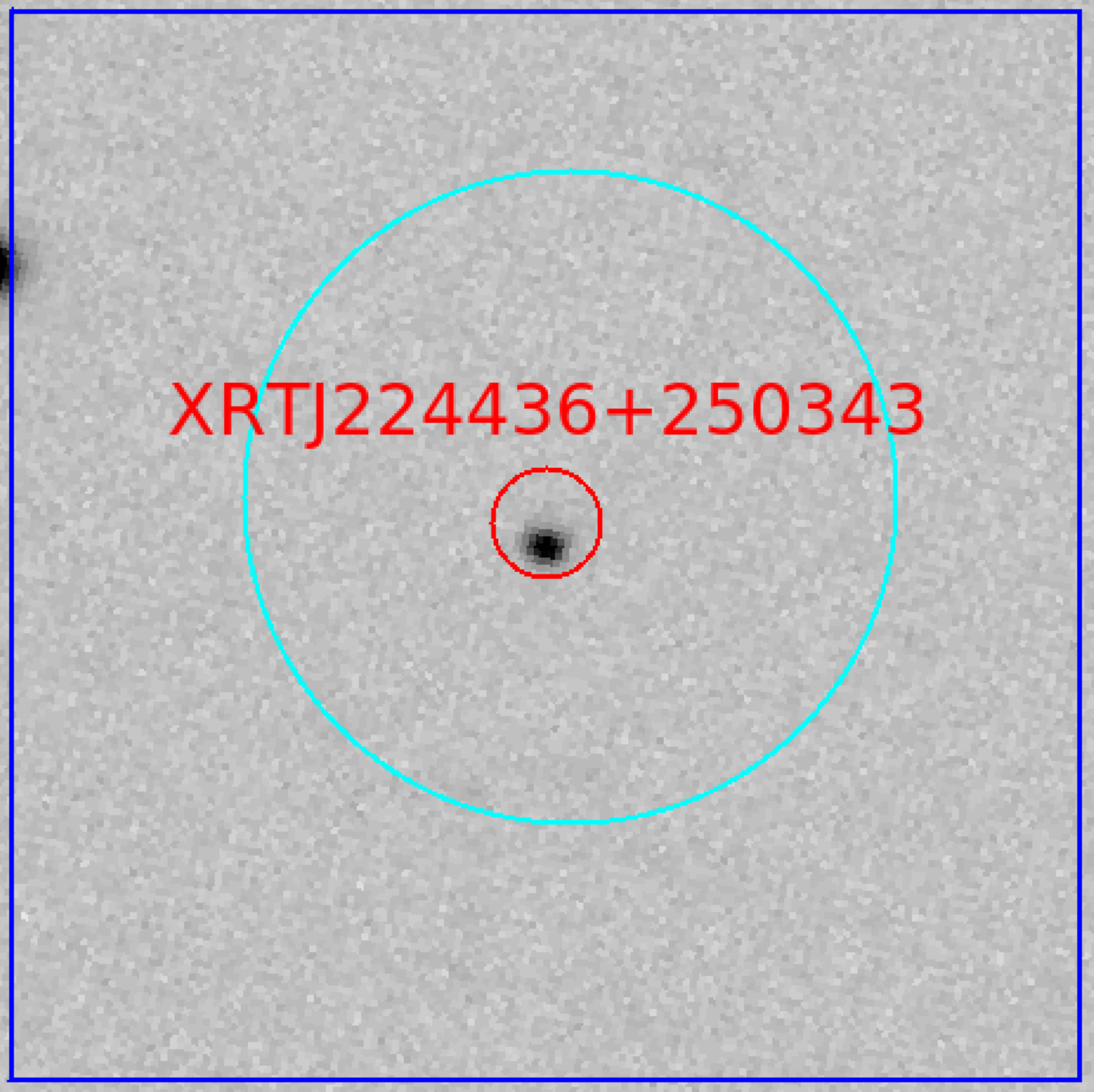}}
\subfigure[\protect\url{3FGLJ2250.3+1747}\label{fig:2250opt}]%
{\includegraphics[width=0.3\textwidth]{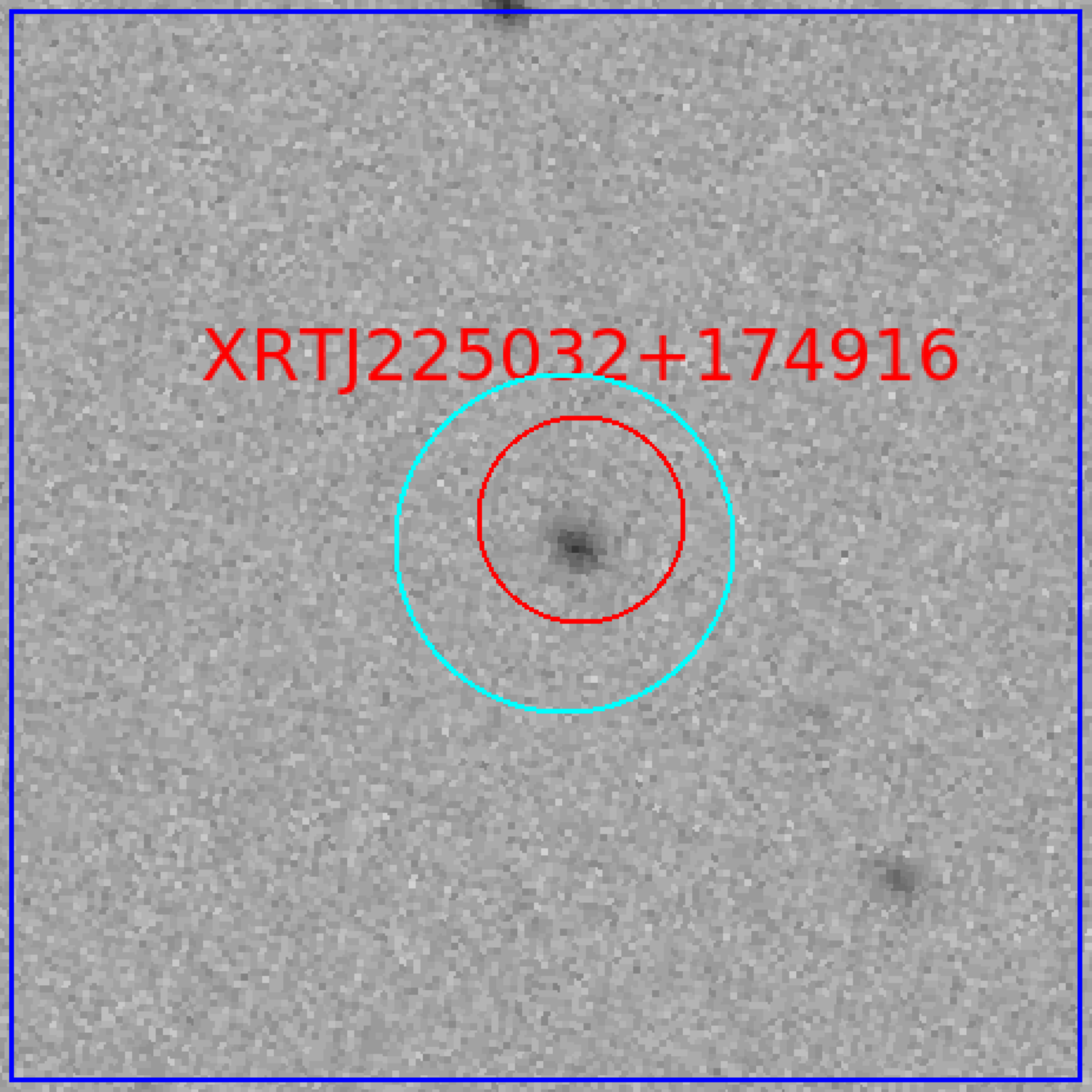}}
\subfigure[\protect\url{3FGLJ2321.6-1619}\label{fig:2321opt}]%
{\includegraphics[width=0.3\textwidth]{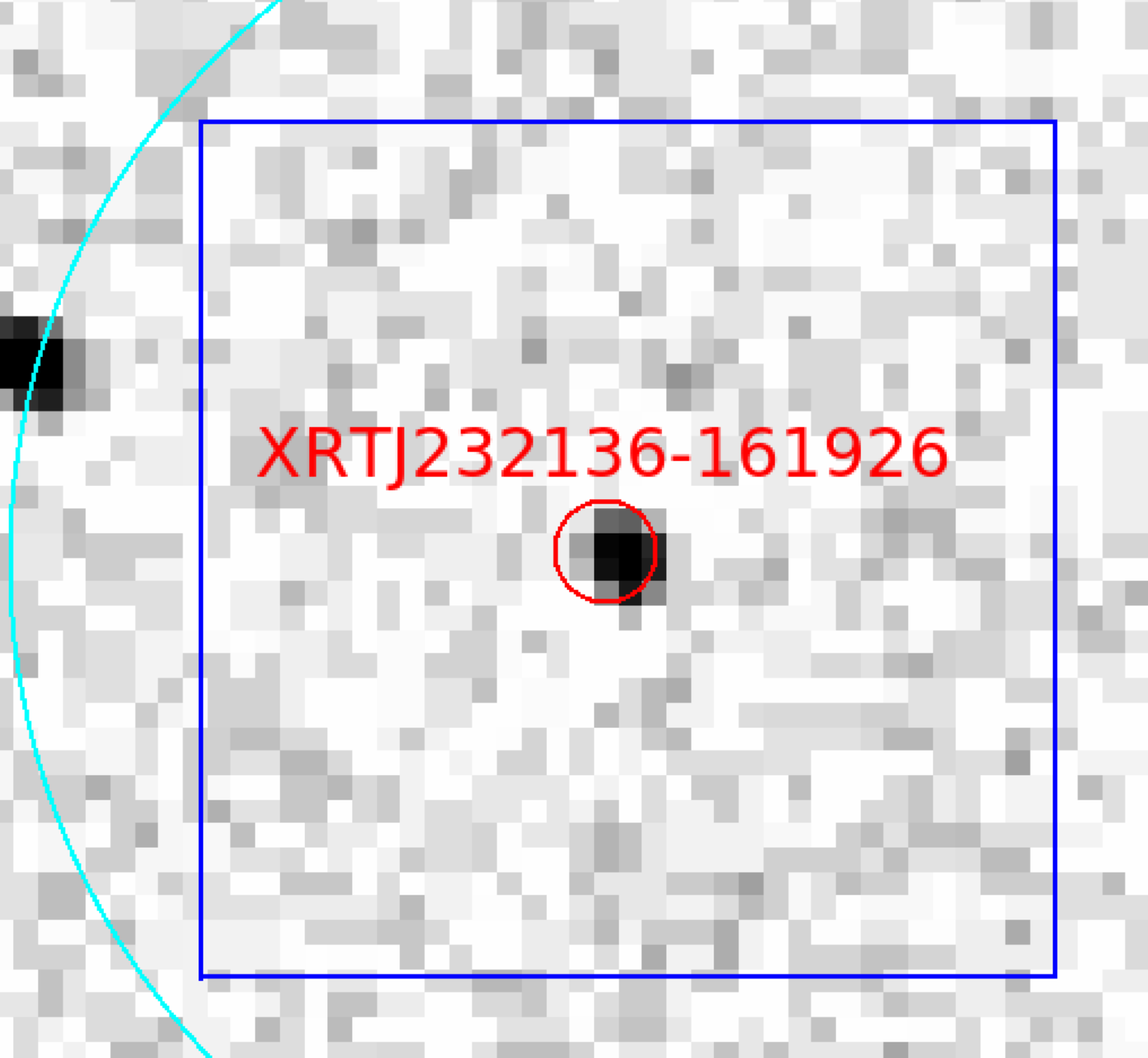}}
\subfigure[\protect\url{3FGLJ2358.6-1809}\label{fig:2358m18opt}]%
{\includegraphics[width=0.3\textwidth]{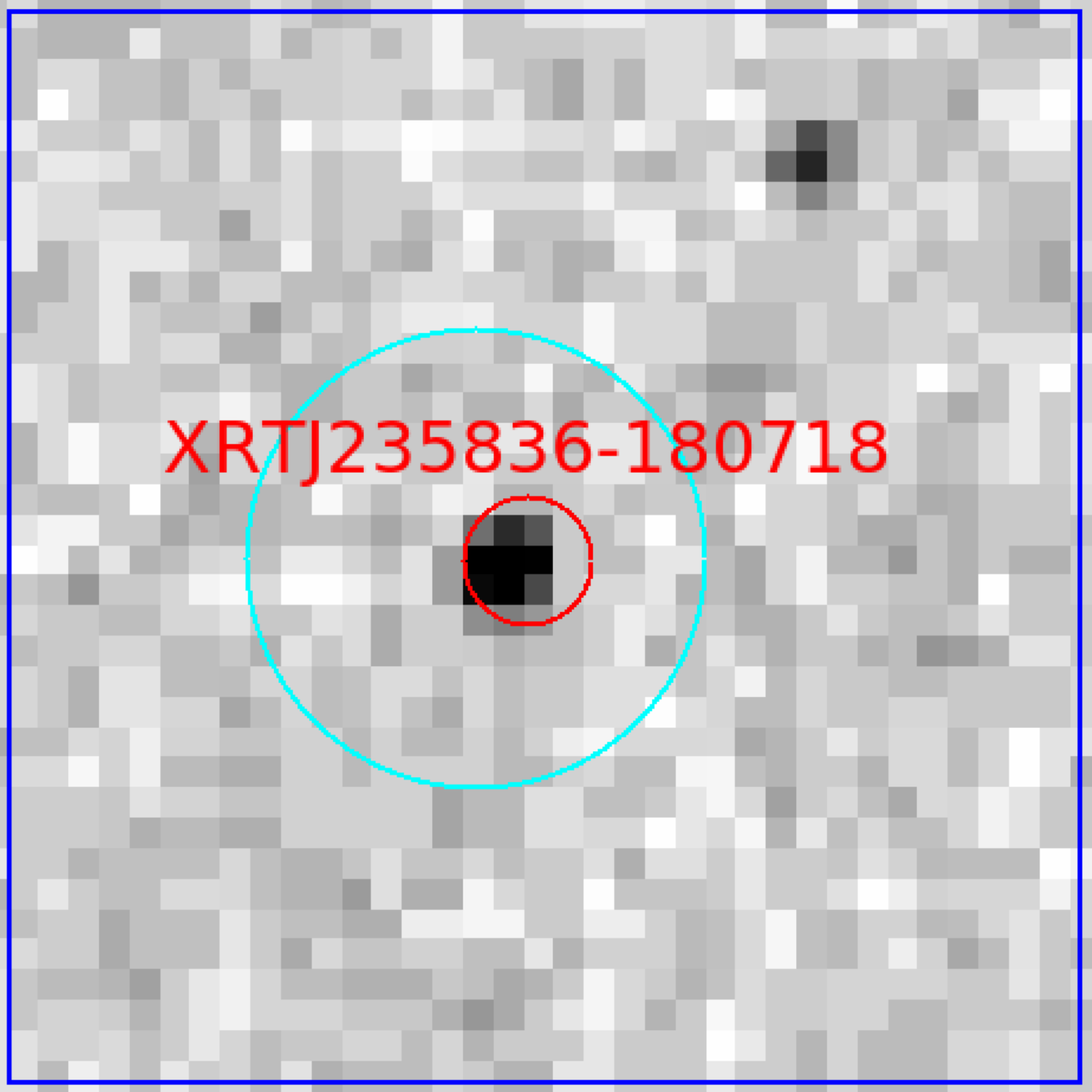}}
\subfigure[\protect\url{3FGLJ2358.5+3827}\label{fig:2358p38opt}]%
{\includegraphics[width=0.3\textwidth]{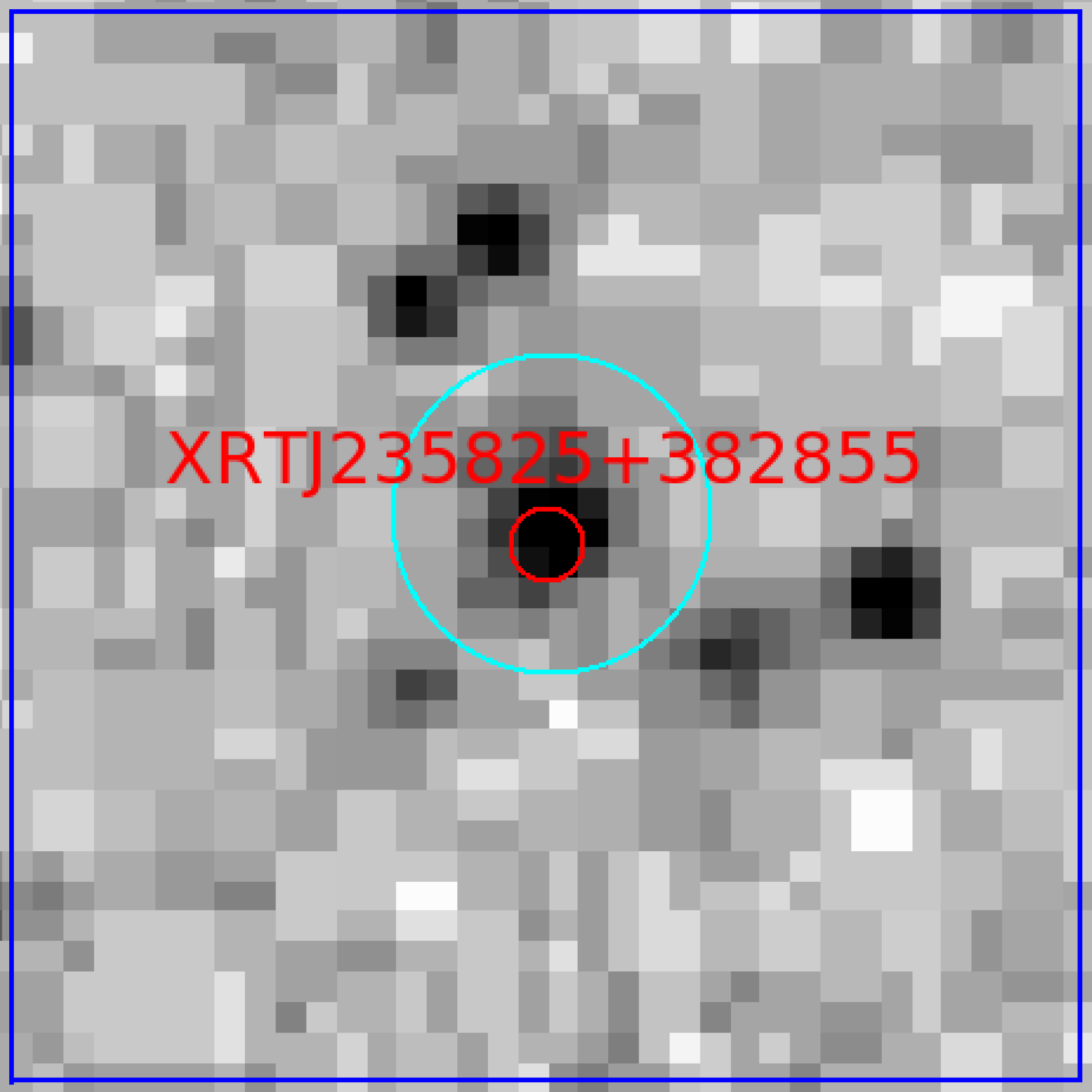}}
\caption{Continued from Fig. \ref{fig:fc1}.}
\end{figure*}

\newpage
\setcounter{figure}{1}
\begin{figure*}
\includegraphics[width=0.4\textwidth, angle=-90]{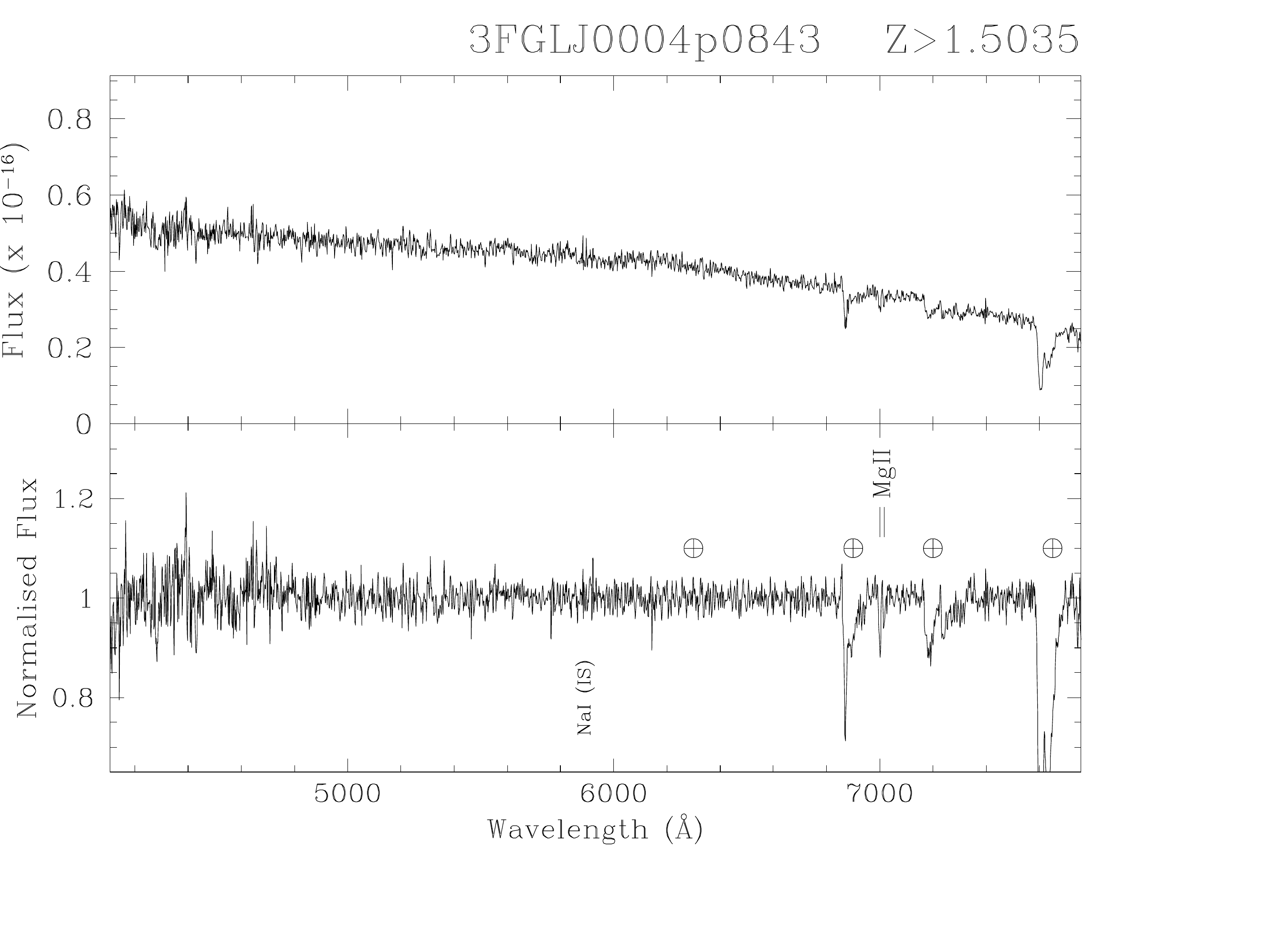}
\includegraphics[width=0.4\textwidth, angle=-90]{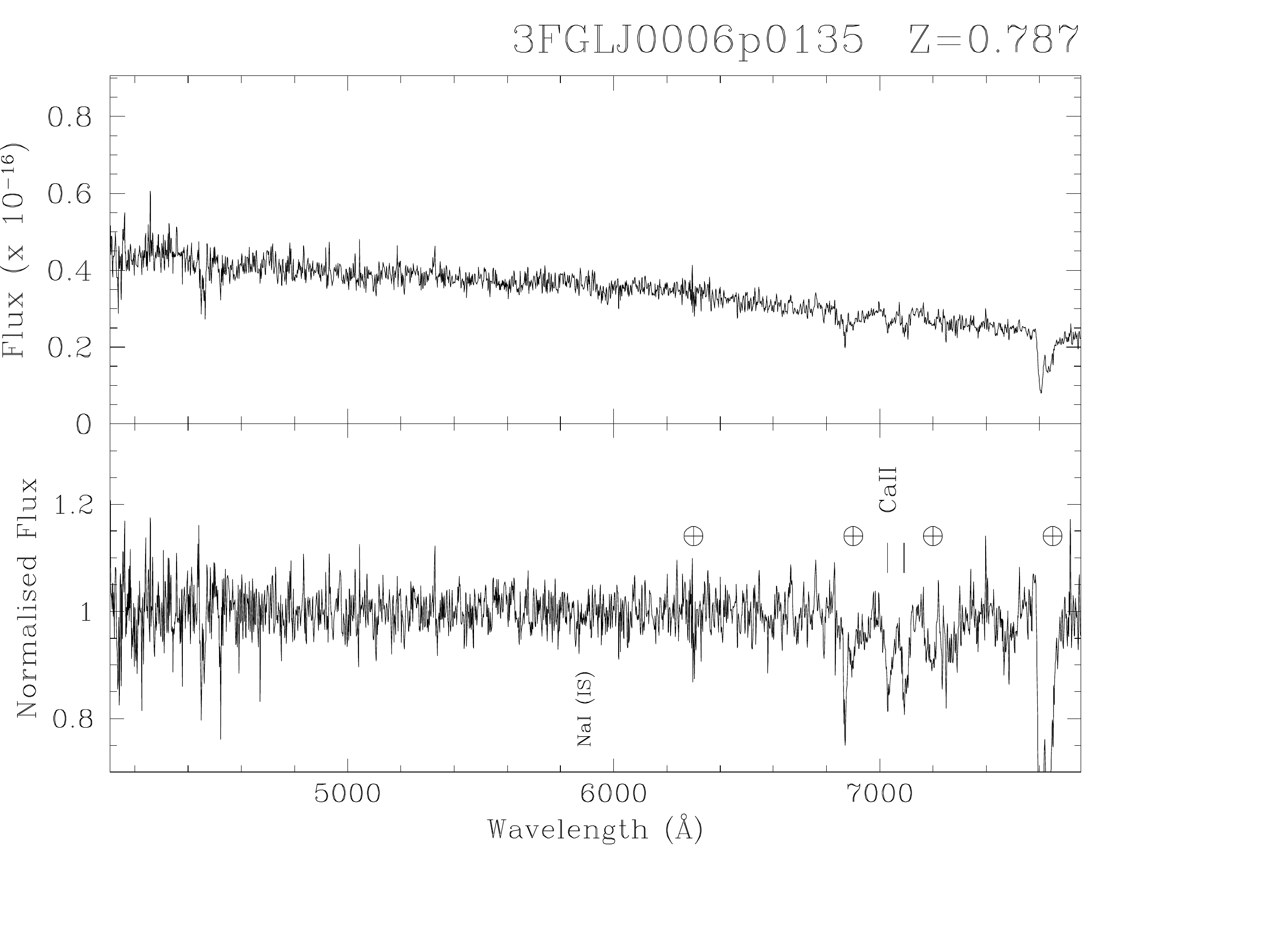}
\includegraphics[width=0.4\textwidth, angle=-90]{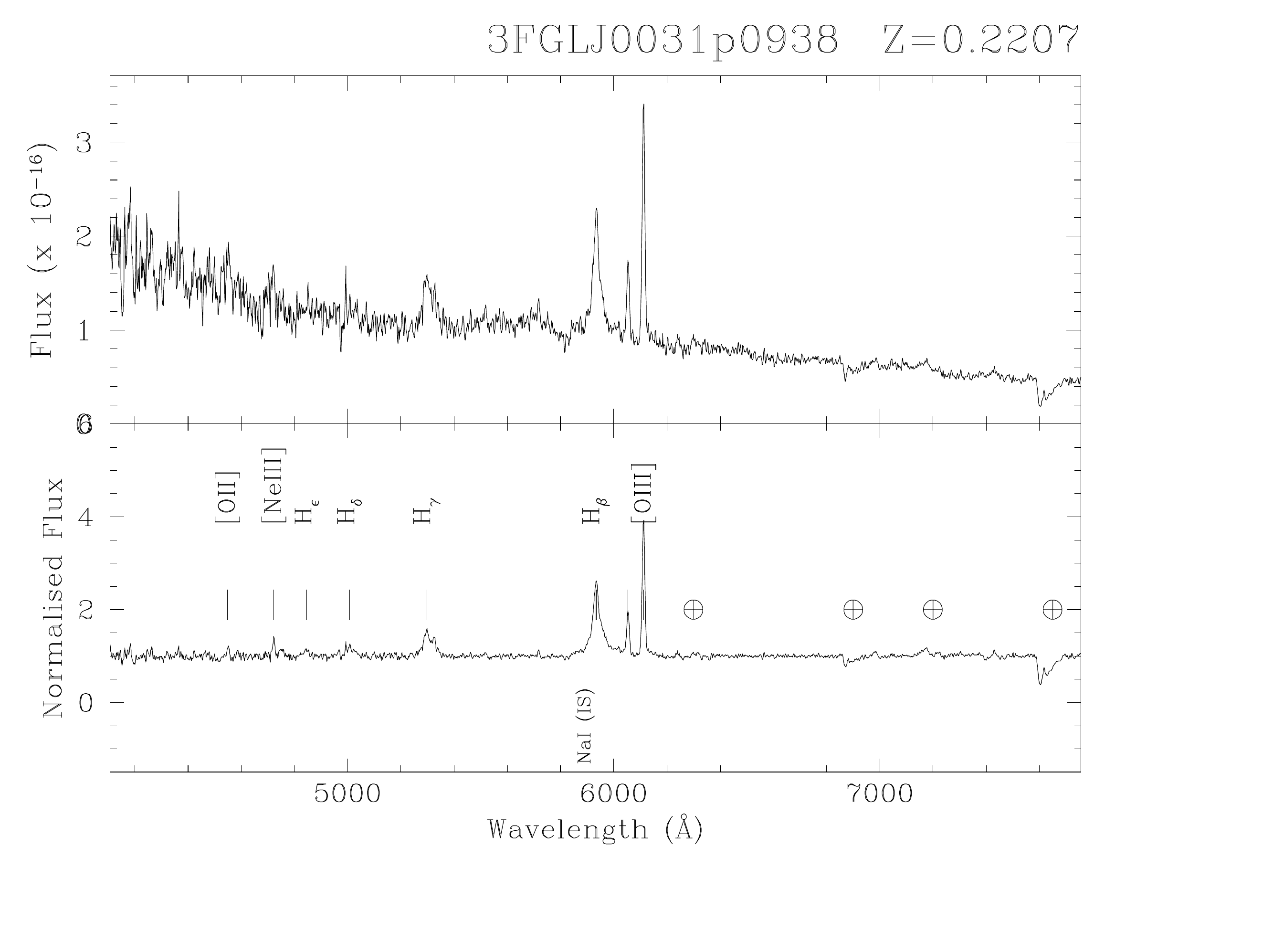}
\includegraphics[width=0.4\textwidth, angle=-90]{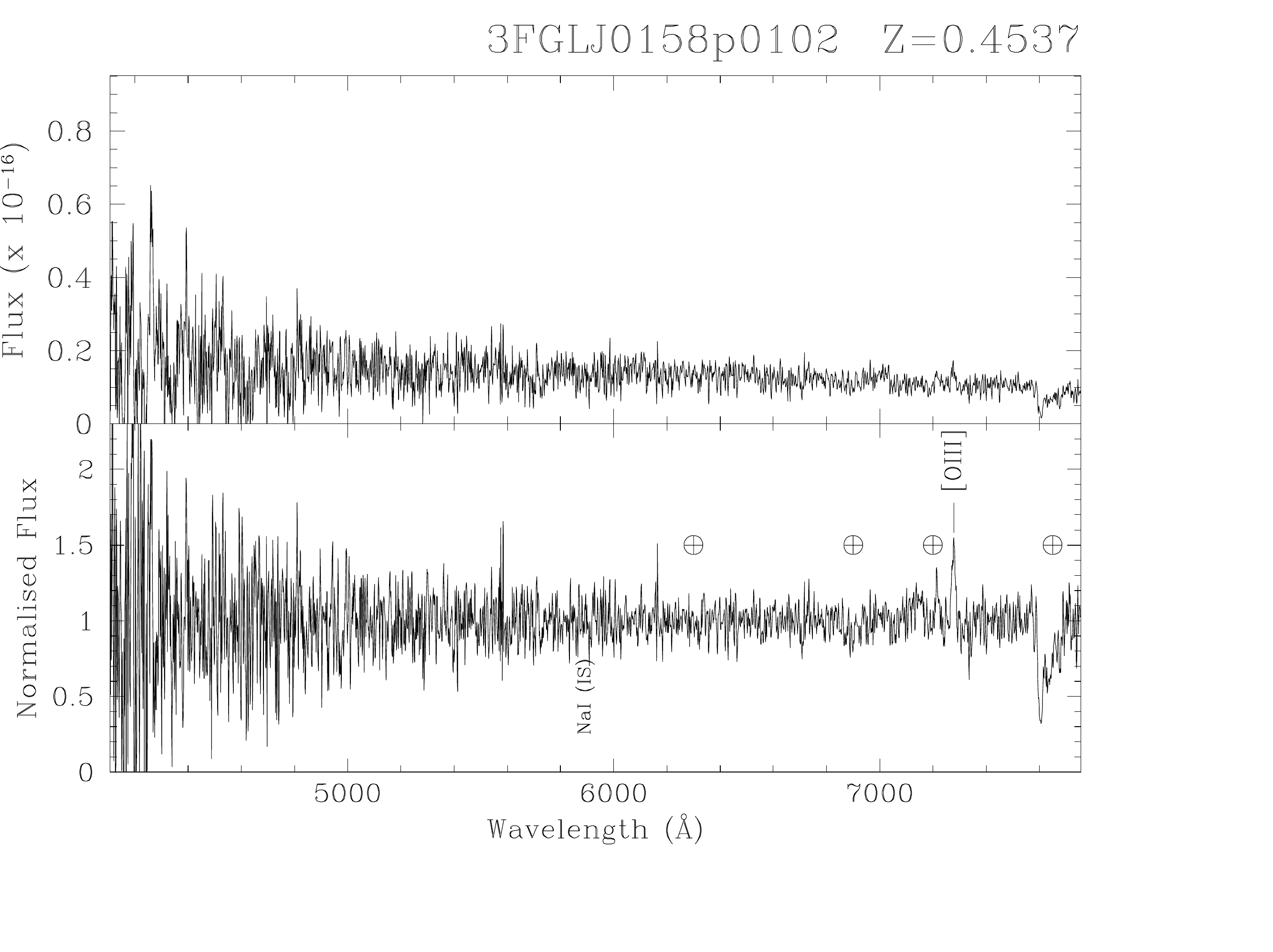}
\includegraphics[width=0.4\textwidth, angle=-90]{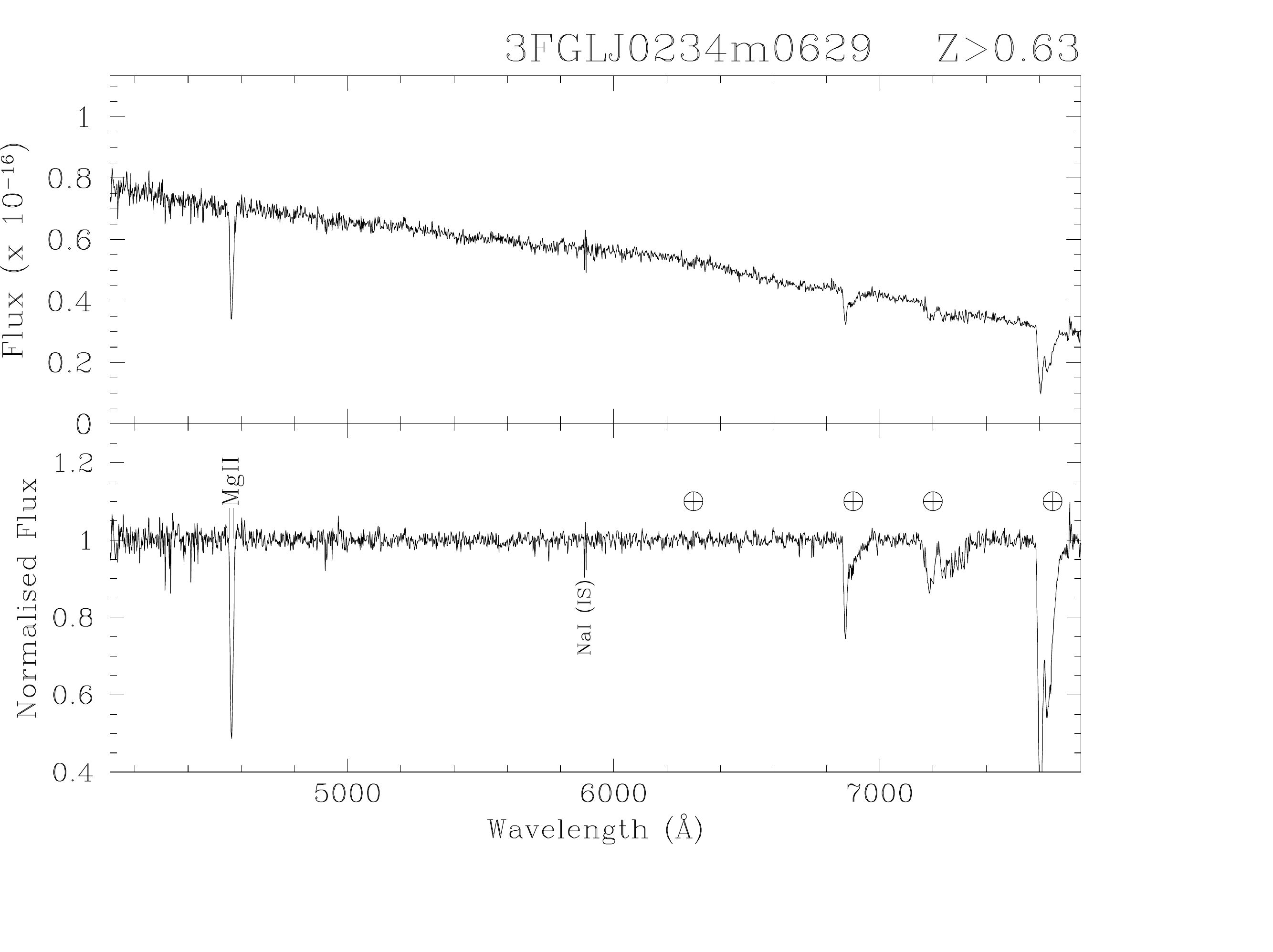}
\includegraphics[width=0.4\textwidth, angle=-90]{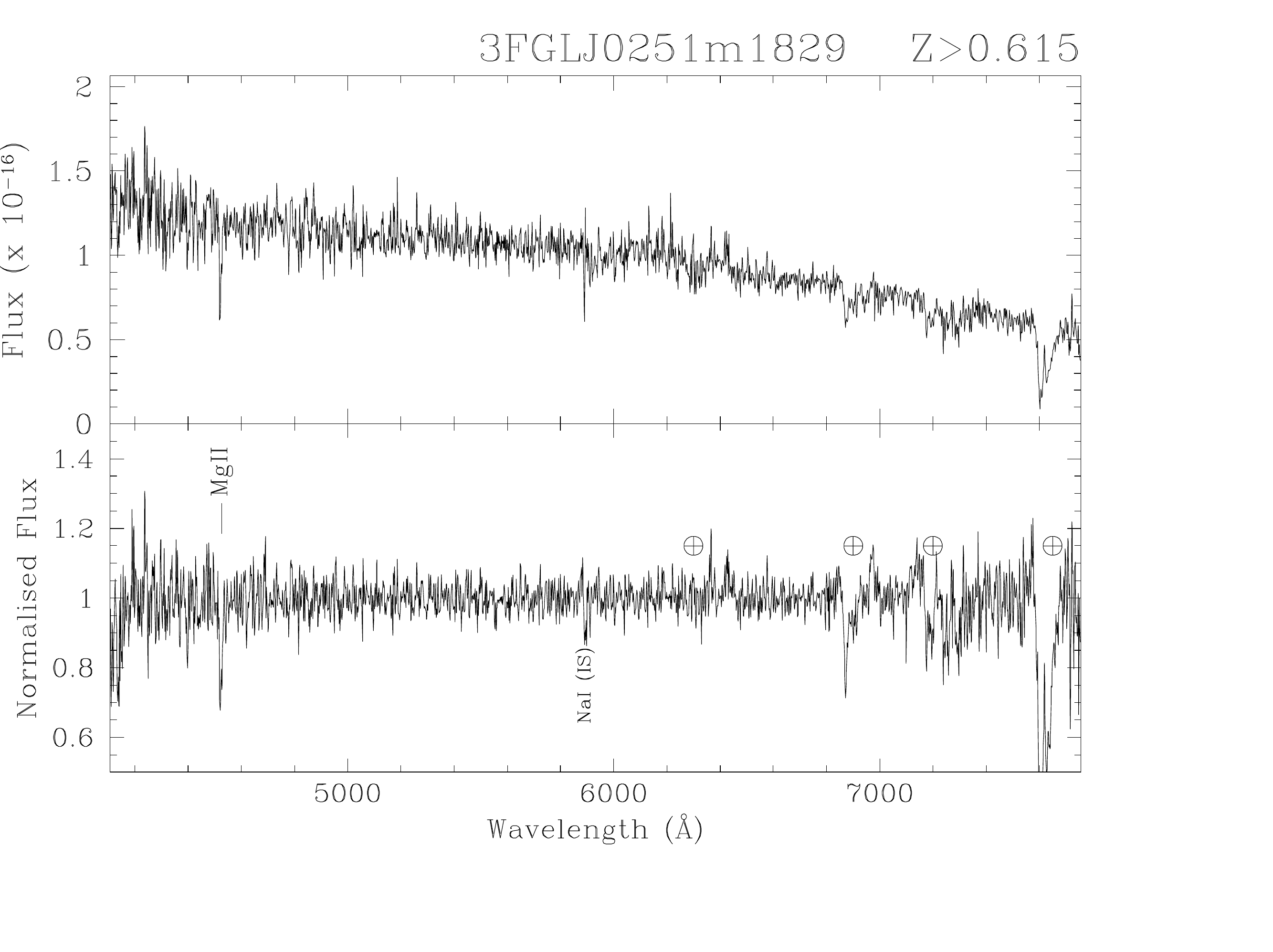}
\caption{Spectra of the UGSs obtained at GTC. \textit{Top panel}: Flux calibrated and dereddered spectra. \textit{Bottom panel}: Normalized spectra. The main telluric bands are indicated by $\oplus$, the absorption features from interstellar medium of our galaxies are labelled as IS (Inter-Stellar)}
\label{fig:fig1}
\end{figure*}

\setcounter{figure}{1}
\begin{figure*}
\includegraphics[width=0.4\textwidth, angle=-90]{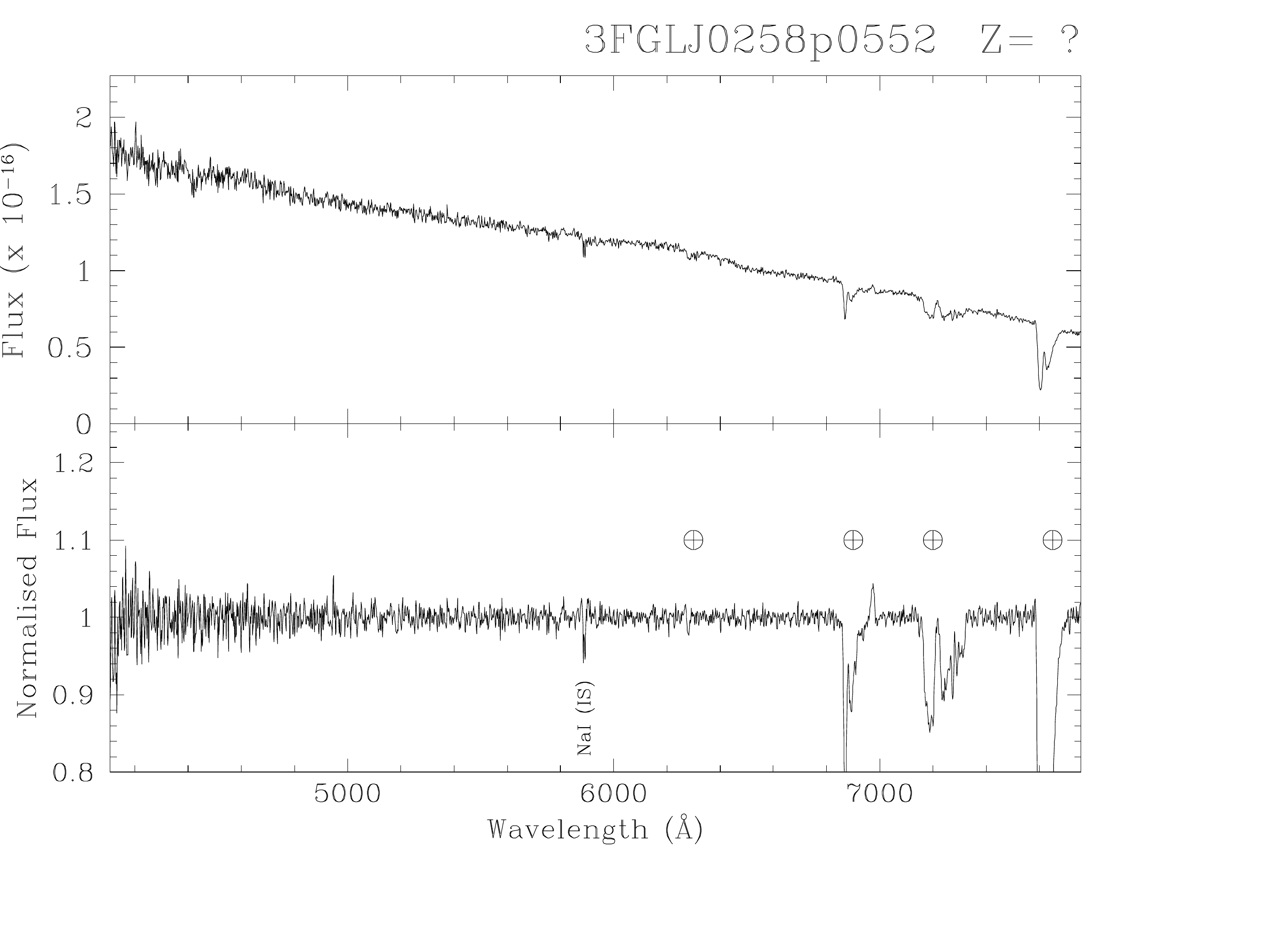}
\includegraphics[width=0.4\textwidth, angle=-90]{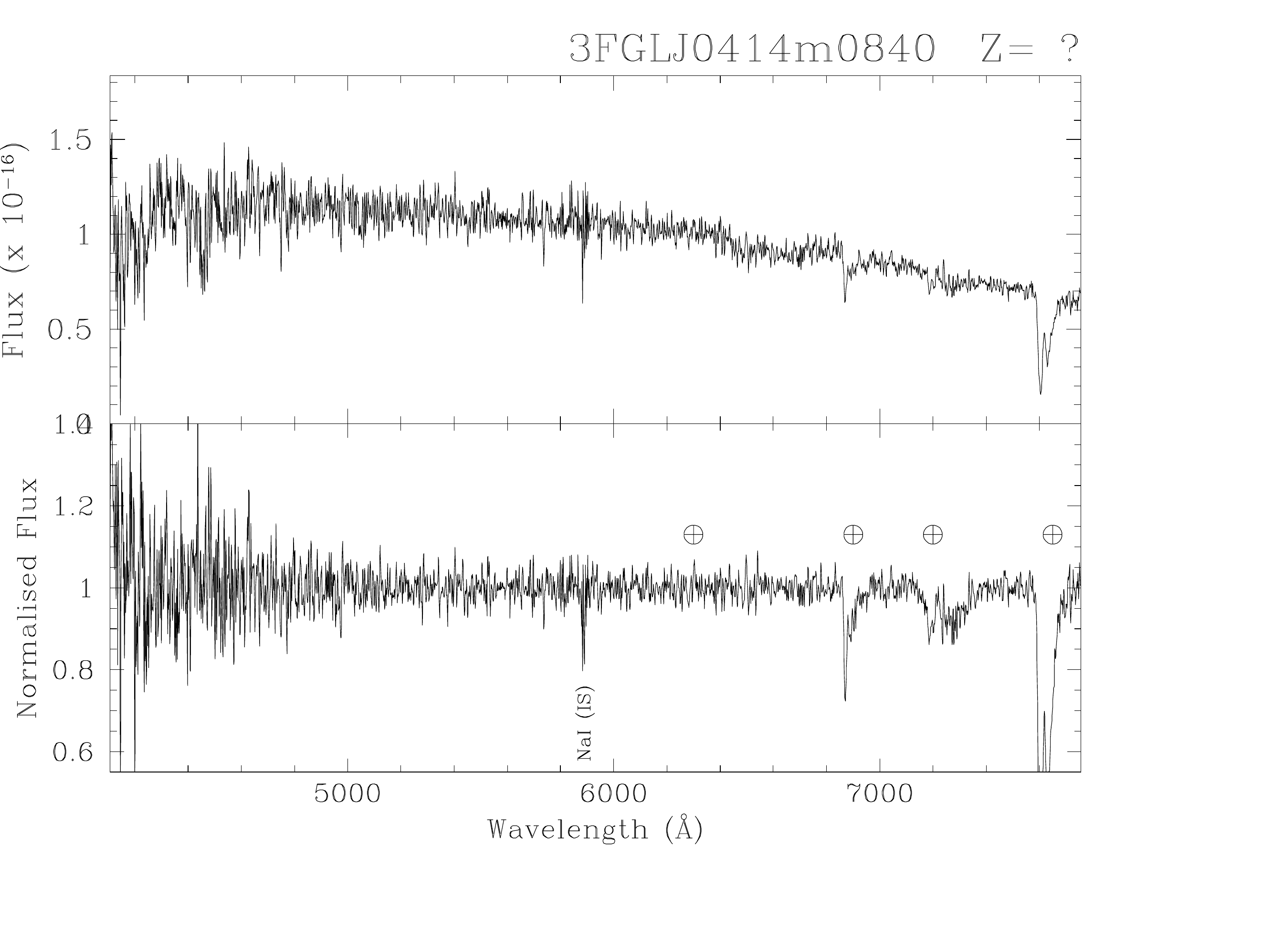}
\includegraphics[width=0.4\textwidth, angle=-90]{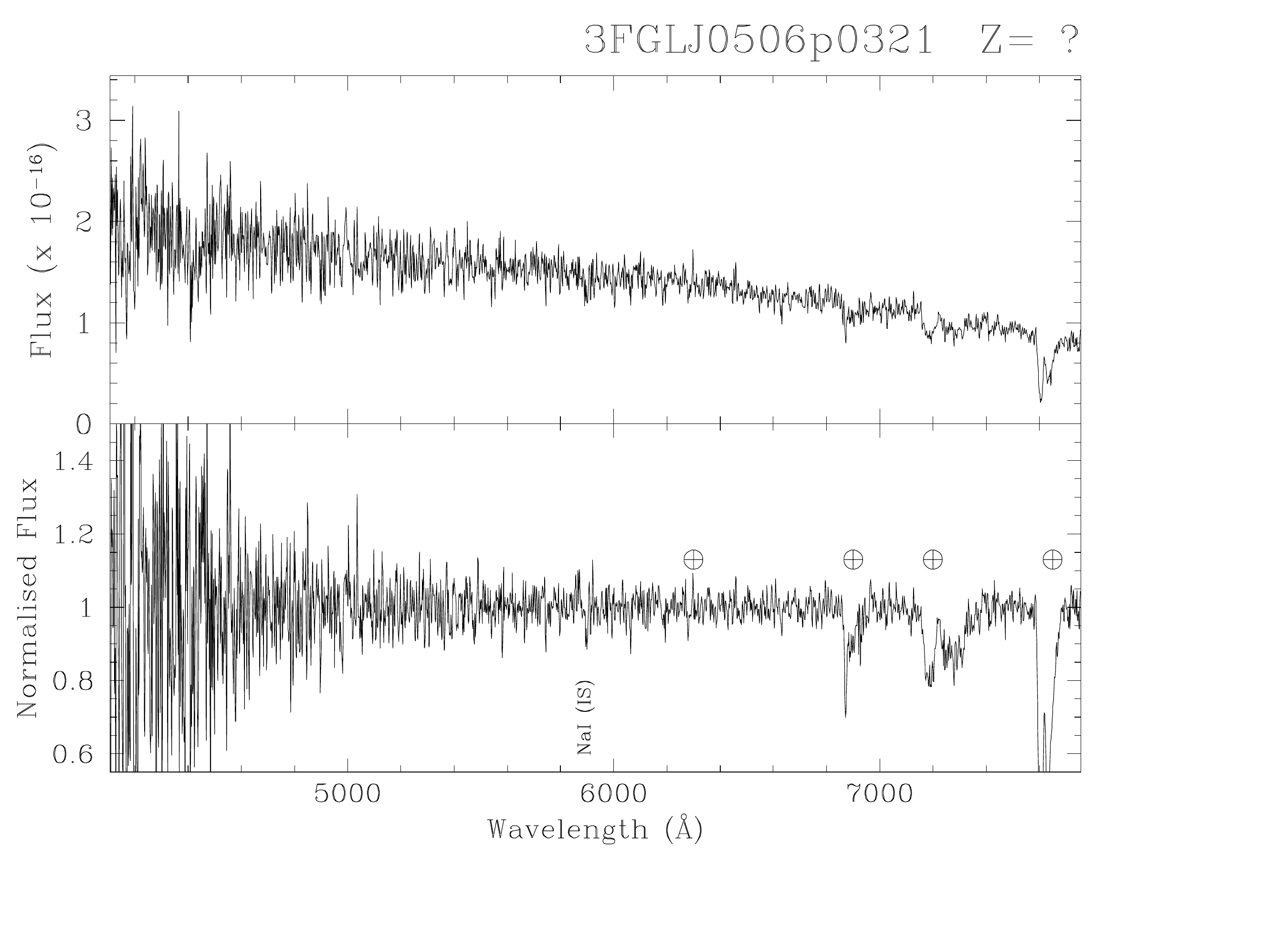}
\includegraphics[width=0.4\textwidth, angle=-90]{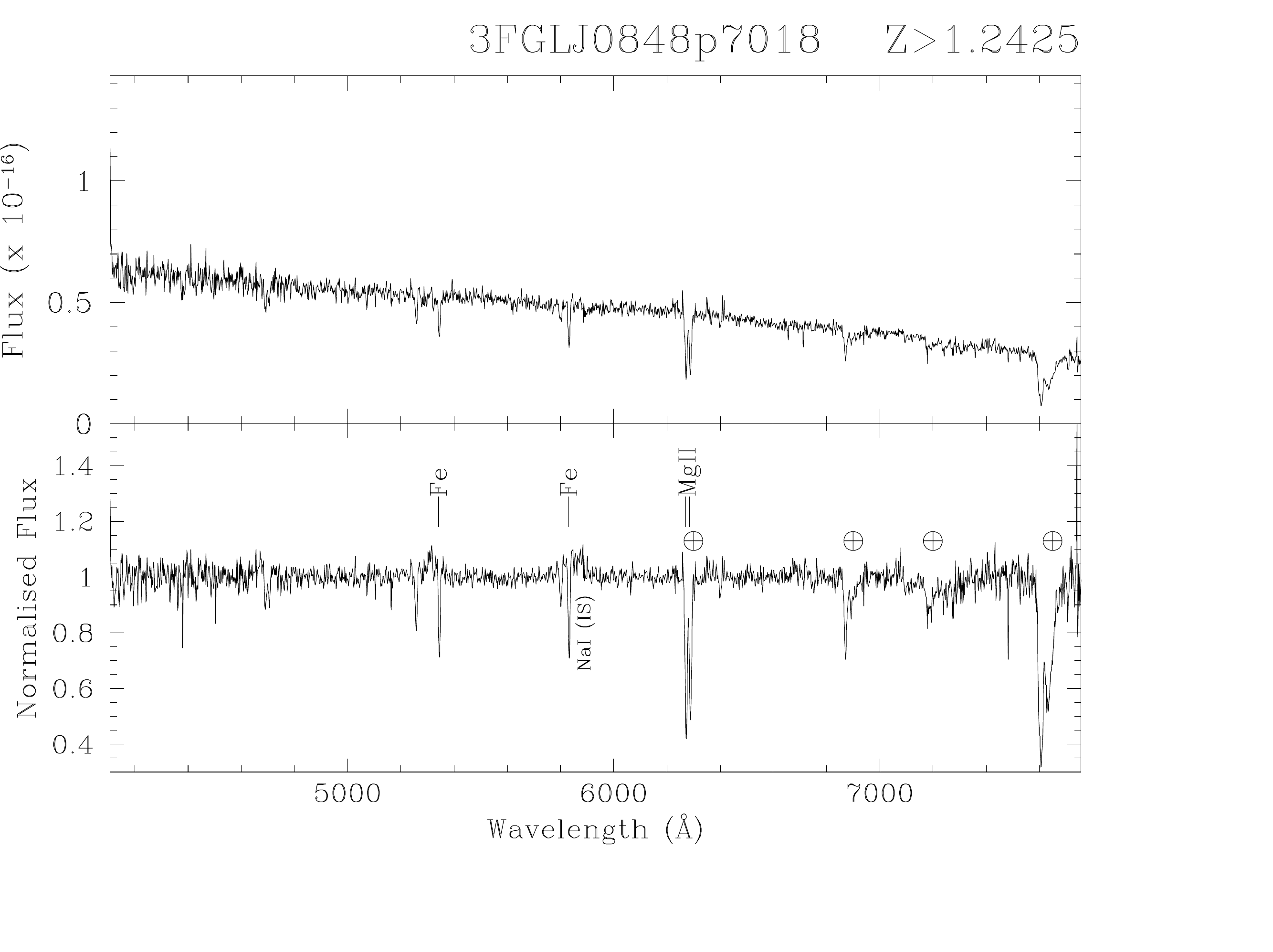}
\includegraphics[width=0.4\textwidth, angle=-90]{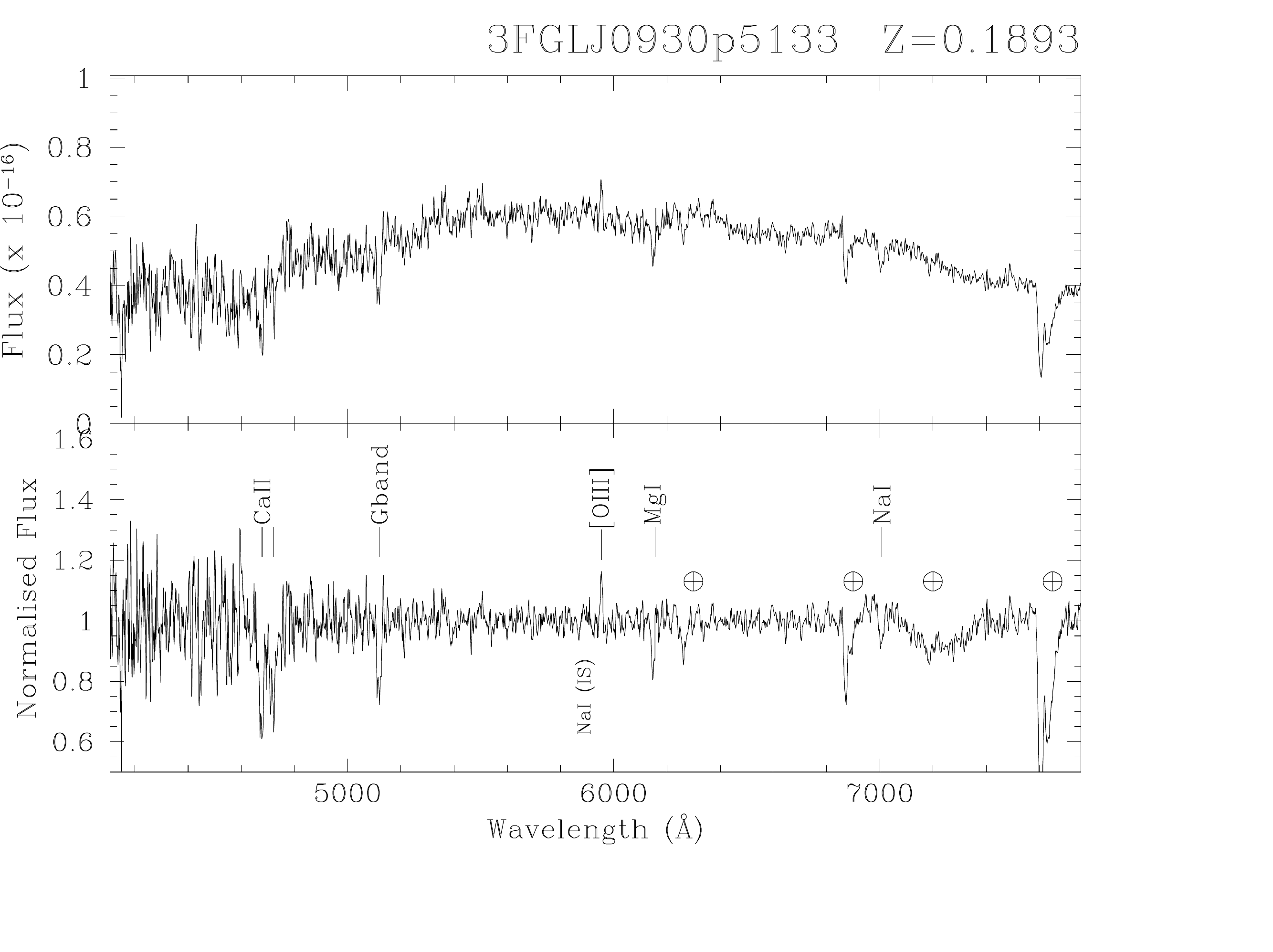}
\includegraphics[width=0.4\textwidth, angle=-90]{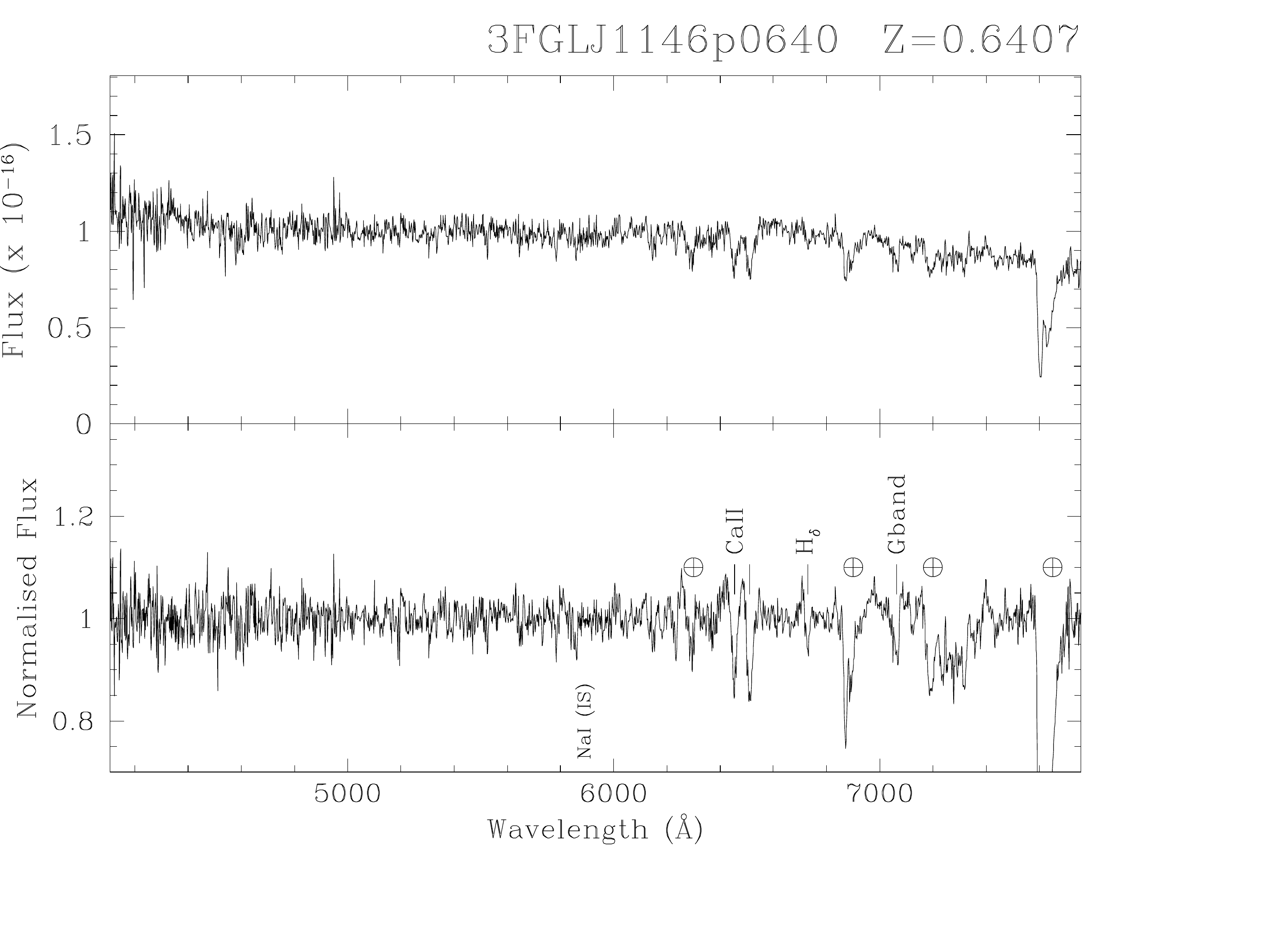}
\caption{Continued from Fig. \ref{fig:fig1}.}
\end{figure*}

\setcounter{figure}{1}
\begin{figure*}
\includegraphics[width=0.4\textwidth, angle=-90]{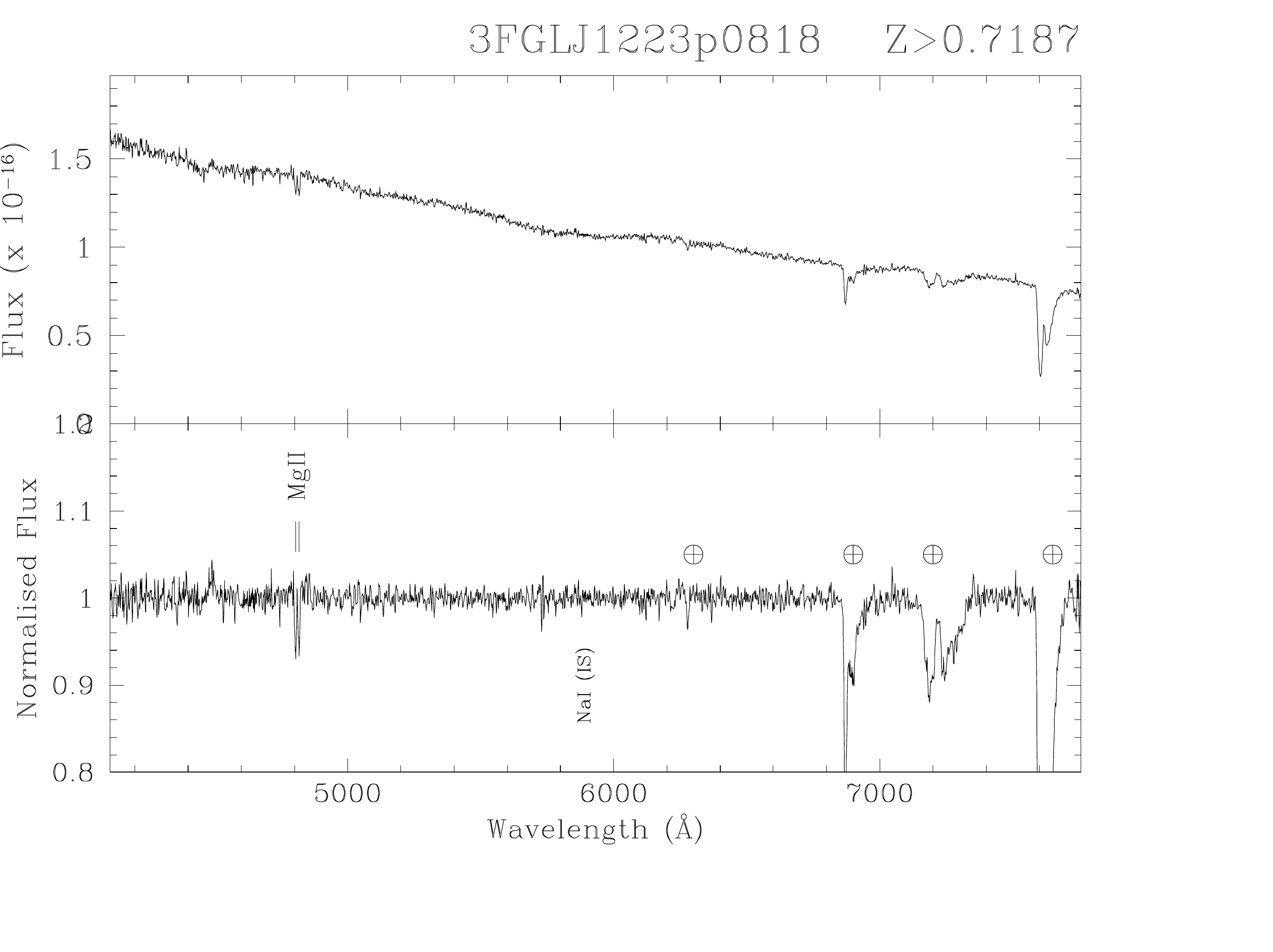}
\includegraphics[width=0.4\textwidth, angle=-90]{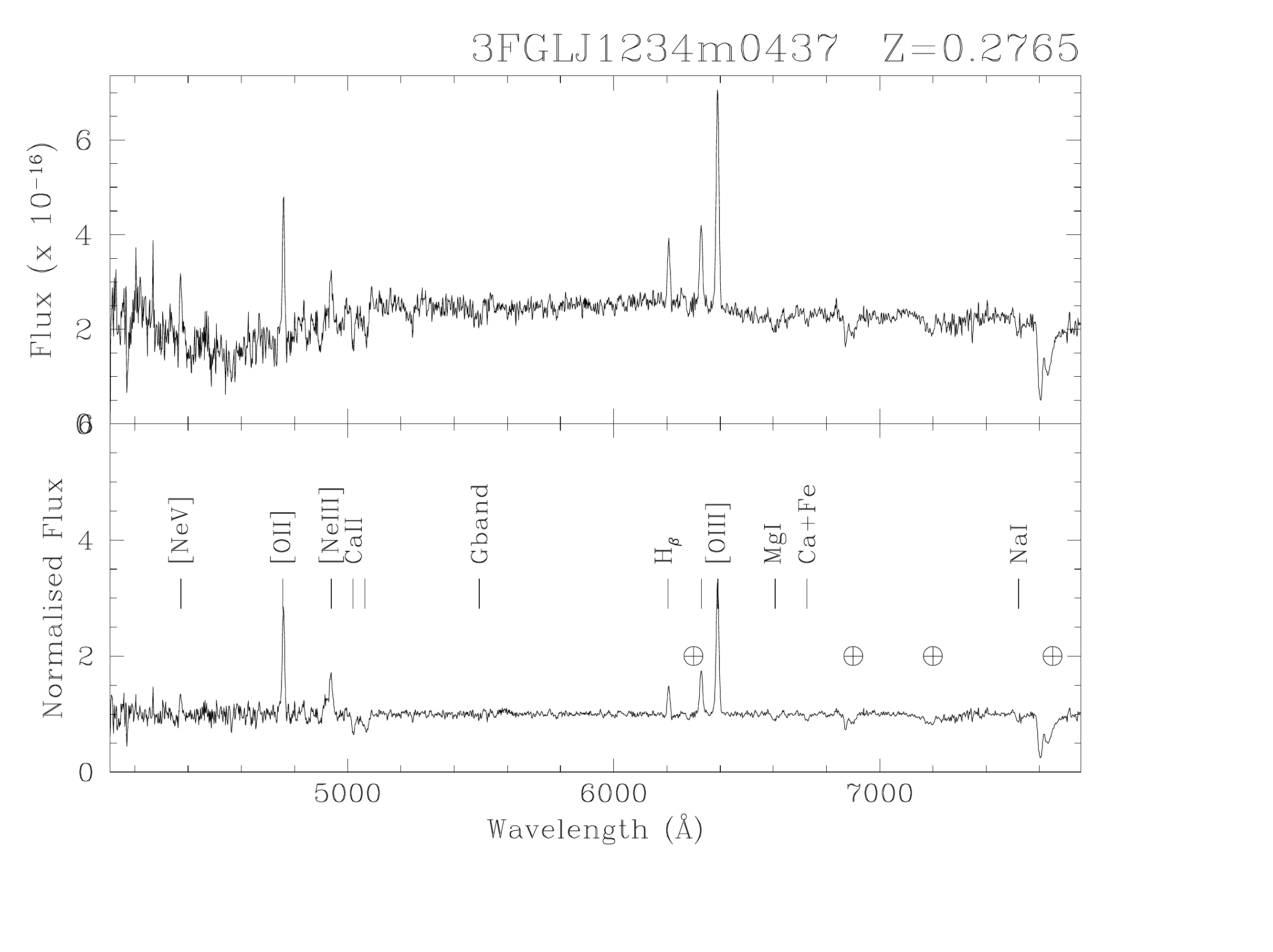}
\includegraphics[width=0.4\textwidth, angle=-90]{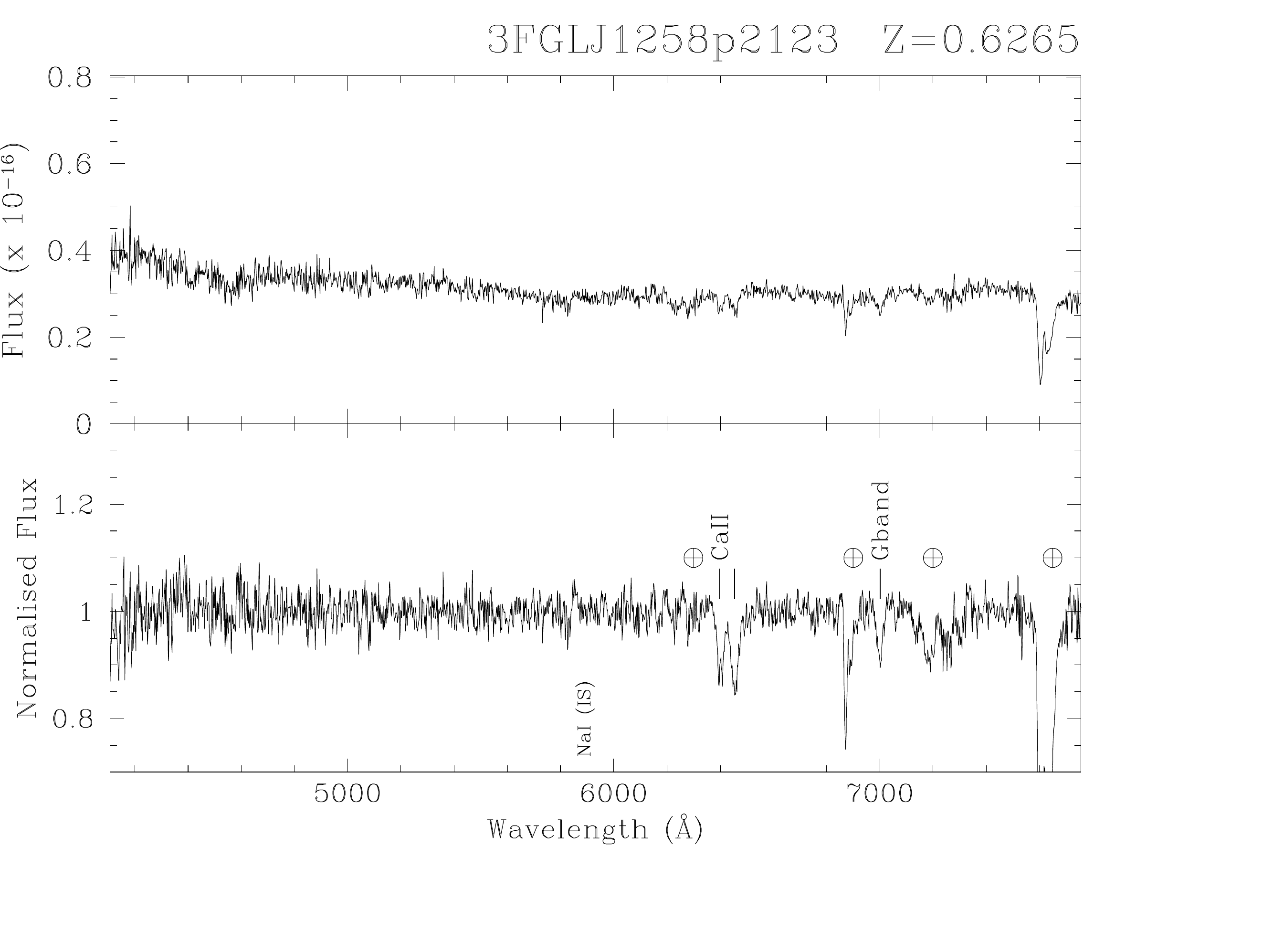}
\includegraphics[width=0.4\textwidth, angle=-90]{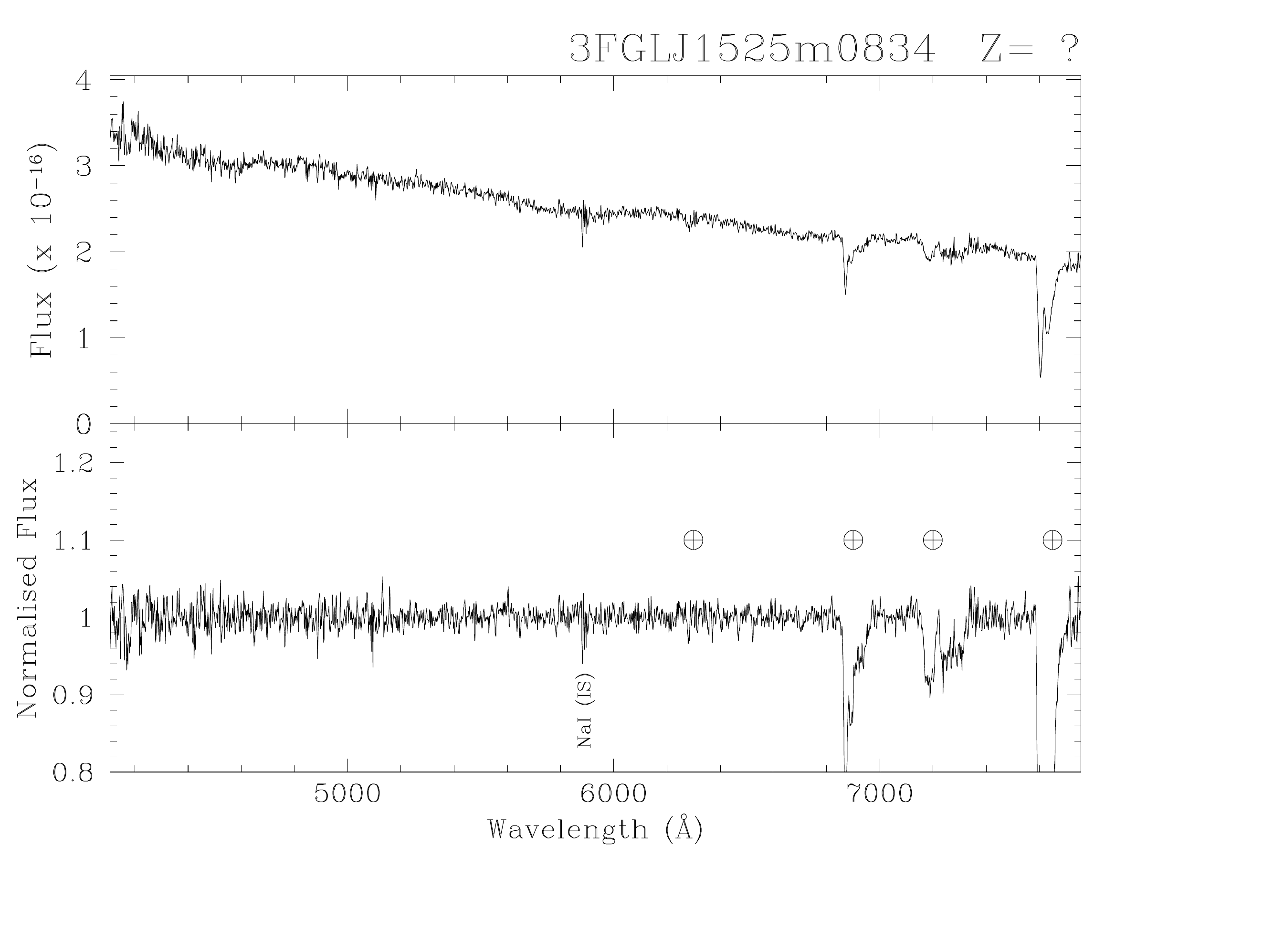}
\includegraphics[width=0.4\textwidth, angle=-90]{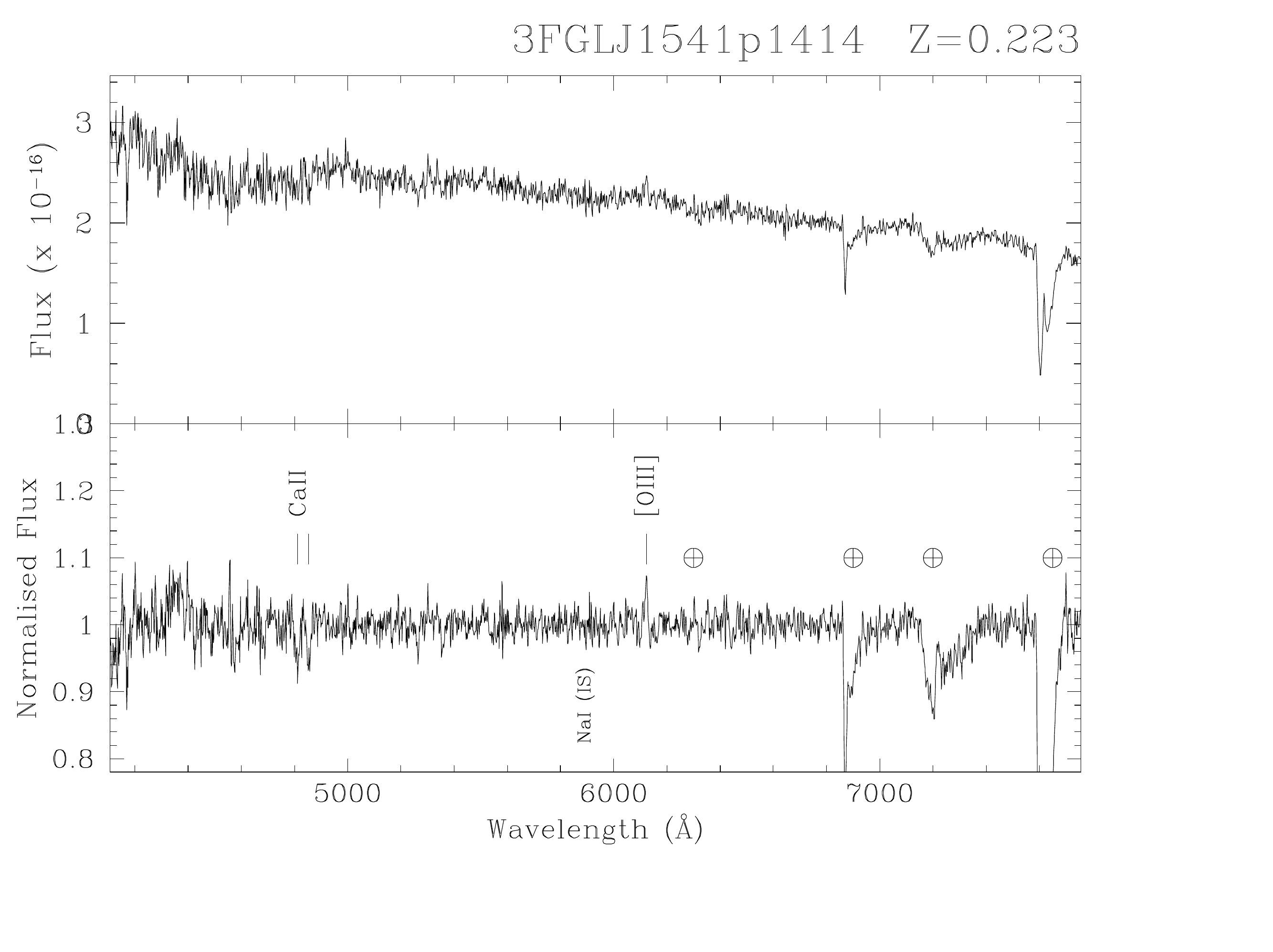}
\includegraphics[width=0.4\textwidth, angle=-90]{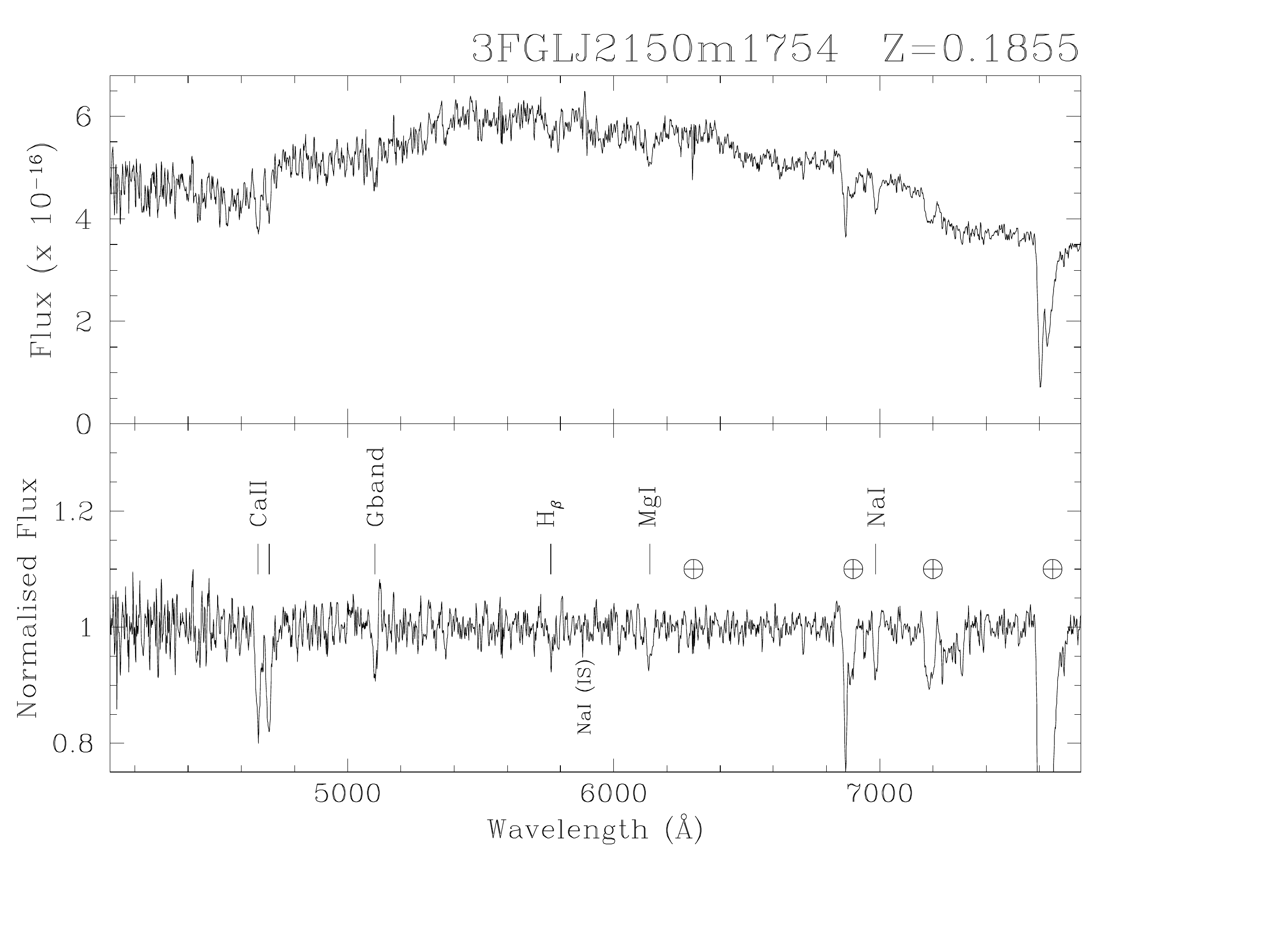}
\caption{Continued from Fig. \ref{fig:fig1}.}
\end{figure*}

\setcounter{figure}{1}
\begin{figure*}
\includegraphics[width=0.4\textwidth, angle=-90]{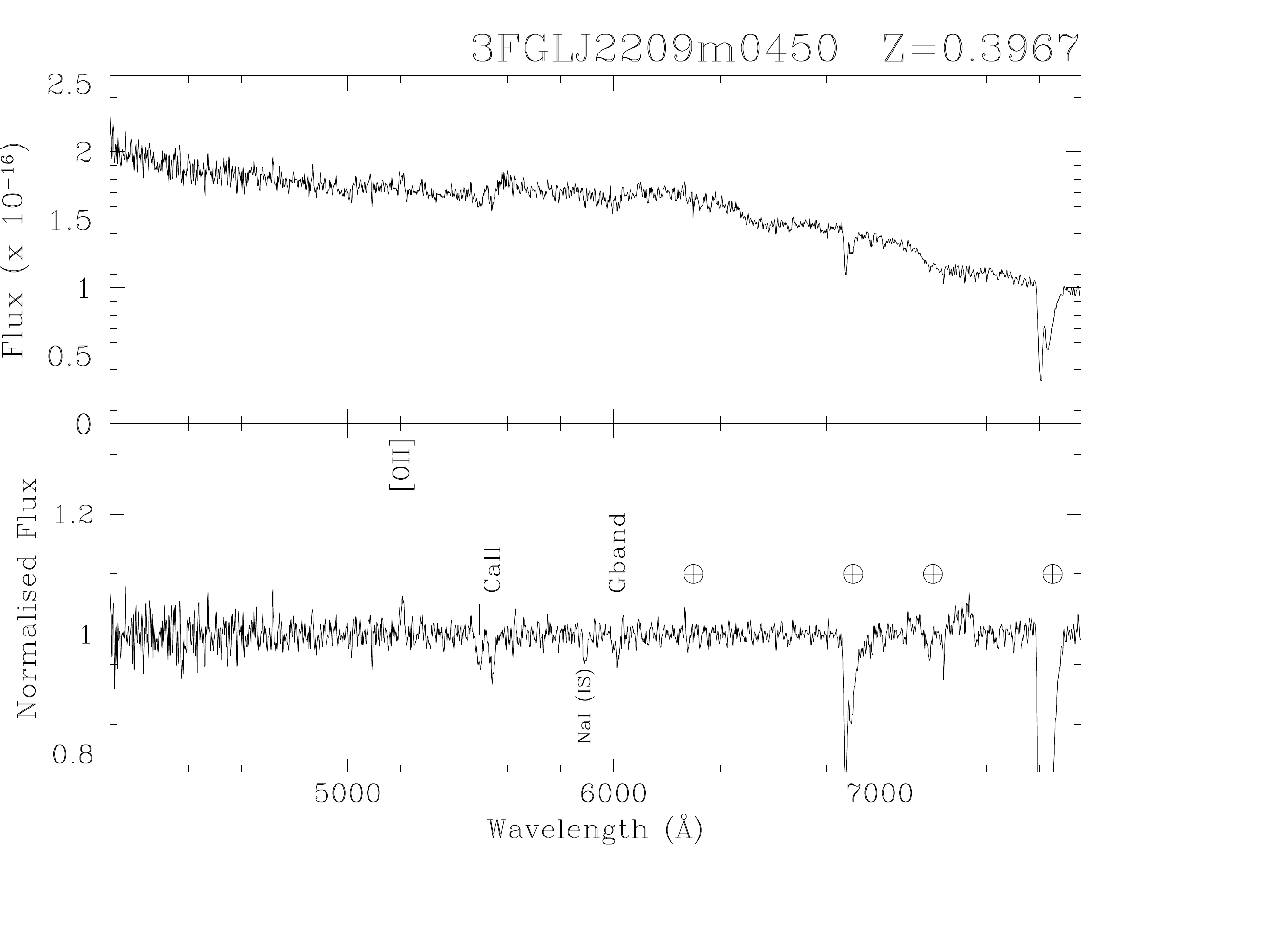}
\includegraphics[width=0.4\textwidth, angle=-90]{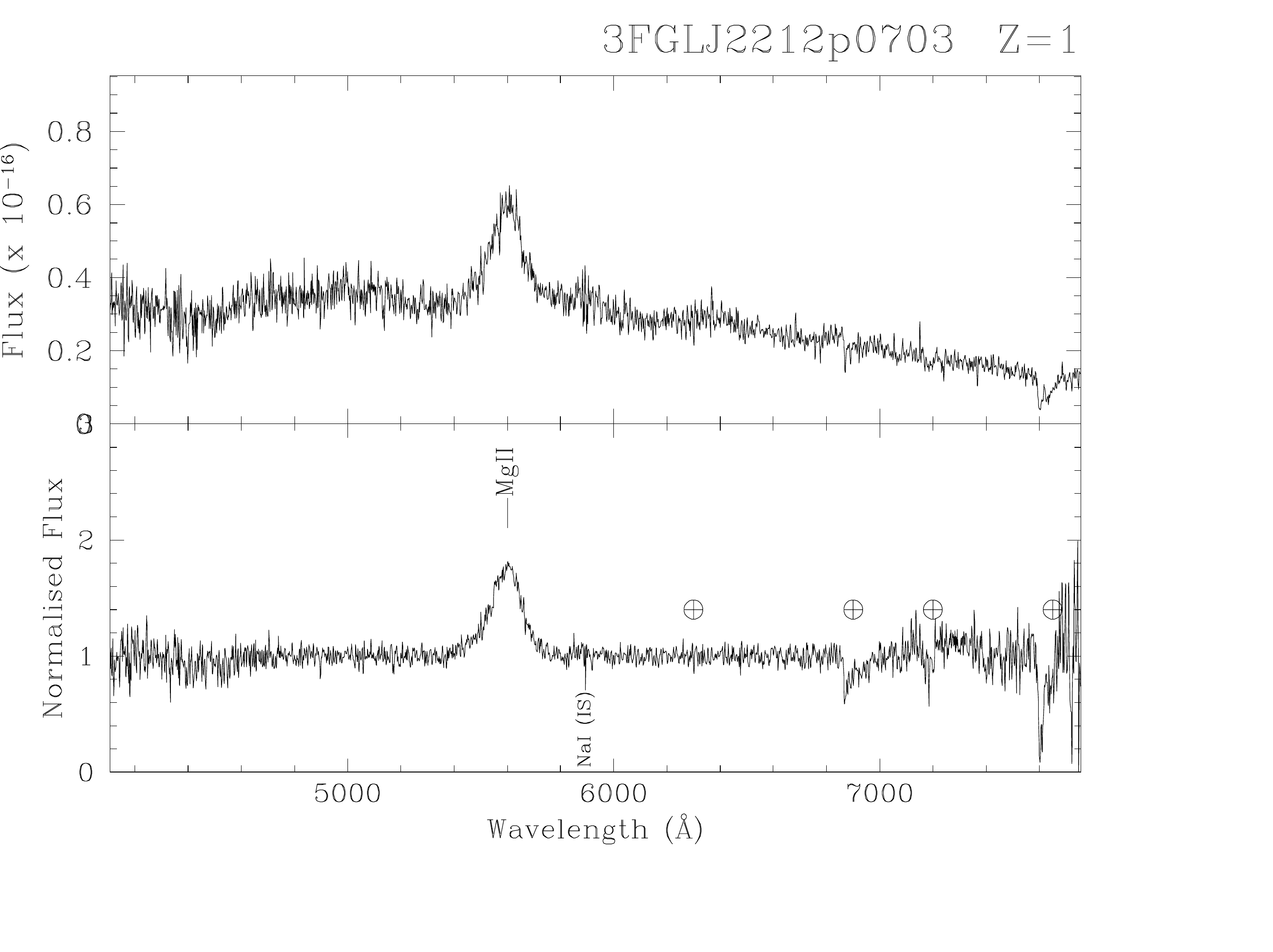}
\includegraphics[width=0.4\textwidth, angle=-90]{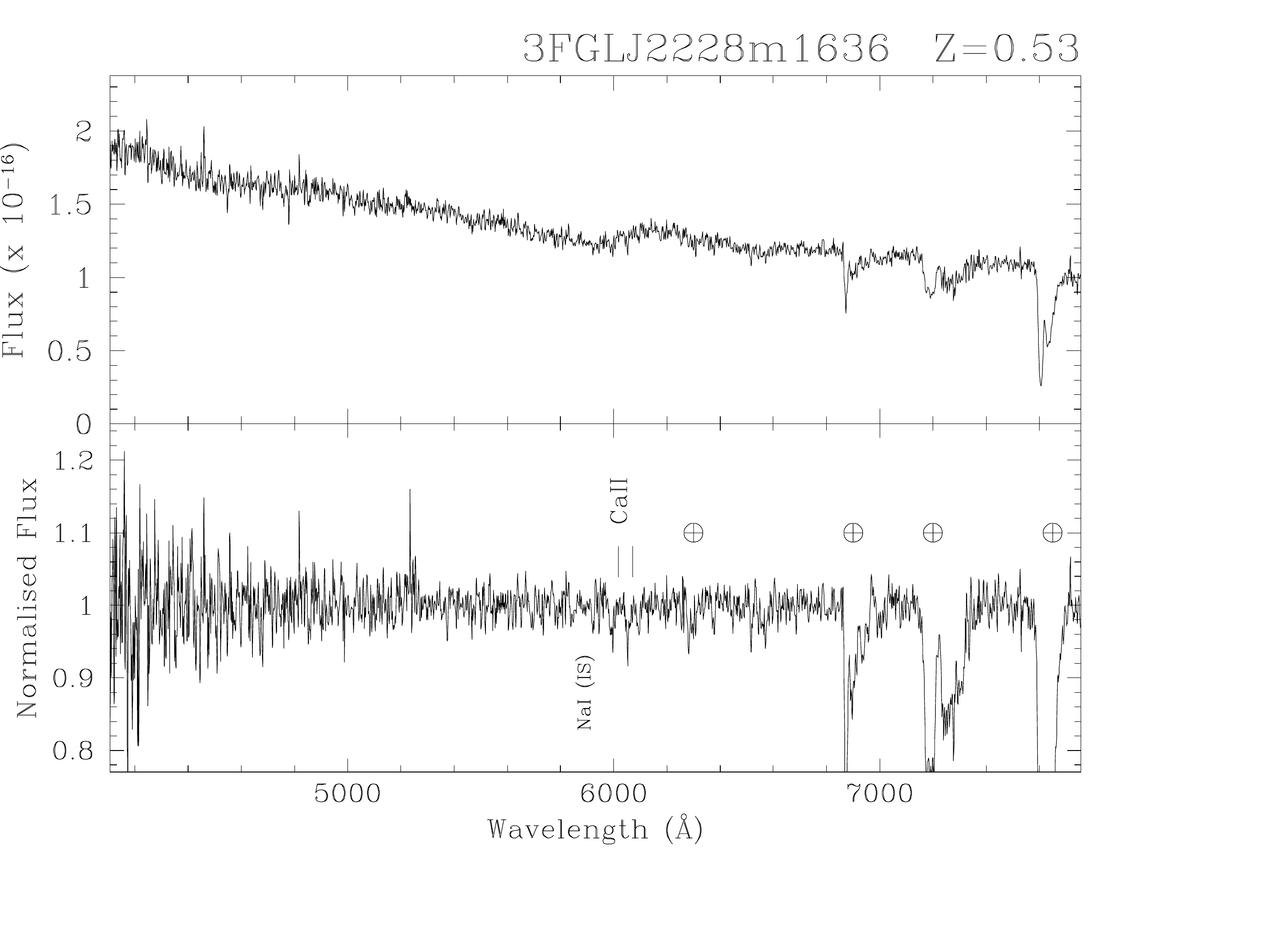}
\includegraphics[width=0.4\textwidth, angle=-90]{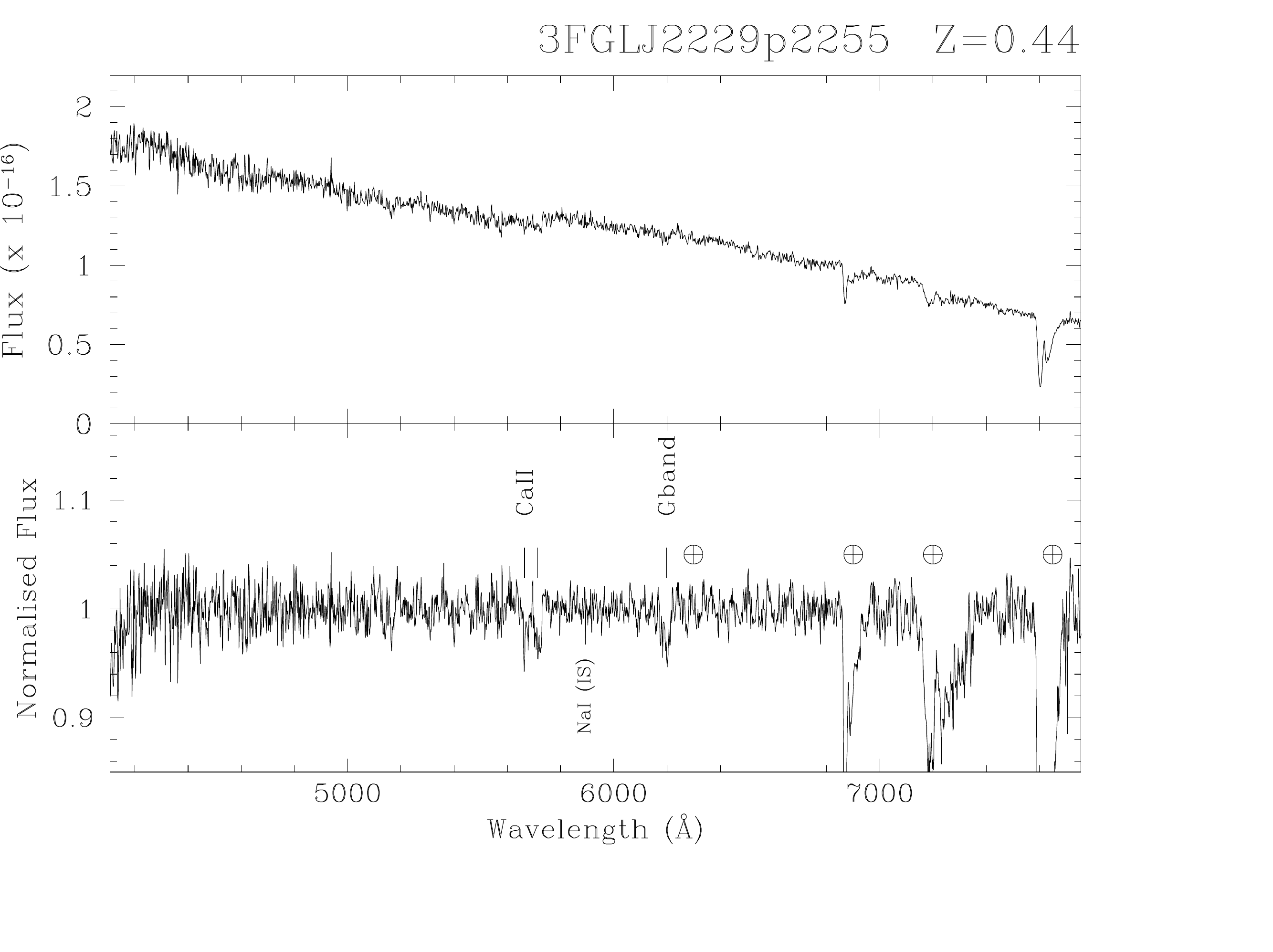}
\includegraphics[width=0.4\textwidth, angle=-90]{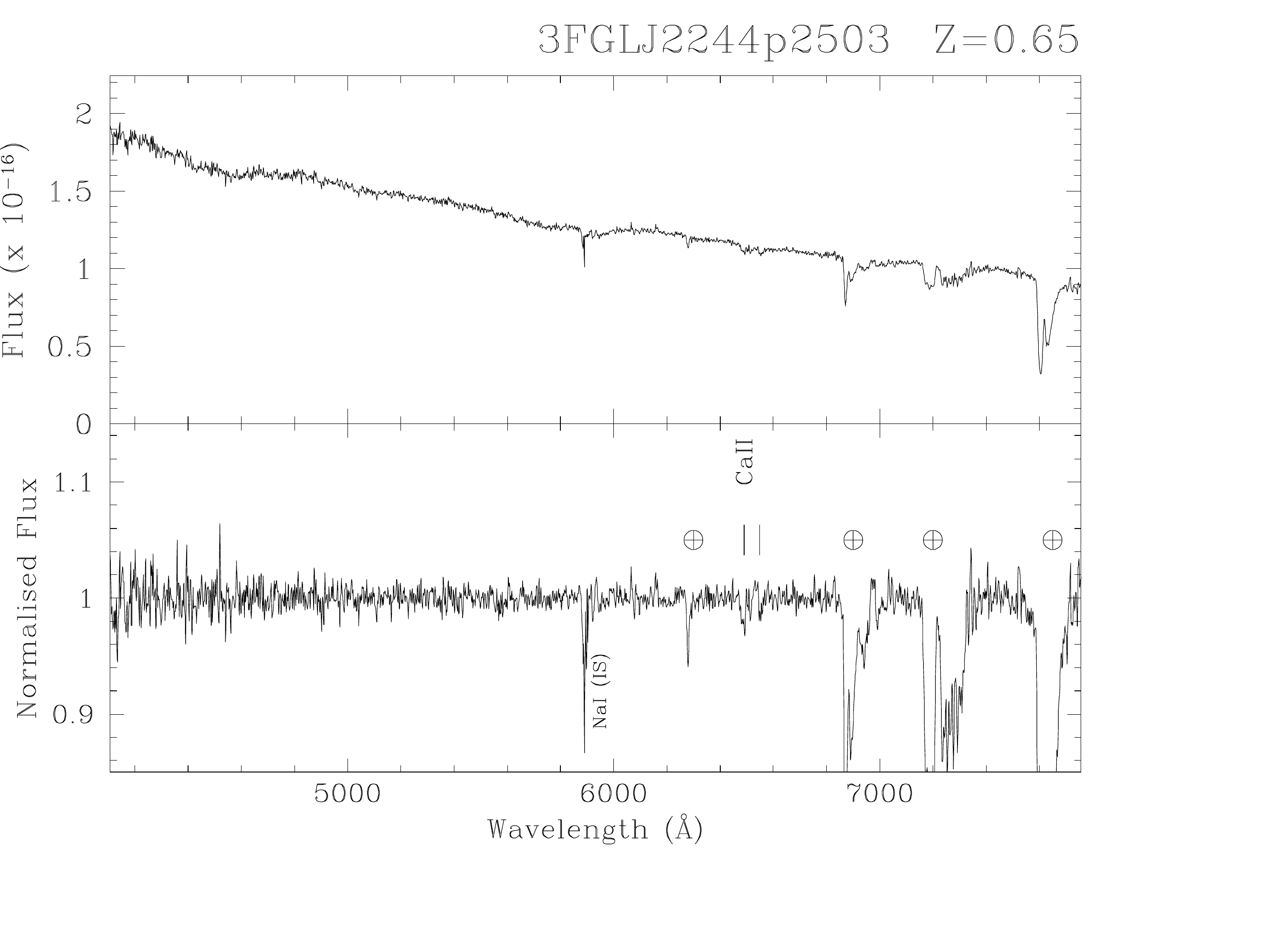}
\includegraphics[width=0.4\textwidth, angle=-90]{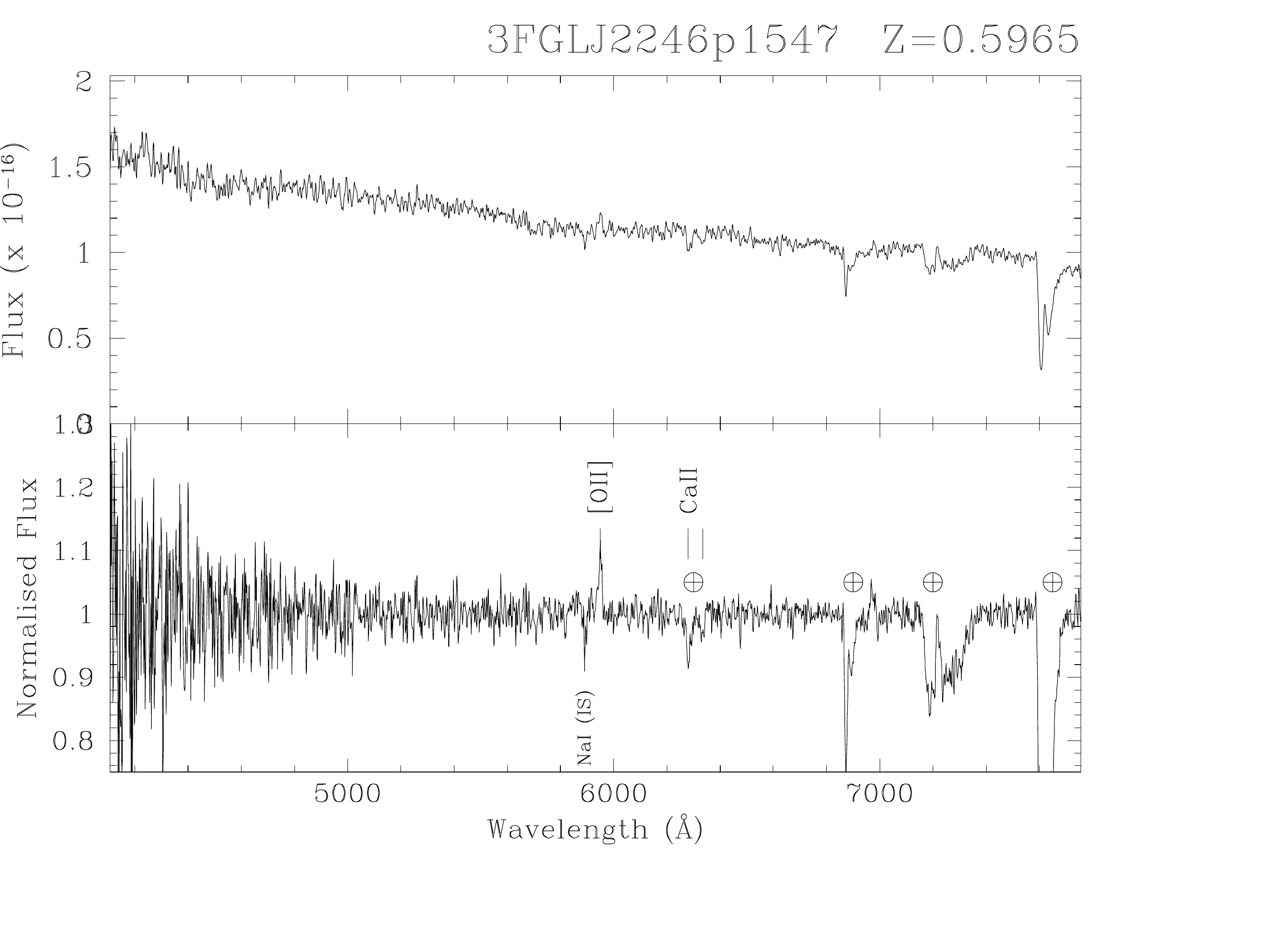}
\caption{Continued from Fig. \ref{fig:fig1}.}
\end{figure*}

\setcounter{figure}{1}
\begin{figure*}
\includegraphics[width=0.4\textwidth, angle=-90]{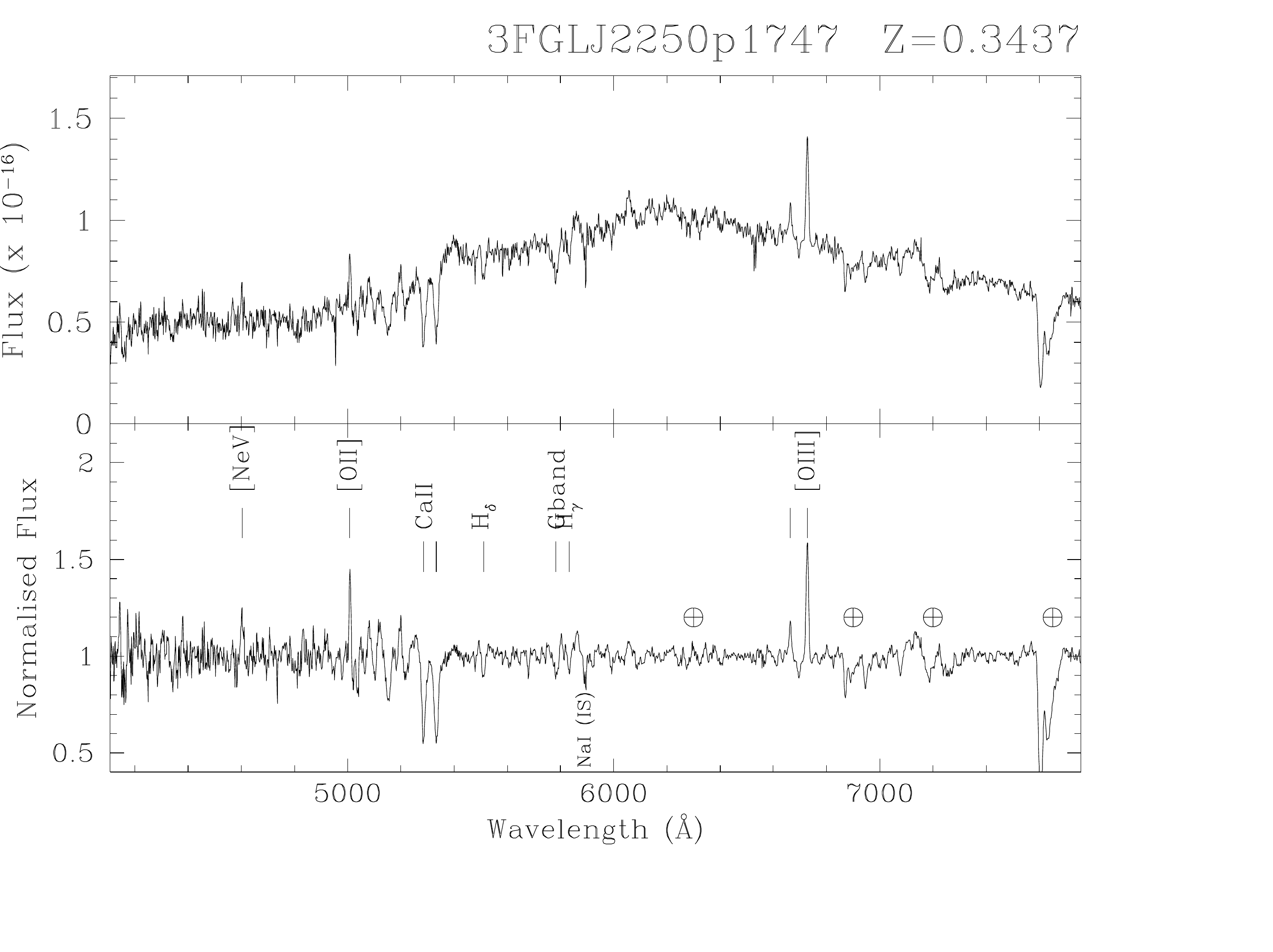}
\includegraphics[width=0.4\textwidth, angle=-90]{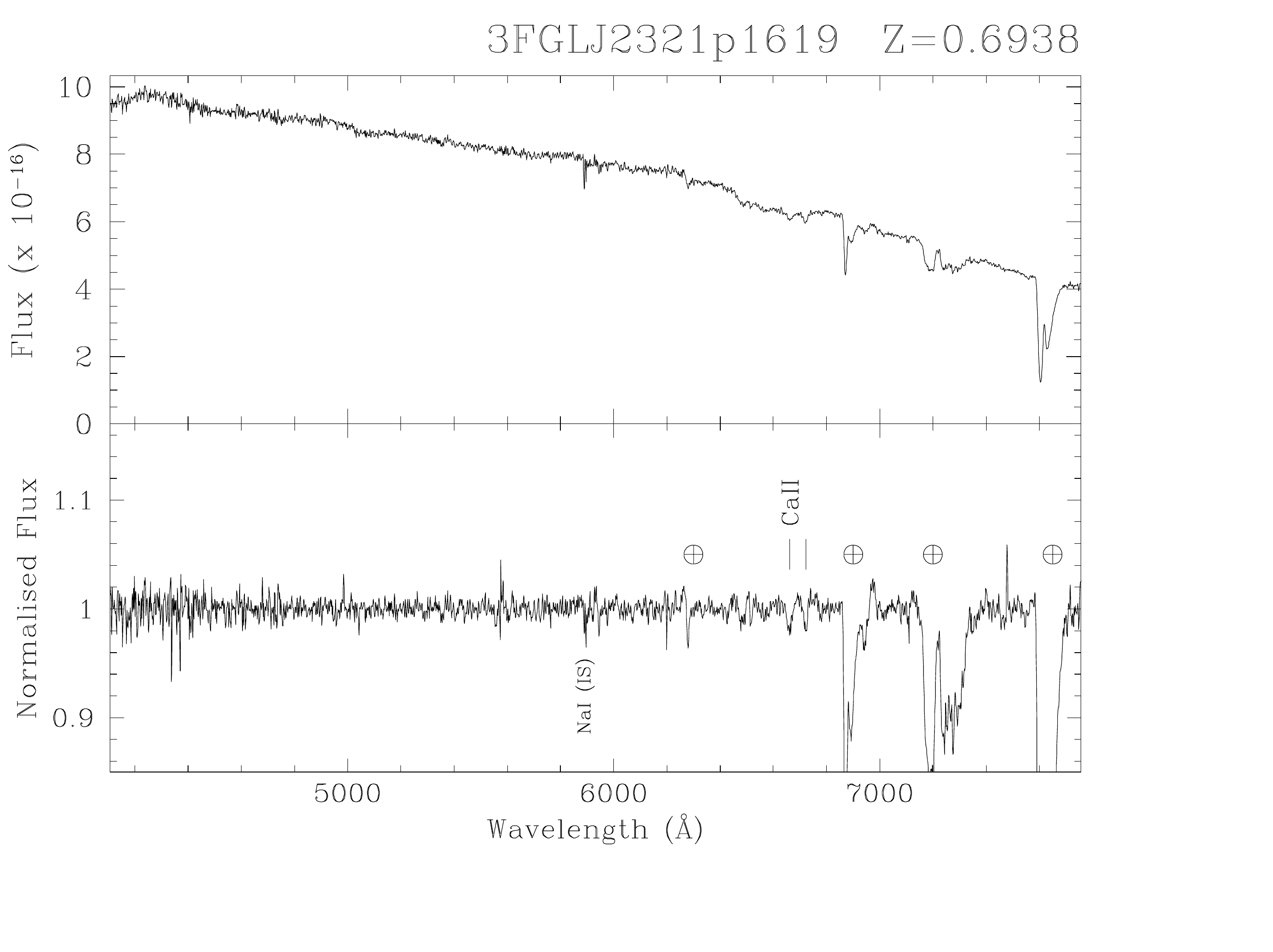}
\includegraphics[width=0.4\textwidth, angle=-90]{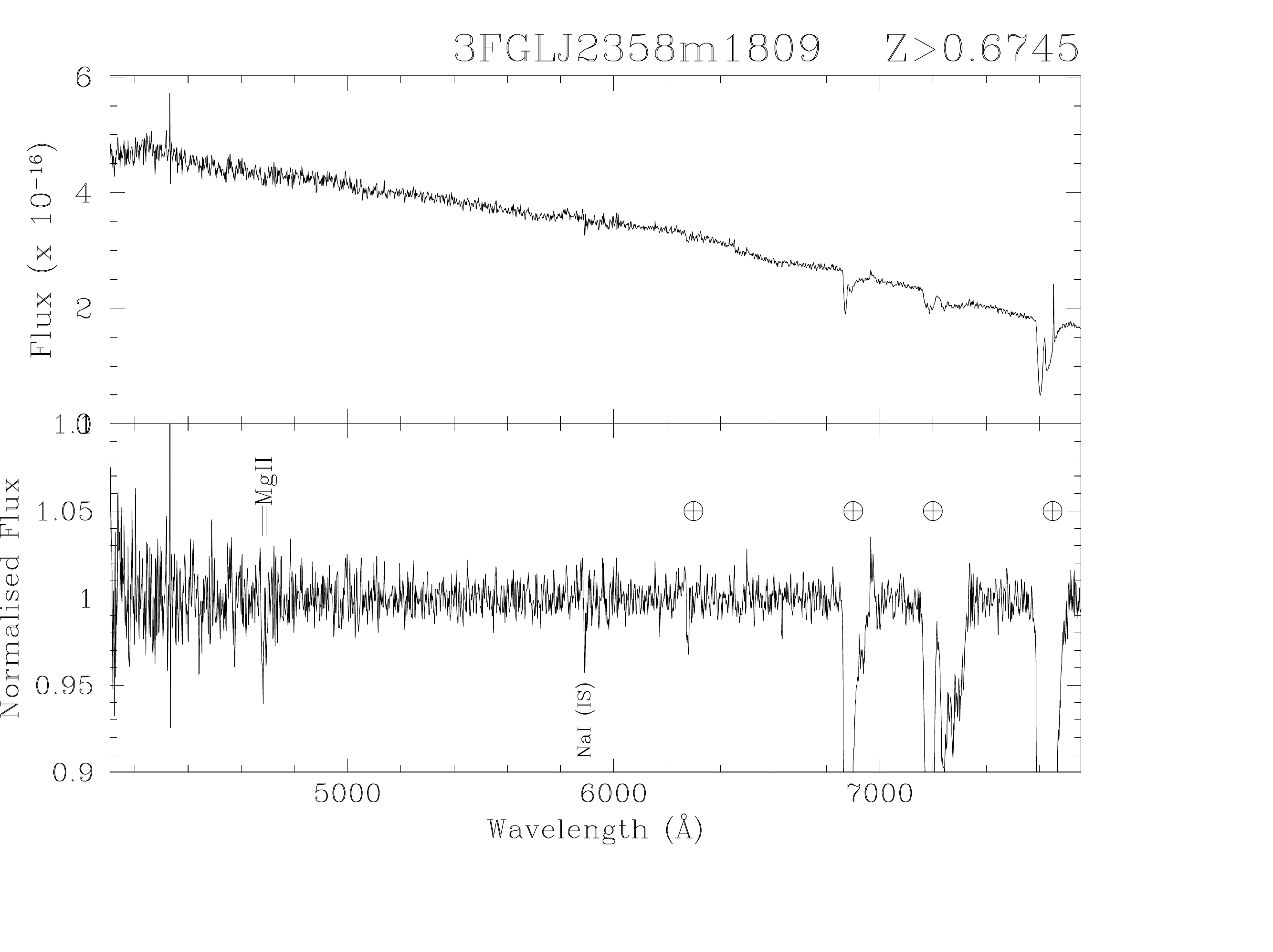}
\includegraphics[width=0.4\textwidth, angle=-90]{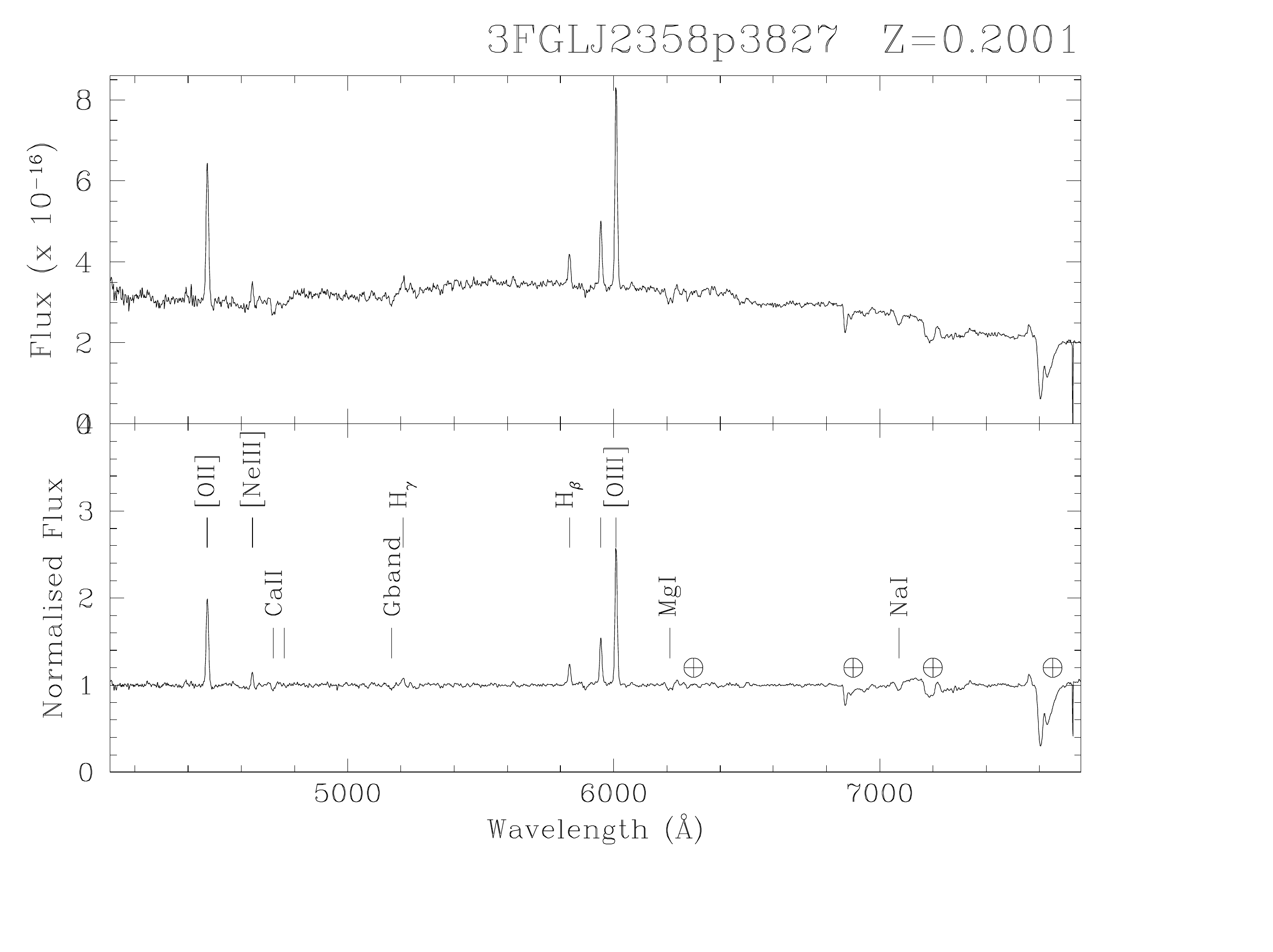}
\caption{Continued from Fig. \ref{fig:fig1}.}
\end{figure*}

\newpage

\setcounter{figure}{2} 
\begin{figure*}
 \includegraphics[width=1.0\textwidth, angle=-90]{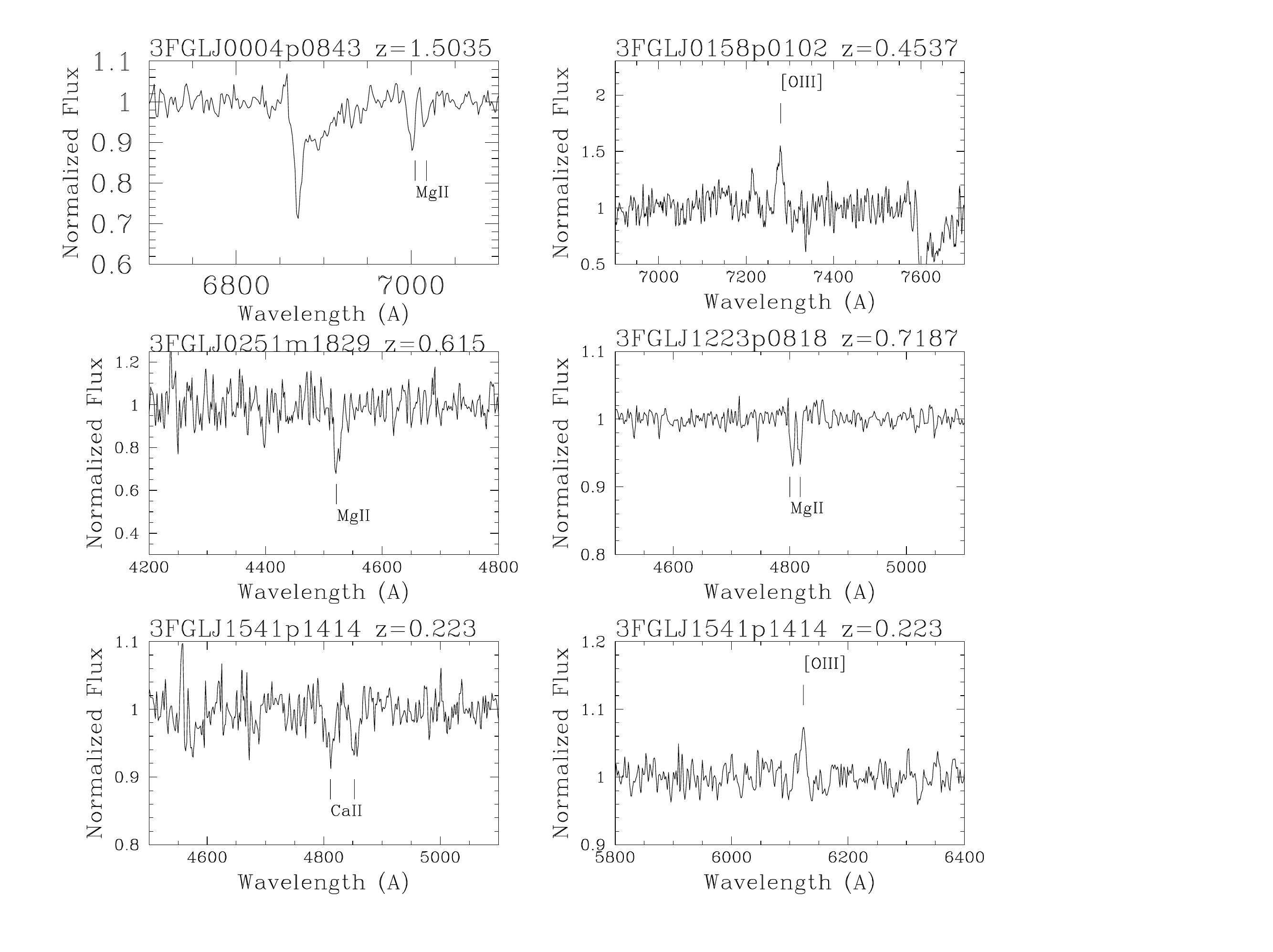}
 \caption{Close-up of the normalized spectra around the faintest detected spectral features of the UGSs obtained at GTC. Main telluric bands are indicated as $\oplus$, spectral lines are marked by line identification.} 
   \label{fig:fig2}
\end{figure*}

\newpage
\setcounter{figure}{2}
\begin{figure*}
 \includegraphics[width=1.0\textwidth, angle=-90]{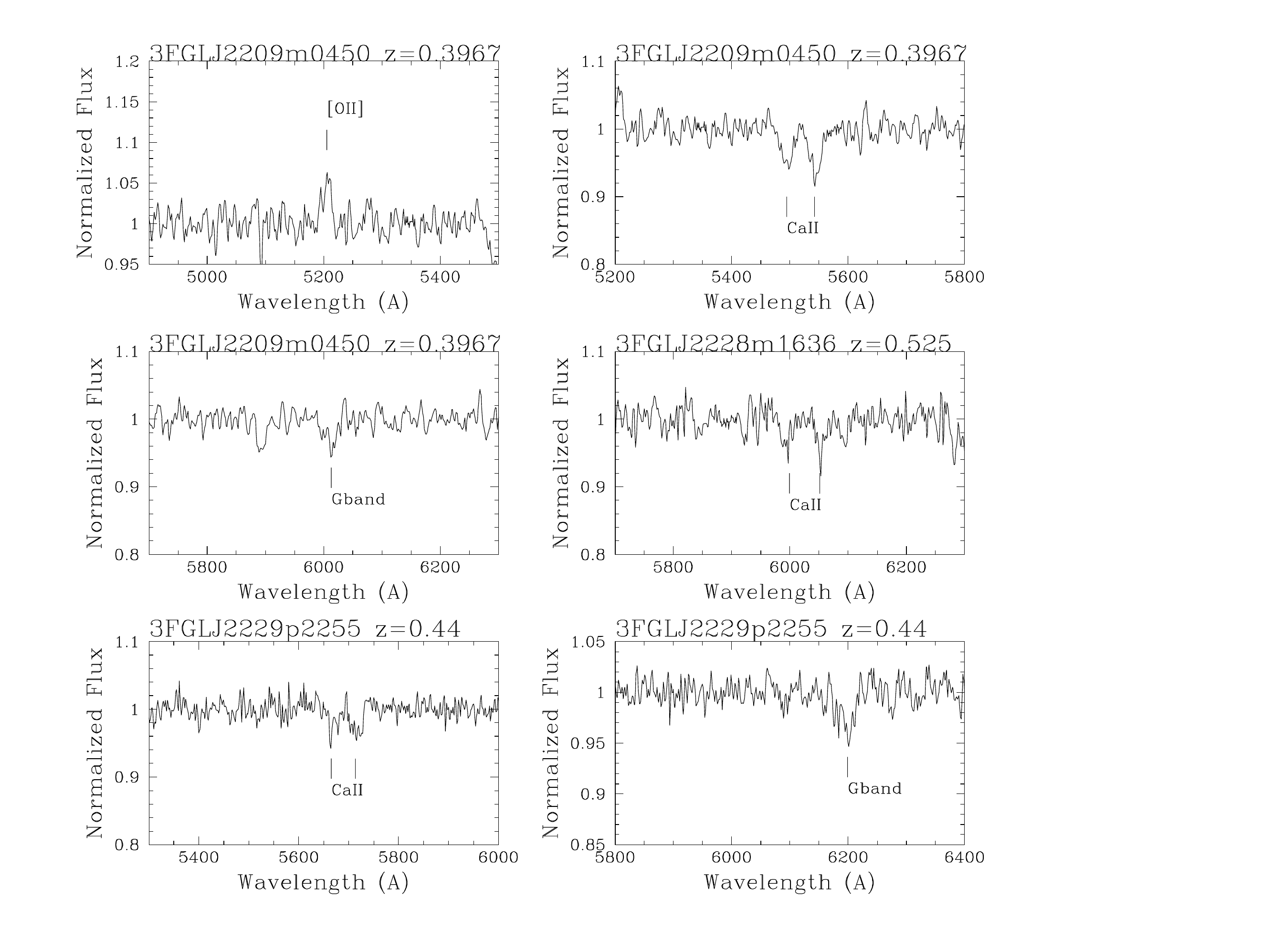}
\caption{Continued from Fig. \ref{fig:fig2}.}
  \end{figure*}

\newpage
\setcounter{figure}{2}
\begin{figure*}
 \includegraphics[width=1.0\textwidth, angle=-90]{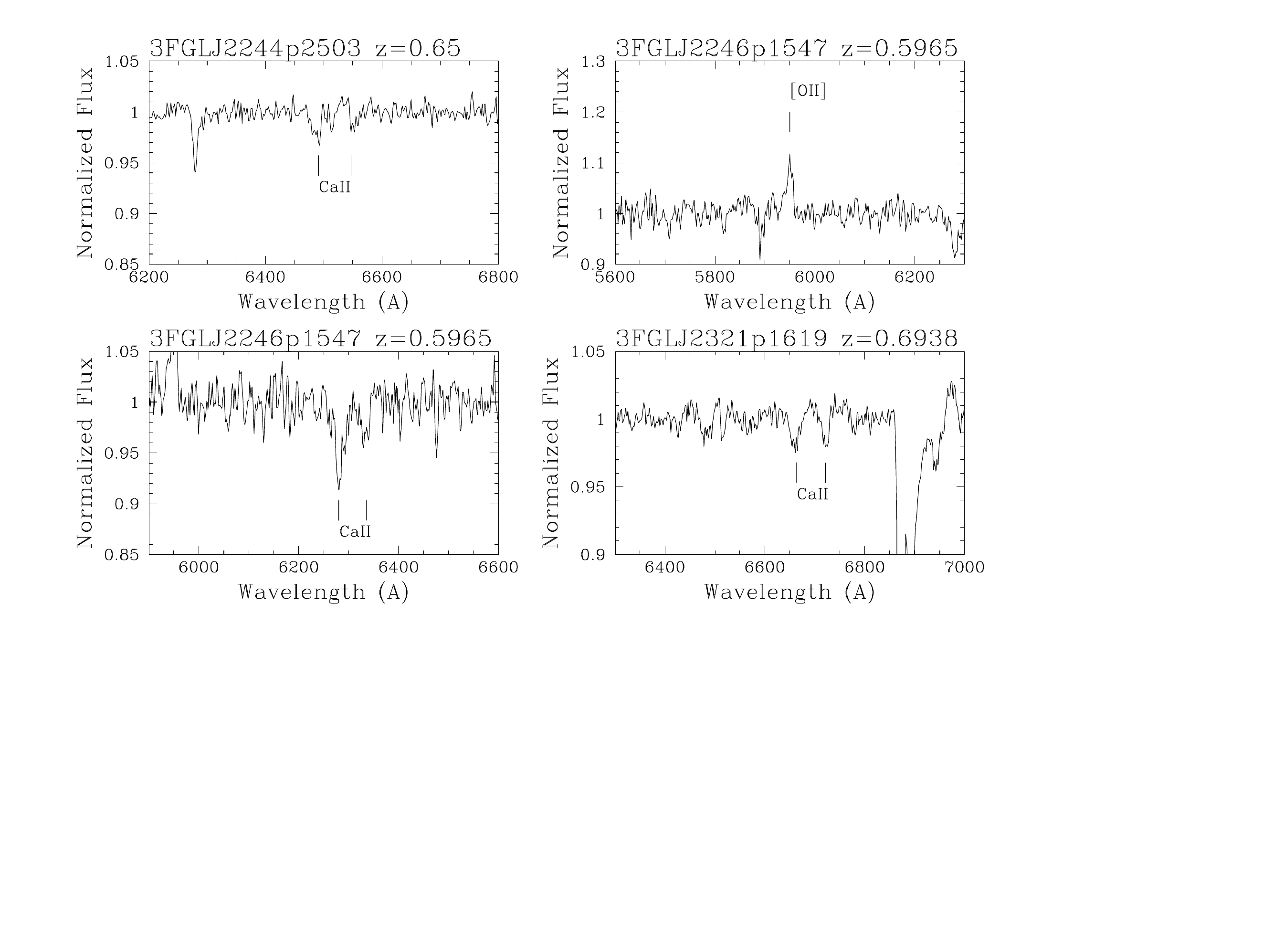}
\caption{Continued from Fig. \ref{fig:fig2}.}
  \end{figure*}

\clearpage
\newpage

\acknowledgments
We acknowledge an anonymous referee for her/his constructive suggestions and comments that allowed us to improve our paper. 

We acknowledge support from INAF under PRIN SKA/CTA ``FORECaST''

The financial contribution by the contract \textit{Studio e Simulazioni di Osservazioni (Immagini e Spettri) con MICADO per E-ELT} (DD 27/2016 - Ob. Fun. 1.05.02.17)  of the INAF project \textit{Micado simulazioni casi scientifici} is acknowledged.
%

\facilities{GTC-OSIRIS, \citep{cepa2003}}

\software{IRAF \citep{tody1986, tody1993}}

\bibliographystyle{aasjournal}
\bibliography{UFO2_GTC_biblio}




\end{document}